\documentclass[10pt]{iopart}

\usepackage{graphicx}

\usepackage{amsfonts}
\usepackage{enumerate}
\newcommand{\argsup}{\mathrm{argsup}}

\newcommand{\rmD}{{\rm D}}

\usepackage{color}

\newcommand{\bb}{\ensuremath{\mathbf{b}}}
\newcommand{\bsigma}{{\mbox{\boldmath $\sigma$}}}

\begin{document}

\title[Statistical mechanics of  lymphocyte networks modelled with slow and fast variables]{Statistical mechanics of clonal expansion\\ in  lymphocyte networks modelled with\\ slow and fast variables} %Statistical Physics/Mechanics,  clonal expansion,  clonal distribution, adaptive immune system, interacting system , Fast-Slow system, Langevin dynamics, lymphocyte network

\author{Alexander Mozeika$^{\dag}$ and Anthony CC Coolen$^{\dag\ddag}$}
\address{$\dag$Institute for Mathematical and Molecular Biomedicine, King's College London, Hodgkin Building, London SE1 1UL, UK\\
$\ddag$ Department of Mathematics,  King's College London, The Strand, London WC2R 2LS, UK }

% 05.50.+q	Lattice theory and statistics (Ising, Potts, etc.) 
%05.70.Fh	Phase transitions: general studies
%02.50.-r	Probability theory, stochastic processes, and statistics
%87.18.Vf	Systems biology
%75.10.Nr	Spin-glass and other random models (for spin glasses and other random magnets, see 75.50.Lk)

\ead{alexander.mozeika@kcl.ac.uk, ton.coolen@kcl.ac.uk}
\begin{abstract}
We  use statistical mechanical techniques to model the adaptive immune system, represented by lymphocyte networks in which B cells interact with T cells and  antigen.  We assume that B- and T-clones evolve in different thermal noise environments and on different timescales, and derive stationary distributions and  study  expansion of B clones for the case where these timescales are  adiabatically separated.   We compute characteristics of B-clone sizes, such as average  concentrations,  in parameter regimes where T-clone sizes are modelled as binary variables.  This analysis is independent of the network topology, and its results are qualitatively consistent with experimental observations. To obtain the full distributions of B-clone sizes we assume further that the network topologies are random and locally equivalent to trees. This allows us to compete these  distributions via the Bethe-Peierls approach.  As an example we calculate B-clone distributions for immune models defined on random regular networks. 
\end{abstract}

%the former (1st of two) and the latter (2nd of two)
\section{Introduction\label{section:Intro}}
 %Immune System:
 The main task of the  immune system is to defend an organism from invading pathogens such as viruses, bacteria, parasites, etc.  In complicated  multicellular organisms, such as vertebrates, the immune system is usually divided into two subsystems: the innate immune system and  the adaptive immune system~\cite{Janeway2012}.  The former can be seen as a first line of defence on which the organism relies for protection in the first hours and days of infection with a new pathogen, but its immune response is not specific to this particular pathogen. The latter is a second line of defence, which is usually triggered by the innate immune system,  whose immune response is more specific and also offers  a more 
 long-term protection. These properties of the adaptive immune system arise from its ability  to  learn and memorise a wide range of pathogens,  an important part of which is learning to  recognise  the molecules of the organism~\cite{Borghans1999},  which emerges from  the interactions of its cells,  mediated by signalling proteins (\emph{cytokines}).
 
 %Ingredients and interactions of  immune system:
 Interactions between B cells and T cells are dominant in  the adaptive immune response~\cite{Janeway2012}.   The main feature of a B cell is its B cell receptor  (BCR) which is used to recognise  antigen (Ag).  \emph{Antigen} is a (unique) protein on the surface of the pathogen.  However,  a BCR recognises only a specific part of the  Ag, which is called \emph{epitope}, so potentially a single BCR can recognise many different Ags~\cite{Notkins2004, Dimitrov2013}.  An Ag that binds to the BCR is subsequently internalised  and is  broken by the cell into  \emph{peptides} (short proteins) which are then displayed on its surface.   The T cell also has its T cell receptor (TCR) to recognise Ag, however,  in contrast to B cells, a T cell binds to the peptides on the surface of  antigen presenting cells (APCs), such as dendritic cells, B cells, etc.
The adaptive immune response  is triggered when T \emph{helper} cells that are attached by their TCRs to B cells,  activate these B cells (by cytokine-mediated signals)  that recognise the same Ag. The strength of this interaction between the BCR (TCR) and the Ag is called  \emph{affinity}.  Upon activation, the B cells with highest affinity begin to  proliferate in a process known as clonal expansion.  A group of B cells (or T cells) with the same BCR (or TCR) is called a \emph{clone}.  While B cell clones (B clones) are expanding the T cells,  activated by Ag, they are also going through the process of clonal expansion.  We note that the mode of B cell activation described here is called ``T-dependent B cell activation'', however for some ``simple'' Ags with repetitive patterns, such as polysaccharides, the B cell can also be activated directly by  Ag and without the help from a T  cell (``T-independent activation''). 

Some proliferating B cells differentiate into \emph{plasma} cells, which secrete large quantities of  \emph{antibodies} (BCRs in soluble form).  The antibodies (Ab) protect the host from infection by binding to the pathogen and thus, by blocking its external parts, make  the  attack ineffective or ``mark'' the pathogen  for ingestion by other host cells.   However,   the Abs produced in this initial clonal expansion are of very low affinity, especially if the host is infected with this pathogen for the first time. 
The other proliferating B cells are used to seed the \emph{germinal centres} (GCs).  GCs are special micro-environments, formed within secondary lymphoid organs, which are divided into the light and dark zones. In the dark zone  B cells undergo \emph{somatic hypermutation}  which introduces random mutations into the genes of BCRs.  This process is followed by the interaction of B cells with the helper cells and Ag in the light zone  of the GC. Here B cells with higher affinity are selected and differentiated into plasma cells  and \emph{memory} B cells.  Presence of high affinity memory B cells in the organism leads to a more vigorous immune response on the subsequent infections by the same Ag.

%Statistical Physics of the adaptive immune system:
Statistical physics (SP) approaches  to modelling the adaptive immune system~\cite{Perelson1997, Chakraborty2010} are predominantly  based on the idea that the immune system can be viewed as a network (or graph).  The nodes in such networks are \emph{lymphocyte clones} (B cells and T cells),  antibodies, antigens,  etc.  and the edges  model interactions such as B cells receiving ``signals'' from T cells, antibodies binding to antigen, etc.  One of the first such studies was~\cite{Parisi1990}, describing a network of interacting antibodies~\cite{Sulzer1999}, i.e. an \emph{idiotypic} network.  Despite being very simple (binary variables assigned to the nodes represented  Ab concentrations, and the interactions  were assumed  to be random), this model exhibited  memory, which was interpreted as the (immunological) memory of the past exposure to the Ag.
A more detailed model with lymphocytes and cytokines was introduced in~\cite{Agliari2011}.  Here the (bipartite) network was formed by interactions between effector clones, formed by  B cells and killer T cells,   and helper clones (T cells), mediated by cytokines (strengths of (random) interactions in the model).  The immunological memory in this model arises from the ability of T clones to memorise an extensive number of  cytokine signalling ``strategies'' (or patterns), allowing to cope with possible Ag stimulation. However,  the fully connected network topology used in this model restricted the immune memory to retrieval of  only one pattern at a time, limiting the abilities of the immune system to fight against multiple Ags~\cite{Agliari2011}.  The finitely connected network topology, which is more realistic from the biological point of view~\cite{Agliari2013},  allows to retrieve an extensive number of patterns in parallel. 

 The above studies all used the equilibrium SP framework, and assumed existence of an energy function, governing behaviour of Abs, lymphocytes, etc., from the very beginning. However, whenever  it is not clear that such a function exists, or if we want to study relaxation to equilibrium,  we must study dynamics. Examples of studies of the dynamics of immune idiotypic networks  and  bipartite lymphocyte networks are~\cite{Uezu2006}  and~\cite{ Bartolucci2015}.  
 Finally,  we note that SP models and concepts  are also used in statistical inference of  immune system data. The maximum entropy model of amino acid sequences was used to study the  repertoire of memory B cell receptors in zebrafish~\cite{Mora2010}.  Recently, similar approaches were applied to study the repertoires of T  cells~\cite{Murugan2012} and B cells~\cite{Elhanati2015} in humans. 
 
 %Motivation and results of this work: 
 In this paper we  develop further the lymphocyte network model~\cite{Agliari2011, Agliari2013}, in both qualitative and quantitative directions.  The main difference between the current and previous versions of the model is that we relax the assumption that B clones, represented by log-concentrations of B cells, and T cell clones (T clones),  represented by concentrations of T cells, are subject to the same thermal noise~\cite{ Agliari2013}.  We note that both populations of B cells and T cells are affected by various random events such as stochasticity in cell division and cell death~\cite{Duffy2012},  thermal fluctuations in the TCR-peptide~\cite{Qi2001, Raychaudhuri2003, Merwe2000, Yokosuka2005} bond strengths~\cite{Bush2014}, etc.  Also,  these are populations of biochemically, and possibly physically~\cite{Strokotov2009}, distinct cells.  Furthermore, in the T-dependent immune response the magnitude of fluctuations in the populations of B cells and T cells are expected to be very different due to the fact that the former are subject to the process of somatic hypermutation and the latter are not~\cite{Janeway2012}. For the above reasons  it is implausible  for  the random noise in  B  and T  clones to be of identical  strength.
 
 Having different thermal noise levels, i.e. different ``temperatures'',  in the B and T clone evolution introduces some  technical difficulties in  the analysis of the problem. In contrast to previous studies~\cite{Agliari2011, Agliari2013},  it prevents us from using the equilibrium framework directly.   However, we can make progress by assuming that  either B  clones or T clones are ``fast'' variables~\cite{Freidlin2012}, i.e. they are evolving on different time-scales.  The latter  allows us to obtain the stationary distribution of a  two-temperature system in its explicit form~\cite{Penney1993},  unlike the situation with  the same or comparable time-scales~\cite{Dotsenko2013}.  
 
 Furthermore, previous studies did not distinguish the  T helper  cells from the T \emph{regulator}  cells~\cite{Josefowicz2012}.  The latter play an important role in preventing autoimmunity by suppressing  the growth of  self-reactive B clones which could be produced during the T-dependent immune responses~\cite{Vanderleyden2014}.   Also, B cells infiltrate tumours  and there is an evidence for \emph{in situ} immune responses~\cite{Coronella2002}, to tumour-associated (self) Ags,  which are thought to be strongly influenced by the  presence of regulator  T  cells ~\cite{Zou2006}. 

Finally,  we note that  our main focus in this paper will be on the properties of B clone distributions and how these are affected by the parameters of the model, such as network  topology, concentration of Ag, etc.  Existing models  of clone distributions usually disregard such details  (see~\cite{Desponds2016} and references therein) and do not explicitly include interactions between the B cells and T cells dominant in the  T-dependent immune response.   Recently,  the importance of these properties  in our understanding of the ageing immune system was emphasised in~\cite{Wu2012} and we also envisage that  these properties are important in tumour immunology~\cite{Coronella2002}.  

\section{Dynamics\label{section:Dynamics}}
 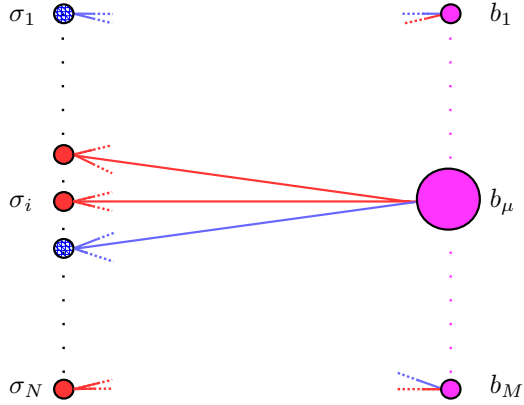
\begin{figure}[t]
 %\vspace*{-7mm}
 \setlength{\unitlength}{0.9mm}
 \begin{center}{
 %\hspace*{50mm}
 \begin{picture}(100,71)
 \put(0,0){\includegraphics[height=75\unitlength]{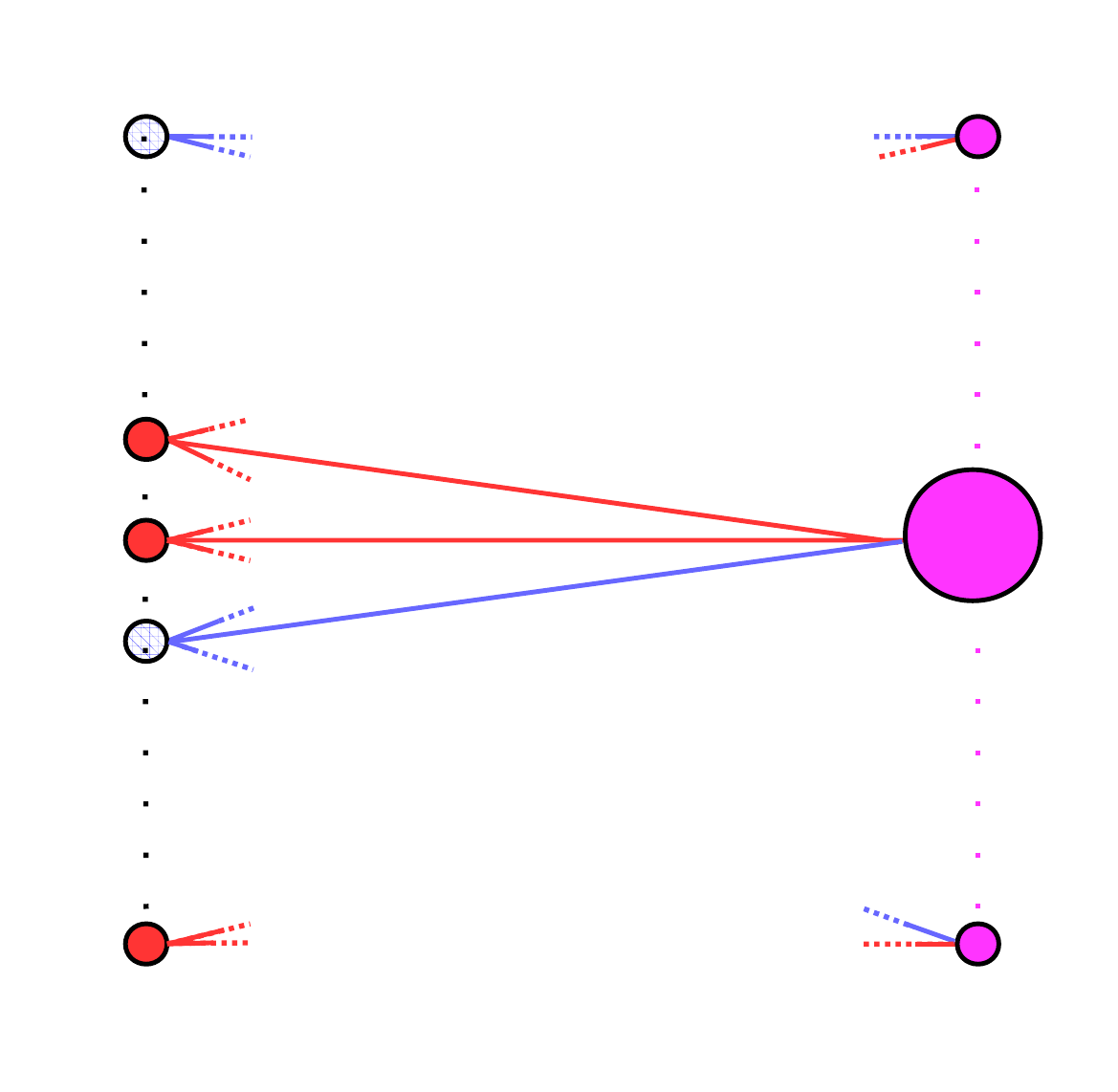}}
 
  %\put(5, 70 ){T-clones} 
 
 \put(2, 64.5 ){$\sigma_1$} 
 
\put(2 ,37){$\sigma_i$}
 
 \put(2 ,9.5){$\sigma_N$}

 \put(73, 64.5 ){$b_1$} 
 
\put(73 ,37){$b_\mu$}
 
 \put(73 ,9.5){$b_M$}
 \end{picture}
 }\end{center}
 \vspace*{-7mm}
 \caption{Bi-partite network of  T-clones and B-clones, generated by the T-dependent  immune response. A red link between a (helper)  T-clone  $i$   and a B-clone $\mu$ represent that this B-clone receiving a signal  to expand. A blue link between a (regulator) T-clone  $i$   and B-clone $\mu$ represents a signal  to contract.}
 \label{figure:model} 
 \end{figure}
We consider $M$ B-clones  interacting with $N$  T-clones  on a bipartite graph $\mathcal{G}=(\mathcal{V}, \mathcal{F},  \mathcal{E})$, where $N\!=\!\vert\mathcal{V}\vert$,  $M\!=\!\vert\mathcal{F}\vert$ and $\mathcal{E}$ is the set of edges (see Figure \ref{figure:model}). The set of indices  $\partial \mu$ contains all T-clones $i$ connected to B-clone $\mu$, and the set $\partial i$ defines all  B-clones $\mu$ that are connected to T-clone $i$. The B-clone sizes are specified by the log-concentrations $\bb=(b_1, \ldots, b_M)$, and we assume that these are governed by the  Langevin  equation 
\begin{eqnarray}
\tau_b\frac{\rmd b_\mu}{\rmd t}&=& J_\mu\left(\sum_{i\in \partial \mu}\xi_i^\mu \sigma_i + \theta_\mu\right) - \rho b_\mu  +\chi_\mu(t)\label{eq:B-clone-dynamics}
\end{eqnarray}
where the zero-average Gaussian noise  $\chi_\mu(t)$, with  $\langle\chi_\mu(t)\chi_\nu(t^\prime)\rangle=2\tau_b\tilde{\beta}^{-1}\delta_{\mu\nu}\delta(t-t^\prime)$,   is characterised by a  ``temperature'' parameter $\tilde{T}=\tilde{\beta}^{-1}$. In this dynamics the $\mu$-th B-clone receives the ``signal''  $\sum_{i\in \partial \mu}\xi_i^\mu \sigma_i $ from $\vert\partial\mu\vert$ T-clones,    whose sizes are specified by the concentrations  $\bsigma=(\sigma_1,\ldots, \sigma_N)$.  The $i$-th T-clone is  either formed by T-helper  or by T-regulator cells. The efficacies of the T-helper and T-regulator clones are encoded, respectively, by ``cytokine'' variables $\xi_i^\mu > 0$ and  $\xi_i^\mu < 0$.  The signal $\sum_{i\in \partial \mu}\xi_i^\mu \sigma_i $ from the T-clones   is modulated by the interaction  strength $J_\mu(\mathbf{a})=\sum_{\nu\leq M} S_{\mu\nu} a_\mu$, which depends on the Ags, as represented by the vector of epitope ``concentrations''  $\mathbf{a}=(a_1,\ldots,a_M)$.  Here $S_{\mu\nu}\geq0$ 
is an element of  an ``affinity'' matrix which specifies how  well the  $\nu$-th epitope  is ``matched'' by the  $\mu$-th B-clone.  A very specialised  B-clone will interact with only one epitope (i.e. $S_{\mu\nu}=\delta_{\mu\nu}$), whereas  \emph{poly-reactive} B-clones~\cite{Notkins2004, Dimitrov2013} can interact with many different epitopes (e.g. $S_{\mu\nu}>0$ for all $\nu$). The $\theta_\mu$ term gives the possibility of B-clone activation even in the absence of a  signal (or in the presence of only a weak signal) from the T  clones, i.e. it facilitates T-independent activation. 

A positive (negative)  ``field'' $F_\mu(\bsigma)=J_\mu\big(\sum_{i\in \partial \mu}\xi_i^\mu \sigma_i + \theta_\mu\big)$  has an excitatory (inhibitory) effect on the growth of $\mu$-th B-clone.    The strength of this effect  is increasing  with larger  amounts of Ag:  the interaction  $J_\mu(\mathbf{a})$ is a monotonic non-decreasing function of the antigen $\mathbf{a}$, i.e. $J_\mu(\mathbf{a})\geq J_\mu(\tilde\mathbf{a})$ for all $\mu$ when $a_\nu \geq\tilde{a}_\nu$ for all $\nu$.  Furthermore, the growth of  clone $\mu$ is kept in check by the  ``apoptosis''  therm $-\rho b_\mu$,  which limits its amplitude. For  $J_\mu=0$, i.e. without Ag,  the distribution of  log-concentration $b_\mu$  in equilibrium takes the Gaussian form $p(b_\mu)=(2\pi/\rho\tilde{\beta})^{\frac{1}{2}}\exp[-\frac{1}{2}\rho\tilde{\beta}b^2_\mu]$. 

In order to derive a dynamical equation for the  T-clones  we will follow the ideas of \cite{Agliari2011,Agliari2013}. Firstly,  we note that if we define the energy function  (or Hamiltonian) 
\begin{eqnarray}
\mathcal{H}(\bb, \bsigma)&=&-\sum_{\mu=1}^{M} b_\mu F_\mu(\bsigma) +   \frac{1}{2}\rho \sum_{\mu=1}^{M} b^2_\mu ,\label{def:H} 
\end{eqnarray}
then equation (\ref{eq:B-clone-dynamics}) can be written in the form 
\begin{eqnarray}
\tau_b\frac{\rmd b_\mu}{\rmd t}&=& -\frac{\partial}{\partial b_\mu}  \mathcal{H}(\bb, \bsigma) +\chi_\mu(t).\label{eq:B-clone-dynamics-H}
\end{eqnarray}
We note that this dynamics is invariant under the transformation $\mathcal{H}(\bb, \bsigma)\rightarrow\mathcal{H}(\bb, \bsigma)+V(\bsigma)$ where $V(\bsigma)$ is any function of $\bsigma$. Secondly, we assume that 
\begin{eqnarray}
\tau_\sigma\frac{\rmd \sigma_i}{\rmd t}&=&  -\frac{\partial}{\partial \sigma_i}    \mathcal{H}(\bb, \bsigma) +\eta_i(t),\label{eq:T-clone-dynamics-H}
\end{eqnarray}
where  the zero-average  Gaussian noise  $\eta_i(t)$, with $\langle\eta_i(t)\eta_j(t^\prime)\rangle=2\tau_\sigma\beta^{-1}\delta_{ij}\delta(t-t^\prime)$, is chaacterised by a ``temperature''    $T=\beta^{-1}$. From the above follows the equation 
\begin{eqnarray}
\tau_\sigma\frac{\rmd \sigma_i}{\rmd t}&=& \sum_{\mu\in \partial i}J_\mu \xi_i^\mu b_\mu  -\frac{\partial}{\partial \sigma_i}V(\bsigma) + \eta_i(t).\label{eq:T-clone-dynamics} %-4\Delta\sigma_i\left(\sigma^2_i-1\right)^2
\end{eqnarray}
For now we will leave the function $V(\bsigma)$ unspecified -- it  will be used later to ``restrict'' the range of $\sigma_i$'s -- while allowing us to define various thermodynamic functions.  We note that the advantage of assuming that T-clones are governed by the same Hamiltonian as  B-clones is that it allows us to use analytical tools from  equilibrium statistical mechanics.  The disadvantage is that wether  this  approach is correct  or not, as  in any other phenomenological  approach,  can be only established \emph{a posteriori}.   One of the consequences of using the present  approach, which is  explicit in the T-clone equation (\ref{eq:T-clone-dynamics}),  is that, in the presence of Ag, the evolution of the T-clones is governed by the B-clones. The latter can be interpreted as  ``B-cells acting as Ag presentation cells (APCs) for the T-cells'' which is a well known immunological fact~\cite{Janeway2012}.  However, besides B-cells there are other dedicated  APCs, such as dendritic cells, etc. The latter can be included via the potential $V$.  Finally, we note that the   relation between the  time-scales of the T- and B-clone subsystems ($\tau_b\gg\tau_\sigma$ versus $\tau_\sigma\gg\tau_b$)   give us two possible scenarios to analyse the equilibrium state~\cite{Coolen1993}. 

\subsection{Fast equilibration of  B-clones\label{section:B-fast}}

We first assume that the B-clone variables $b_\mu$ are ``fast'' variables, and equilibrate on timescales much shorter than those characterising the evolution of the T-clones. In equilibrium the former will then be governed by the distribution 
\begin{eqnarray}
P_{\tilde{\beta}}(\bb\vert \bsigma)&=& \frac{1}{Z_{\tilde{\beta}}(\bsigma)}\rme^{-\tilde{\beta}\mathcal{H}(\bb, \bsigma)},\label{eq:P(b|s)}
\end{eqnarray}
where $Z_{\tilde{\beta}}(\bsigma)=\int\rmd\bb\,\rme^{-\tilde{\beta}\mathcal{H}(\bb, \bsigma)}$ is a partition function. From this it then follows that the dynamic equation  (\ref{eq:T-clone-dynamics-H})  becomes 
\footnote{We note that the `implicit averaging' procedure used here is exact  when $\tau_b\rightarrow 0$ ~\cite{Freidlin2012}. } 
\begin{eqnarray}
\frac{\rmd \sigma_i}{\rmd t}&=&  -\left\langle\frac{\partial}{\partial \sigma_i}    \mathcal{H}(\bb, \bsigma)\right\rangle_{\tilde{\beta}} +\eta_i(t)\label{eq:Langevin_sigma}\\
&=&  -\frac{\partial}{\partial \sigma_i}    \mathcal{F}_{\tilde\beta}(\bsigma)  +\eta_i(t),\nonumber
\end{eqnarray}
where $\mathcal{F}_{\tilde\beta}(\bsigma)=-\tilde{\beta}^{-1}\log Z_{\tilde{\beta}}(\bsigma)$ is (formally) a free energy of  a state $\bsigma$. The above subsequently implies that in equilibrium the T-clones are  governed by the distribution
\begin{eqnarray}
P_{\beta, \tilde{\beta}}(\bsigma)&=& \frac{1}{Z_{\beta,\tilde{\beta}}}\rme^{-\beta \mathcal{F}_{\tilde\beta}(\bsigma)    },\label{eq:P(s)}
\end{eqnarray}
where $Z_{\beta,\tilde{\beta}}=\int\rmd\bsigma\,\rme^{-\beta \mathcal{F}_{\tilde\beta}(\bsigma)}$.  

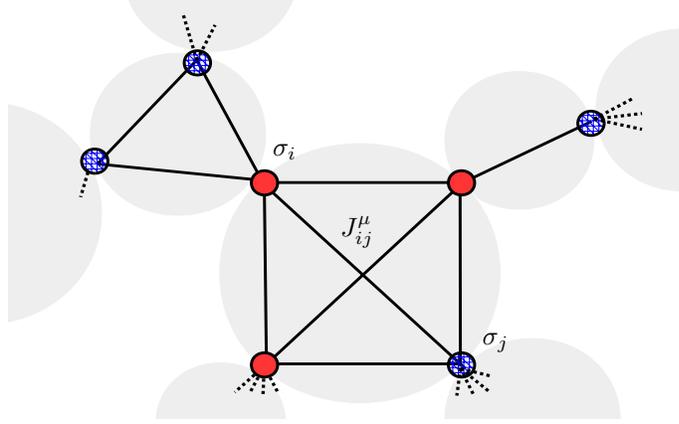
\begin{figure}[t]
 %\vspace*{-7mm}
 \setlength{\unitlength}{0.9mm}
 \begin{center}{
 %\hspace*{50mm}
 \begin{picture}(100,65)
 \put(0,0){\includegraphics[height=62\unitlength, width=100\unitlength]{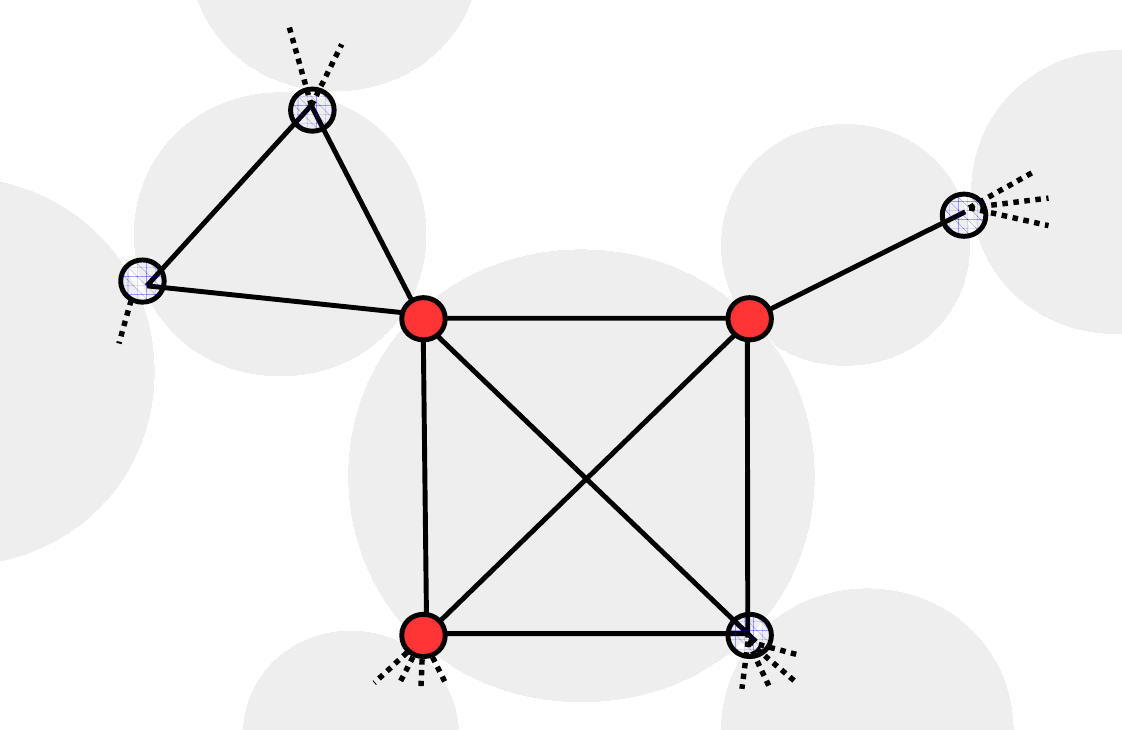}}%{Tsystem_shadow}}
% \put(40 ,40){$\sigma_i$} 
% \put(49 ,27){$J^{\mu}_{ij}$}
%  \put(73 ,9.5){$\sigma_j$}
\put(39 ,39){$\sigma_i$} 
 \put(49 ,27){$J^{\mu}_{ij}$}
  \put(70 ,11){$\sigma_j$}  
   \end{picture}
 }\end{center}
 \vspace*{-3mm}
 \caption{Example of a local topology in the  T-clone network generated by the system with  fast B-clone equilibration.  The evolution of B-clone $\mu$ (large grey circle) is governed by three T helper clones (red nodes) and one T regulator clone (blue node), forming a clique of four nodes. }
 \label{figure:Tsystem} 
 \end{figure}

From (\ref{eq:P(b|s)}) and (\ref{eq:P(s)}) we can construct the joint distribution 
\begin{eqnarray}
P_{\beta, \tilde{\beta}}(\bb, \bsigma)&=& P_{\tilde{\beta}}(\bb\vert \bsigma)P_{\beta, \tilde{\beta}}(\bsigma)\nonumber\\
&=&\Bigg\{\prod_{\mu=1}^{M}\frac{                    \rme^{-\frac{1}{2}\rho\tilde\beta\left(b_\mu - \frac{F_\mu(\bsigma)    }{\rho}\right)^2}                    }{\sqrt{2\pi/\rho\tilde\beta}} \Bigg\} 
\frac{\rme^{\frac{\beta   }{2\rho}\sum_{\mu=1}^{M}F^2_\mu(\bsigma)   }}{  \int \rmD\tilde{\bsigma}\,  \rme^{\frac{\beta   }{2\rho}\sum_{\mu=1}^{M}F^2_\mu(\tilde{\bsigma})    }         }\label{eq:P(b,s)-B-fast }
\end{eqnarray}
which in turn  allow us to construct the average density
\begin{eqnarray}
P(b)&=&\frac{1}{M}\sum_{\mu=1}^{M}\left\langle\delta(b- b_\mu )\right\rangle_{\beta, \tilde{\beta}}
=
\int\!\rmd F ~P(F)\frac{  \rme^{-\frac{1}{2}\rho \tilde\beta\left(b - \frac{F}{\rho}\right)^2}  }{\sqrt{2\pi/\rho \tilde\beta}}, 
\end{eqnarray}
where
\begin{eqnarray}
P(F)&=&
\int\! \rmD\bsigma
\frac{\rme^{\frac{\beta   }{2\rho}\sum_{\nu=1}^{M}F^2_\nu(\bsigma) }}{ \int\! \rmD\tilde{\bsigma}\,    \rme^{\frac{\beta   }{2\rho}\sum_{\nu=1}^{M}F^2_\nu(\tilde{\bsigma}) }         }\frac{1}{M}\sum_{\mu=1}^{M}\delta\left(F-F_\mu(\bsigma)\right)\label{def:P(F)}
\end{eqnarray}
is a distribution B-clone activation fields. It follows from the above that the number of B cells in a volume $V$, their concentration $c$, is governed by the distribution\footnote{For any distribution $P(b)$ we have $P(c)=\int P(b)\, \delta\!\left(c -\rme^b\right)\! \rmd b=\frac{1}{c}P\left(\log\left(c\right)\right)$.}
\begin{eqnarray}
P(c)&=&\int\!\rmd F~P(F)~ \frac{  \rme^{-\frac{1}{2}\rho \tilde\beta\left(\log\left(c\right)- \frac{F}{\rho}\right)^2}  }{c\sqrt{2\pi/\rho \tilde\beta}}.\label{eq:P(c)-B-fast}
\end{eqnarray}
We note that in equation (\ref{eq:P(b,s)-B-fast })  the ``Boltzmann'' factor $\exp[\frac{\beta   }{2\rho}\sum_{\mu\leq M}F^2_\mu(\bsigma)]$ can be written, up to an irrelevant  constant, as $\exp[-\beta \sum_{\mu\leq M}E_\mu(\bsigma)]$.  The ``energy'' function  $E_\mu(\bsigma)=-\frac{1}{2}\sum_{i\in\partial\mu}\sum_{j\in\partial\mu}J^\mu_{ij}\sigma_i\sigma_j -\sum_{i\in\partial\mu}\theta_i^\mu\sigma_i$, where $J^\mu_{ij}=(J_\mu^2/\rho) \xi_i^\mu\xi_j^\mu$ and  $\theta^\mu_{i}=(J_\mu^2/\rho) \xi_i^\mu\theta_\mu$,  can be  represented as a fully connected weighted graph (or clique) formed by those the T-clones that control B-clone $\mu$. Thus the B-clone dynamics (\ref{eq:B-clone-dynamics}), given the assumptions of this section,  leads us to the result that  in equilibrium the T-clones are interacting via this network of cliques  (see Figure  \ref{figure:Tsystem}).   

\subsection{Fast equilibration of  T-clones\label{section:T-fast}}

Here we assume, contrary to the previous subsection, that the T-clones  equilibrate first, and are upon equilibration governed by the distribution 
\begin{eqnarray}
P_{\beta}(\bsigma\vert \bb)&=& \frac{1}{Z_{\beta}(\bb)}\rme^{-\beta \mathcal{H}(\bb, \bsigma)}, \label{eq:P(s|b)}
\end{eqnarray}
where $Z_{\beta}(\bb)=\int\rmd\bsigma\,\rme^{-\beta \mathcal{H}(\bb, \bsigma)}$. The dynamics of B-clones  is then given by
\begin{eqnarray}
\frac{\rmd b_\mu}{\rmd t}&=& -\left\langle\frac{\partial}{\partial b_\mu}  \mathcal{H}(\bb, \bsigma)\right\rangle_\beta +\chi_\mu(t)\label{eq:Langevin_b}\\
&=& -\frac{\partial}{\partial b_\mu}  \mathcal{F}_\beta(\bb) +\chi_\mu(t),\nonumber
\end{eqnarray}
where $\mathcal{F}_\beta(\bb)=-\beta^{-1}\log Z_{\beta}(\bb)$.  The latter thus evolve towards the equilibrium state 
\begin{eqnarray}
P_{\tilde{\beta}, \beta}(\bb)&=& \frac{1}{Z_{\tilde{\beta}, \beta}}\rme^{-\tilde\beta \mathcal{F}_\beta(\bb)    }=  \frac{ Z^n_{\beta}(\bb)  }{ \int\rmd\tilde{\bb} ~Z^n_{\beta}(\tilde{\bb})},    \label{eq:P(b)}        
\end{eqnarray}
where $n=\tilde\beta/\beta$.  We can use  (\ref{eq:P(b)})     and  (\ref{eq:P(s|b)}) to construct the joint distribution  
\begin{eqnarray}
P_{\tilde{\beta}, \beta}(\bb, \bsigma)&=& \frac{\rme^{ -\frac{1}{2}\rho n \beta \sum_{\mu=1}^{M}b^2_\mu}\,\rme^{ \beta \sum_{\mu=1}^{M} b_\mu F_\mu(\bsigma)} W^{n-1}_{\beta}(\bb) }{
\int\!\rmd\tilde{\bb}\,   \rme^{ -\frac{1}{2}\rho n \beta \sum_{\mu=1}^{M}{\tilde{b}_\mu}^2} W^{n}_{\beta}(\tilde{\bb})  }, \label{eq:P2.2(b,s)}
\end{eqnarray}
where $W_{\beta}(\bb)=\int\! \rmD\bsigma\,\rme^{ \beta \sum_{\mu=1}^{M} b_\mu F_\mu(\bsigma)}$. 

\begin{figure}[t]
 %\vspace*{-7mm}
 \setlength{\unitlength}{0.9mm}
 \begin{center}{
 %\hspace*{50mm}
 \begin{picture}(100,48)
 \put(0,0){\includegraphics[height=45\unitlength, width=95\unitlength]{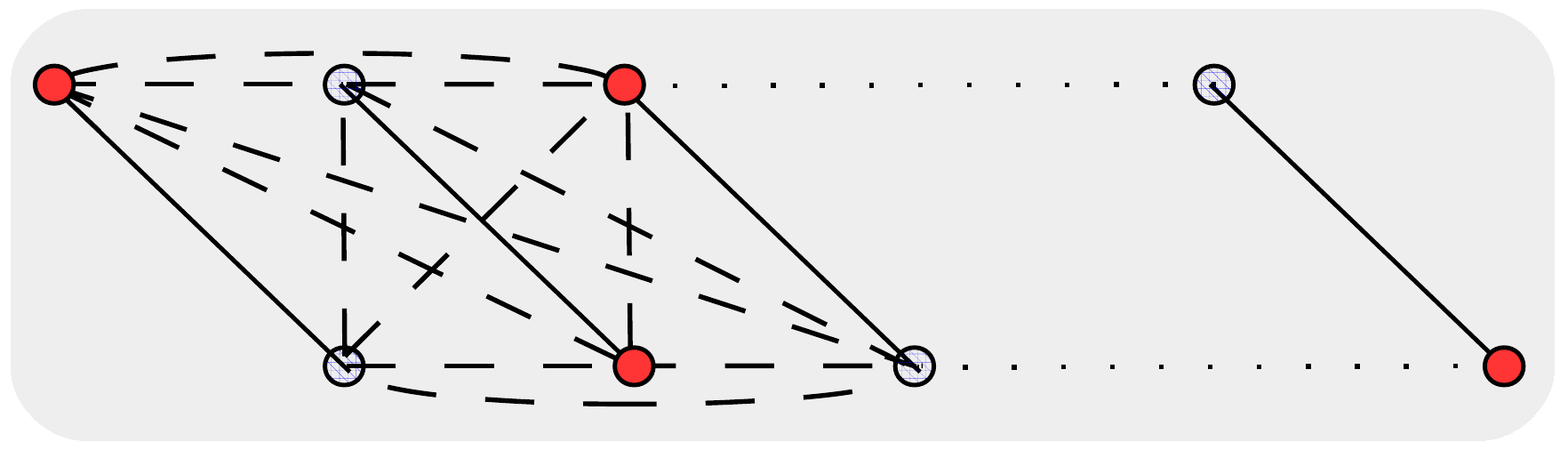}}%Tsystem_repl
\put(5 ,19){   $\frac{J^{\mu}_{ij}}{n}$   }
  \put(1 ,45){$\sigma^1_i$}
  \put(24 ,-2.5){$\sigma^1_j$}
  
   \put(17 ,45){$\sigma^2_i$}
  \put(41 ,-2.5){$\sigma^2_j$}
  
    \put(33 ,45){$\sigma^3_i$}
  \put(57 ,-2.5){$\sigma^3_j$}
  
   \put(68 ,45){$\sigma^n_i$}
  \put(92 ,-2.5){$\sigma^n_j$}
   \end{picture}
 }\end{center}
 %\vspace*{-6mm}
 \caption{Enlargement of  the T-clone network in the system with fast T-clone equilibration and $n\in \mathbb{Z}^{+}$.  All $n$ copies of a T-clone clique (shown here only for one link in this clique) associated with the $\mu$-th B-clone are interconnected in a such way that they form a fully connected (weighted) network of $n\times\vert\partial\mu\vert$ nodes. }
 \label{figure:Tsystem_n} 
 \end{figure}
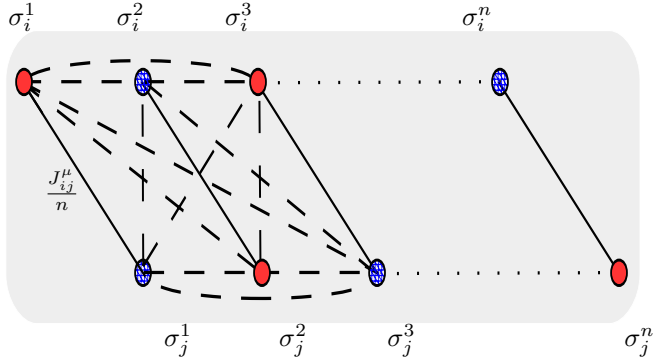

Let us first consider the case when  $n\in \mathbb{Z}^{+}$. Here
\begin{eqnarray}
 W^{n}_{\beta}(\bb)=\int\!\rmD{\bsigma^1}\cdots\int\!\rmD{\bsigma^n}\, \rme^{ \beta \sum_{\mu=1}^{M} b_\mu \sum_{\alpha=1}^{n} F_\mu(\bsigma^\alpha)}\label{def:W-integer-n}
 \end{eqnarray}
 and the thermal average $\langle f(\bsigma)\rangle=\int\!\rmD\bsigma P_{\tilde{\beta}, \beta}(\bsigma) f(\bsigma)$ of any function $f(\bsigma)$ will be given by the following ``replica'' relation 
\begin{eqnarray}
\langle f(\bsigma)\rangle
&=& \frac{\int\!\rmd\bb\, \rme^{ -\frac{1}{2}\rho n \beta \sum_{\mu=1}^{M}b^2_\mu} \int \{D\bsigma^\alpha\}\, \rme^{ \beta \sum_{\mu=1}^{M} b_\mu \sum_{\alpha=1}^{n} F_\mu(\bsigma^\alpha)}   }{ % \left(2\pi/\rho n \beta\right)^{ -\frac{M}{2}}
\int\!\rmd\tilde{\bb}\,  \rme^{ -\frac{1}{2}\rho n \beta \sum_{\mu=1}^{M}{\tilde{b}_\mu}^2} W^{n}_{\beta}(\tilde{\bb})  } f(\bsigma^1)\nonumber\\
%
%&=&\sum_\bsigma 
%\frac{\rme^{\frac{\beta   }{2n\rho}\sum_{\nu=1}^{M}\left(\sum_{\alpha=1}^{n}F_\nu(\bsigma^\alpha)\right)^2 }}{ \sum_{\tilde\bsigma}\rme^{\frac{\beta   }{2n\rho}\sum_{\nu=1}^{M}\left(\sum_{\alpha=1}^{n}F_\nu(\tilde{\bsigma}^\alpha) \right)^2}         }  f(\bsigma^1)
%\nonumber\\
%
&=&\Big\langle  \frac{1}{n}\sum_{\alpha=1}^{n}f(\bsigma^\alpha)       \Big\rangle_n, \label{eq:replica-identity}
\end{eqnarray}
where we have defined the average
\begin{eqnarray}
\left\langle\{\cdots\}\right\rangle_n&=&
\int \{\rmD\bsigma^\alpha\}
\frac{  \rme^{\frac{\beta   }{2n\rho}\sum_{\mu=1}^{M}\left(\sum_{\alpha=1}^{n}F_\mu(\bsigma^\alpha)\right)^2 }     }{ \int \{\rmD\tilde\bsigma^\alpha\}\,\rme^{\frac{\beta   }{2n\rho}\sum_{\mu=1}^{M}\left(\sum_{\alpha=1}^{n}F_\mu(\tilde{\bsigma}^\alpha) \right)^2}         }  \{\cdots\}. \label{def:spin-average-integer-n}
\end{eqnarray}
The Boltzmann weight $\exp[\frac{\beta   }{2n\rho}\sum_{\mu\leq M}(\sum_{\alpha\leq n}F_\mu(\bsigma^\alpha))^2 ]$ in this expression can be written, up to an irrelevant constant, in the form $\exp[-\beta \sum_{\mu\leq M} E_\mu(\{\bsigma^\alpha\})]$. The energy function $E_\mu(\{\bsigma^\alpha\})=-\frac{1}{2}\sum_{\alpha, \gamma} \sum_{i\in\partial\mu}\sum_{j\in\partial\mu}(J^{\mu}_{ij}/n)\sigma_i^\alpha\sigma_j^\gamma-\sum_{\alpha=1}^n\sum_{i\in\partial\mu}\theta_i^\mu\sigma_i^\alpha$, where  $J^\mu_{ij}=(J_\mu^2/\rho) \xi_i^\mu\xi_j^\mu$ and  $\theta^\mu_{i}=(J_\mu^2/\rho) \xi_i^\mu\theta_\mu$, can be represented as a clique constructed from those $n\times\vert\partial\mu\vert$  T-clones that control B-clone $\mu$. Thus for $n\in \mathbb{Z}^{+}$ the fast T-clone equilibration  leads to the enlargement  of the `local' T-clone system (see Figure  \ref{figure:Tsystem_n}).

The B-clone density $P(b)=\frac{1}{M}\sum_{\nu=1}^{M}\left\langle\delta(b- b_\nu )\right\rangle_{\tilde{\beta}, \beta}$ can now be computed by using the identity  (\ref{def:W-integer-n}) in distribution (\ref{eq:P2.2(b,s)}), which  gives us
\begin{eqnarray}
P(b)&=&\int\!\rmd F~P(F)~\frac{  \rme^{-\frac{1}{2}\rho n\beta\left(b - \frac{F}{n \rho}\right)^2}  }{\sqrt{2\pi/\rho n\beta}}\label{eq:P(b)-T-fast}
\end{eqnarray}
with the distribution of (replicated) fields 
\begin{eqnarray}
P(F)&=&  \frac{1}{M}\sum_{\nu=1}^{M}\Big\langle\delta\Big(F - \sum_{\alpha=1}^{n}F_\nu(\bsigma^\alpha)\Big)\Big\rangle_n. \label{def:P(F)_replica}
\end{eqnarray}
From this also follows  the distribution of  B-cell concentrations 
\begin{eqnarray}
P(c)&=&\int\!\rmd F~P(F)~ \frac{  \rme^{-\frac{1}{2}\rho n\beta\left( \log\left(c\right) - \frac{F}{n \rho}\right)^2}  }{c\sqrt{2\pi/\rho n\beta}}.\label{eq:P(c)-T-fast-integer-n}
\end{eqnarray}

Let us now compute the distribution (\ref{eq:P2.2(b,s)}) for the more general case where $n\in \mathbb{R}^{+}$, not necessarily integer. In order to do this we assume that $V(\bsigma)=\sum_i V(\sigma_i)$, and use the short-hand 
$\rmD\bsigma=\rme^{-\beta \sum_{i=1}^N V(\sigma_i)}\rmd \bsigma$. 
We then consider the integral 
\begin{eqnarray}
W_{\beta}(\bb)&=& \int\! \rmD\bsigma\, \rme^{ \beta \sum_{\mu=1}^{M} b_\mu F_\mu(\bsigma)}  \nonumber\\
%
%&=&  \rme^{ \beta \sum_{\mu=1}^{M} b_\mu J_\mu \theta_\mu    } \int\! \rmD\bsigma\, \rme^{ \beta \sum_{i=1}^{N}\sigma_i \sum_{\mu\in\partial i} b_\mu J_\mu \xi_i^{\mu}}\nonumber\\
%
&=& \rme^{ \beta \sum_{\mu=1}^{M} b_\mu J_\mu \theta_\mu + \sum_{i=1}^{N}\log\int\! \rmD\sigma\, \rme^{ \beta\sigma\sum_{\mu\in\partial i} b_\mu J_\mu \xi_i^{\mu}    }}\label{eq:W(b)},
\end{eqnarray}
 where $\rmD\sigma=\rme^{-\beta V(\sigma)}\rmd \sigma$.
Upon inserting this result into (\ref{eq:P2.2(b,s)}) we can extract the marginal distributions 
\begin{eqnarray}
\hspace*{-15mm}
P_{\tilde{\beta}, \beta}(\bb)&=& \frac{1}{Z_{\tilde{\beta}, \beta}} \rme^{   -\frac{1}{2}\rho n \beta \sum_{\mu=1}^{M}\left(b_\mu\! -\!\frac{J_\mu \theta_\mu}{\rho}\right )^2 +n\sum_{i=1}^{N}\log\int\! \rmD\sigma\, \rme^{ \beta\sigma\sum_{\mu\in\partial i} b_\mu J_\mu \xi_i^{\mu}    }}  \label{eq:Prob(b)-real-n}
\\
\hspace*{-15mm}
P_{\tilde{\beta}, \beta}(\bsigma)
%&=& Z^{-1}_{\tilde{\beta}, \beta} \int\!\rmd\bb\,\rme^{    -\frac{1}{2}\rho n \beta \sum_{\mu=1}^{M}\left(b_\mu -\frac{J_\mu \theta_\mu}{\rho}\right )^2\!+(n-1)\sum_{i=1}^{N}\log\int\! D\sigma\, \rme^{ \beta\sigma\sum_{\mu\in\partial i} b_\mu J_\mu \xi_i^{\mu}    }} \nonumber\\[-2mm]
%&&\hspace*{50mm}
%\times \rme^{ \beta \sum_{i=1}^{N}\sigma_i \sum_{\mu\in\partial i} b_\mu J_\mu \xi_i^{\mu} } 
%\nonumber\\[2mm]  
&=&\int\! \rmd\bb~P_{\tilde{\beta}, \beta}(\bb)  \left\{\prod_{i=1}^N \frac{   \rme^{ \beta \sigma_i\sum_{\mu\in\partial i} b_\mu J_\mu \xi_i^{\mu} }    }{
\int\!\rmD\tilde\sigma\,
  \rme^{ \beta\tilde\sigma\sum_{\mu\in\partial i} b_\mu J_\mu \xi_i^{\mu} } 
 %\rme^{ \beta \tilde\sigma_i \sum_{\mu\in\partial i} b_\mu J_\mu \xi_i^{\mu} } 
 }\right\}, \label{eq:P2.2(s)-real-n}
\end{eqnarray}
with $Z_{\tilde{\beta}, \beta}=\int\!\rmd\tilde{\bb}~\exp[-\frac{1}{2}\rho n \beta \sum_{\mu=1}^{M}(\tilde{b}_\mu -\frac{J_\mu \theta_\mu}{\rho})^2\! +n\sum_{i=1}^{N}\log\int\! D\sigma\, \rme^{ \beta\sigma\sum_{\mu\in\partial i} \tilde b_\mu J_\mu \xi_i^{\mu}    } ]$.  The distributions (\ref{eq:Prob(b)-real-n},\ref{eq:P2.2(s)-real-n}) can be used to construct the densities $P(b)=M^{-1} \sum_{\mu=1}^{M}\int\rmd\bb\, P_{\tilde{\beta}, \beta}(\bb)\,\delta\!\left(b-b_\mu\right)=M^{-1} \sum_{\mu=1}^{M}P_\mu(b)$ (the concentration density $P(c)$ is given by $P(b)/c$ with $b=\log c$ ) and $P(\sigma)=N^{-1}\sum_{i=1}^{N}\int\! \rmD\bsigma\, P_{\tilde{\beta}, \beta}(\bsigma)\,\delta\!\left(\sigma-\sigma_i\right)=N^{-1} \sum_{i=1}^{N}P_i(\sigma)$, where  the marginal distribution $P_i(\sigma)$ is given by 
\begin{eqnarray}
P_i(\sigma)&=& \int\! \rmd\bb~P_{\tilde{\beta}, \beta}(\bb) \frac{ \int\!\rmD\sigma_i\,\rme^{ \beta \sigma_i\sum_{\mu\in\partial i} b_\mu J_\mu \xi_i^{\mu} } \delta\!\left(\sigma-\sigma_i\right)}{ \int\!\rmD\tilde\sigma_i\,\rme^{ \beta \tilde\sigma_i \sum_{\mu\in\partial i} b_\mu J_\mu \xi_i^{\mu} } }
\label{eq:Pi(s)-real-n},
\end{eqnarray}
respectively.  In a similar manner, using the definition $$P_\mu(F)= \int\! \rmD\bsigma\,P_{\tilde{\beta}, \beta}(\bsigma)\delta\left(F-F_\mu(\bsigma)\right),$$ we can derive the distribution of fields\footnote{Similarly to a B-clone, the  $i$-th T-clone experiences the field $\sum_{\mu\in\partial j} b_\mu J_\mu \xi_j^{\mu}$; see equation (\ref{eq:Pi(s)-real-n}).}
\begin{eqnarray}
P_\mu(F)&=& \!\int\!\rmd\bb~P_{\tilde{\beta}, \beta}(\bb)\delta\left(F\!-\!F_\mu(\bsigma)\right) \prod_{i\in\partial \mu} \frac{ \int\!\rmD\sigma_i~   \rme^{ \beta \sigma_i\sum_{\nu\in\partial i} b_\nu J_\nu \xi_i^{\nu} }    }{
\int\!\rmD\tilde\sigma_i~
  \rme^{ \beta\tilde\sigma_i\sum_{\nu\in\partial i} b_\nu J_\nu \xi_i^{\nu} } 
 %\rme^{ \beta \tilde\sigma_i \sum_{\mu\in\partial i} b_\mu J_\mu \xi_i^{\mu} } 
 }\label{def:P(F)-real-n}.
\end{eqnarray}
%

%\newpage

\section{Equilibrium analysis\label{section:Analysis}} 
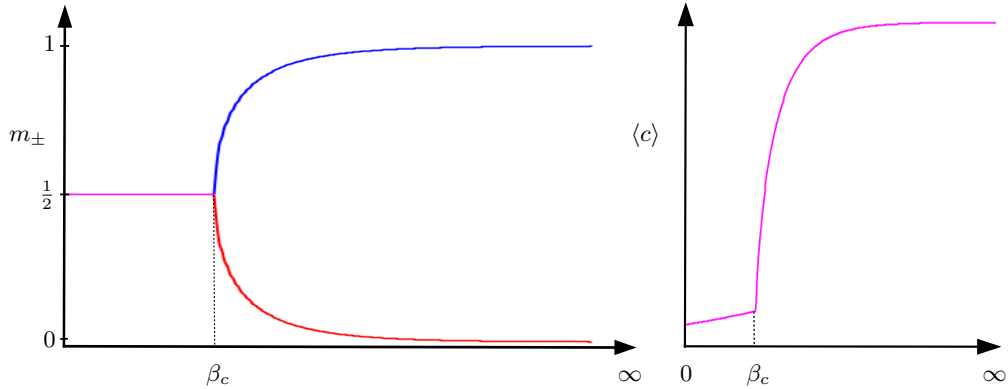
\begin{figure}[t]
%\vspace*{5mm} \hspace*{35mm} 
\setlength{\unitlength}{0.57mm}
%\hspace*{1mm}
\begin{picture}(230,93)
\put(0,0){\includegraphics[height=90\unitlength,width=145\unitlength]{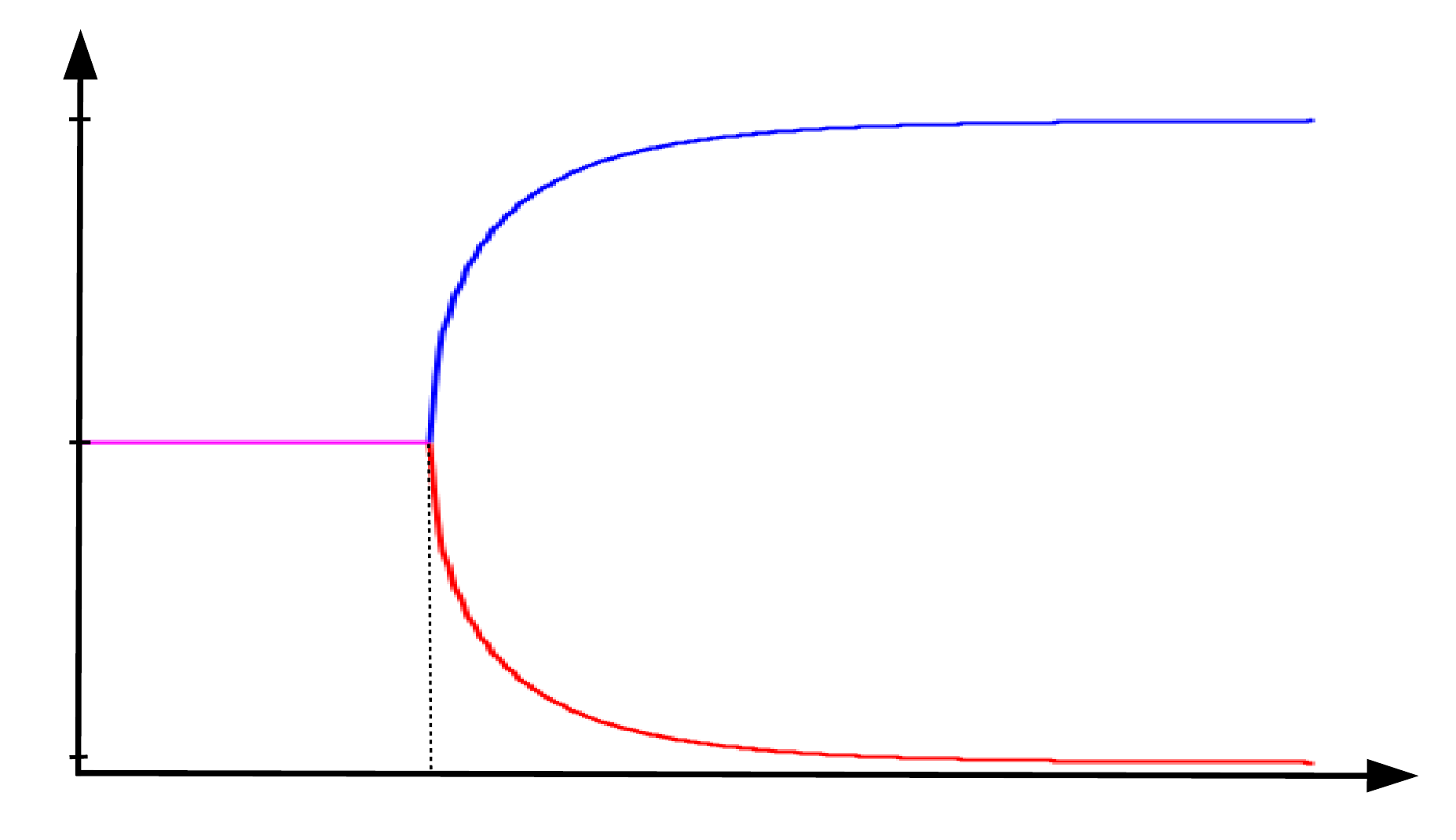}}

\put(140,-2){\includegraphics[height=97\unitlength,width=100\unitlength]{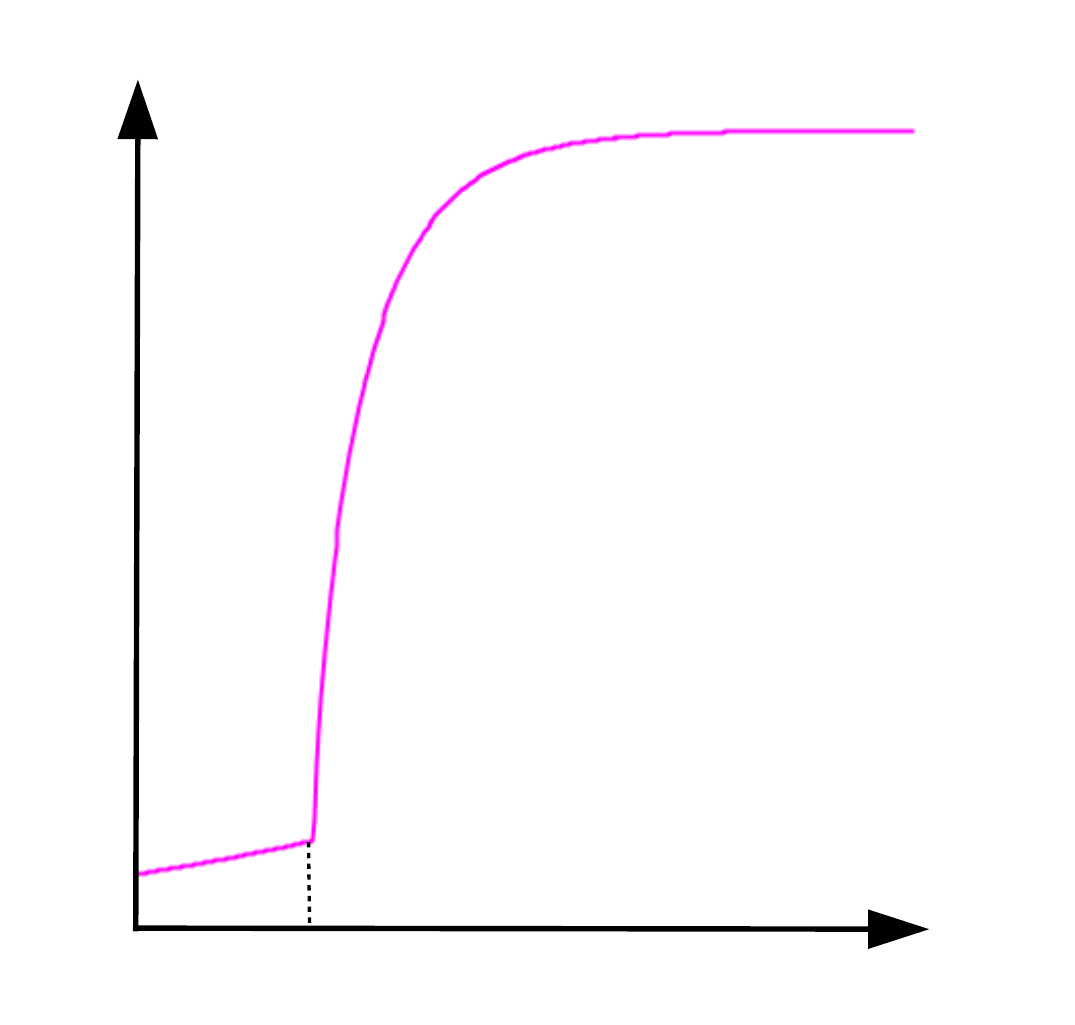}}

%\put(177,25){\includegraphics[width=50\unitlength]{BvsT}}%height=90\unitlength,
%\put(197,24){\small{$m_+$}}

\put(41,-1){\small{$\beta_c$}} \put(137,-1){\small{$\infty$}}

\put(151.5,-1){\small{$0$}} \put(167,-1){\small{$\beta_c$}} \put(222,-1){\small{$\infty$}}

\put(-5,55){\small{$m_{\pm}$}} 

\put(140,55){\small{$\langle c\rangle$}}

%\put(155,65){$(b)$}

\put(3,76){\small{$1$}} \put(2,41){\small{$\frac{1}{2}$}}  \put(3,7){\small{$0$}}
\end{picture}
%\vspace*{-5.0mm} 
\caption{Fraction of regulator  T cells (lower branch) $m_-=\frac{1}{2}(1- m)$, fraction of  helper  T cells (upper branch) $m_+=\frac{1}{2}(1+ m)$, and average B-clone size $\langle c\rangle$, in the fast B-clone equilibration regime, as a function of the inverse temperature $\beta$  in the T-clone system modelled with the binary  variables $\sigma_i\in\{-1,1\}$.  } \label{figure:PD}
\end{figure}
%

%\emph{Choice of variables for T-clones and their ``biological'' meaning:} 
In the remainder of this article we consider the simplest case when the T clones (or single T cells) are modelled by the (binary) Ising variables  $\sigma_i\in\{-1,1\}$, or by the binary variables $\sigma_i\in\{0,1\}$.  The motivation for this choice of variables is as follows.  A single T helper cell can be either active or inactive, i.e. ``on'' or ``off'',  and  is activated by the Ag presenting cell which could be a dendritic cell, a B cell, etc.  Then this T helper cell can activate a B cell if it receives the ``right'' Ag related signal from it.  For the T regulator cells we assume that a similar  mechanism is at work~\cite{Vanderleyden2014}.  On the level of T-clone  we could say that the T-clone is active (inactive) if majority of its  cells are active (inactive). Thus the use of binary variables can be seen either as a crude approximation of T-clones, where we only retain information about the state of a clone but disregard  its  size,  or all $N$ variables are simply treated as single T cells and we no longer distinguish the T clones to which they belong.

Within the analytical  framework of the previous  section the choices $\sigma_i\in\{-1,1\}$ and  $\sigma_i\in\{0,1\}$  can be obtained  by using the double well potentials $V(\bsigma)=\Delta\sum_{i=1}^N\left(\sigma^2_i-1\right)^2$   and $V(\bsigma)=\Delta\sum_{i=1}^N\sigma^2_i\left(\sigma_i-1\right)^2 + \frac{1}{2}\omega\sum_{i=1}^N\sigma^2_i$ respectively. Here the ``chemical'' potential $\omega$ allows us to control the number of activated T cells $\sum_{i=1}^N \delta_{\sigma_i, 1}$, via the integral measure $\rmD\bsigma$. Taking the limit $\Delta\rightarrow\infty$ converts  the $N$-dimensional integral $\int \rmD\bsigma f(\bsigma) $, for example in the equation (\ref{eq:replica-identity}), into a sum $\sum_{\bsigma} f(\bsigma)$ over binary variables.  Furthermore, the choice $\sigma_i\in\{-1,1\}$, combined with $\xi_i^\mu =1$,  gives us the scenario when each  active T cell is  either  a regulator cell  or a helper cell.  The fact that  the $i$-th T cell can change from being a regulator,  $\sigma_i=-1$, to being a helper,   $\sigma_i=1$  (or vice versa), must then  be seen as an assumption that changes in  T cell function  occur on the same time scale as the immune response to the Ag.  However, in  experiments~\cite{Vanderleyden2014, Baumjohann2013} one usually has access only to the number of T cells of either type, and such an assumption may be acceptable to capture the observed phenomena.  For the alternative choice of active or inactive  T cells where $\sigma_i\in\{0,1\}$ for all $i$, with  $\xi_i^\mu \in\{-1,1\}$ (in this case $\xi_i^\mu=\xi_i$), the number of regulating T cells $\sum_{i=1}^N\delta_{\xi_i; -1}$ is independent of the immune response. 

%\emph{T-clone system is a system of  interacting Ising spins (or lattice gas):  ferromagnetic with (or without) uniform external field, with random field:} 
The consequence of our choice  for the measure $V(\bsigma)$ describing the variables $\sigma_i$, when T cells are  fast variables with $n=\in\mathbb{Z}^+$ (see section \ref{section:T-fast}), is that the T cells are governed by the distribution 
\begin{eqnarray}
P(\bsigma^1,\ldots, \bsigma^n)&=&
\frac{  \rme^{\frac{\beta   }{2n\rho}\sum_{\mu=1}^{M}\left(\sum_{\alpha=1}^{n}F_\mu(\bsigma^\alpha)\right)^2 }     }{ \sum_{\{\tilde\bsigma^\alpha\}}    \rme^{\frac{\beta   }{2n\rho}\sum_{\mu=1}^{M}\left(\sum_{\alpha=1}^{n}F_\mu(\tilde{\bsigma}^\alpha) \right)^2}         } , \label{eq:P(s)-integer-n-Ising}
\end{eqnarray}
where $F_\mu(\bsigma^\alpha)= J_\mu\big(\sum_{i\in \partial \mu}\xi_i^\mu \sigma_i^\alpha \!+\! \theta_\mu\big)$,  which is equivalent to an Ising spin model  for $\sigma_i\in\{-1,1\}$ or a ``lattice gas'' for  $\sigma_i\in\{0,1\}$. For the latter we need to add the term $-\frac{1}{2}\beta\omega\sum_{i=1}^N\sum_{\alpha=1}^{n}\sigma^\alpha_i $ to the ``energy'' function in the exponential of (\ref{eq:P(s)-integer-n-Ising}).  We note that $J_\mu(\mathbf{a})=J\sum_{\nu\leq M} S_{\mu\nu} a_\mu$, where $J\geq0$, allows us to control the ``amount''  of Ag $\mathbf{a}$ by increasing or decreasing  $J$. Then the quantity $\beta J^2$  controls either the level of noise in the T clone system for fixed $J$ or the amount of Ag for fixed $\beta$.  Also the distribution of T-clones (\ref{eq:P(b,s)-B-fast }) in the  fast B clone equilibration regime  can be obtained from (\ref{eq:P(s)-integer-n-Ising}) by setting $n=1$, so for T-clones the case of integer $n$ covers both equilibration scenarios.  Let us for now discuss the case of $n=1$ and $\theta_\mu=0$. 

For the Ising case with $\xi_i^\mu=1$, the average ``magnetization''  $m=\frac{1}{N}\sum_{i=1}^N\langle\sigma_i\rangle$,  which is related to  the fraction of activated helper (regulator)  T-cells\footnote{In immune response experiments, as in~\cite{Vanderleyden2014}, such fractions can be computed from the number of activated T helper, T regulator cells and the total number of activated T cells, N.} via the identity $m_+= 1- m_-$ ($m_-=\frac{1}{2}(1-m)$),  has a phase transition at $\beta_c=\beta$ from the disordered paramagnetic (PM) $m=0$  phase to the ordered ferromagnetic  (FM)  phase $m\neq0$  (see Figure \ref{figure:PD}) when $N\rightarrow\infty$~\cite{Baxter1982}.  We note that in this framework the PM and FM phases can be interpreted  respectively,  as  the ``low-dose tolerance'' and  ``vigorous immune response''  phases of a real immune system.  The former is consistent with  its  insensitivity  to ``small'' amounts of Ag and the latter is consistent with its strong reaction to a  larger amounts of Ag~\cite{Janeway2012}. The noise parameter $\beta$ controls this ``sensitivity'' to the Ag: for small (large) $\beta$, i.e. in a high (low) noise  regime, a larger (smaller) amount of Ag is needed to trigger a vigorous immune response.

\begin{figure}[!t]
\setlength{\unitlength}{1mm}

\begin{center}
\hspace*{4mm}
\begin{picture}(130,43)
\put(0,0){\includegraphics[width=120\unitlength]{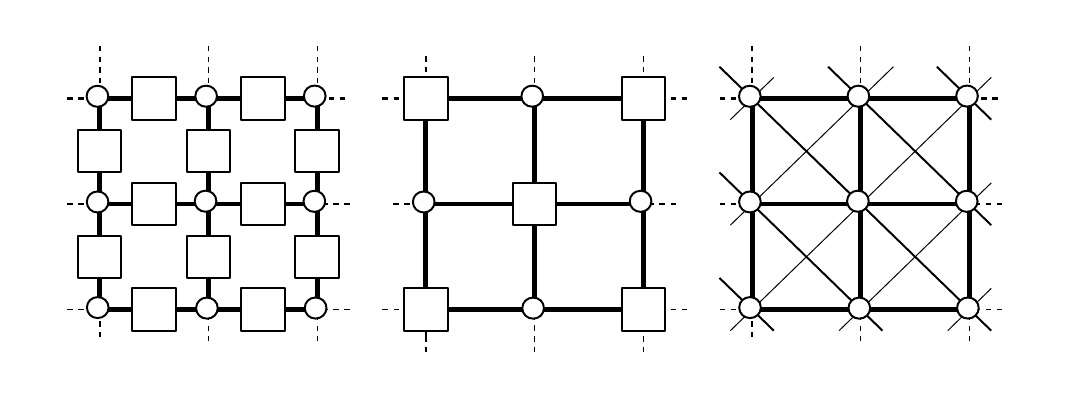}}
%Left
%\put(23,35){$s_0$} \put(33,35){$s_1$}
%\put(23,25){$s_3$} \put(33,25){$s_2$}

%Right
%\put(86,35){$s_0$} \put(101,35){$s_1$}
%\put(86,20){$s_3$} \put(101,20){$s_2$}
\end{picture}
\end{center}
\vspace*{-5mm}
\caption{Two possible topologies for bipartite immunological graphs (left and centre), to illustrate the consequences of eliminating `fast' B-clones. T and B clones are represented by circles and squares respectively.  Upon integrating out the B-clone variables, the bipartite graph on the left gives rise to an effective T-clone system shown on the right, in which the T-clones (circles) interact on the classical square lattice (drawn as {\em thick} solid lines).  Also the bipartite graph in the centre  gives rise to an effective T-clone system  on the square lattice shown on the right,   but now with  extended interaction range (drawn as {\em thin} solid lines). \label{figure:2d} }
\end{figure}

There are many possible topologies of a bipartite graph $\mathcal{G}$ for which the T cell system (\ref{eq:P(s)-integer-n-Ising}) has a finite critical inverse noise level $\beta_c$.  Suppose we choose one such topology (see Figure \ref{figure:2d}), then for $\beta\in(\beta_c, \infty)$ the fraction of helper (regulator) T cells $m_+$ ( $m_-$)  is  either a monotonic non-decreasing  or a monotonic non-increasing function of the noise parameter $\beta$ (see Figure \ref{figure:PD}) and of interaction strength  $J_\mu^2/\rho$.  This follows from  the  Griffiths-Kelly-Sherman (GKS) theorem~\cite{Domb1972}, which holds for any  ferromagnetic Ising spin system.  Furthermore, if we know the average magnetization  $m \equiv m(\{J_\mu^2/\rho\})$  for the  T cell system with uniform interactions, i.e. with $J_\mu=J$ for all $\mu$,   then by the same theorem  $m(J_0^2/\rho)\leq m(\{J_\mu^2/\rho\})\leq  m(J_1^2/\rho) $ , where  $J_0=\inf_{\mu} J_\mu$ and  $J_1=\sup_\mu J_{\mu}$, from which it also follows that   $\beta_c \equiv\beta_c(\{J_\mu^2/\rho\})$ obeys:  $\beta_c(J_0^2/\rho)\leq\beta_c(\{J_\mu^2/\rho\})\leq   \beta_c(J_1^2/\rho)$. 

We note that  the ratio of the number of regulator T cells to the total number of T cells  was observed,  during a normal immune response to the  Ag stimulation,  to be in the range $0.1-0.25$ in experiments on mice~\cite{Vanderleyden2014}.  In the phase diagram (see Figure \ref{figure:PD}) this corresponds to the lower branch of the $m_-$ plot.  Also in this regime, the fraction $m_+$ of helper T-cells  is a  monotonic non-decreasing function of Ag concentration, which is consistent with the experimental data~\cite{Baumjohann2013}. The case of  a ``lattice gas'', i.e. $\sigma_i\in\{0,1\}$ and $\xi_i=1$, which is the scenario where we have only T helper cells which are either active ($\sigma_i=1$) or inactive ($\sigma_i=0$), can be mapped similarly into the ferromagnetic Ising model with (positive) external field~\cite{Baxter1982}. For  example,  by writing $\sigma_i=\frac{1}{2}(1+s_i)$, where $s_i\in\{-1,1\}$. As a consequence, also here the fraction of helper T clones $m_+=1-m_-$ is a continuous and monotonic function of  $\beta$  and $J_\mu^2/\rho$~\cite{Domb1972}. %$\beta\in[0,\beta_c)$ 
%

%\emph{Behavior of B-cells:}
 We now turn to the analysis of B-clone properties in the fast B-clone and fast T-clone ($n\in\mathbb{Z}^+$) equilibration regimes. For the former the average B-clone size (or average B cell concentration)  $\langle c \rangle$ can be computed from (\ref{eq:P(c)-B-fast}), and for the latter the same average can be computed  from   (\ref{eq:P(c)-T-fast-integer-n}). In order to simplify the analysis of both equilibration regimes,  we define the distribution
\begin{eqnarray}
P(c)&=&\int\!\rmd F~P(F) \frac{  \rme^{-\frac{1}{2}\rho \tilde\beta\left(\log\left(c\right)- \frac{F}{n\rho}\right)^2}  }{c\sqrt{2\pi/\rho \tilde\beta}},\label{eq:P(c)-integer-n}
\end{eqnarray}
where 
\begin{eqnarray}
P(F)&=&\sum_{\{\bsigma^\alpha\}}P(\bsigma^1,\ldots, \bsigma^n) \frac{1}{M} \sum_{\nu=1}^{M}\delta\Big(F - \sum_{\alpha=1}^nF_\nu(\bsigma^\alpha)\Big)\label{eq:P(F)-integer-n}.
%\frac{1}{M} \sum_{\nu=1}^{M}
%\frac{ \sum_{\{\bsigma^\alpha\}}  \rme^{\frac{\beta   }{2n\rho}\sum_{\mu=1}^{M}\left(\sum_{\alpha=1}^{n}F_\mu(\bsigma^\alpha)\right)^2 }   \delta\left(F - F_\nu(\bsigma^\alpha)\right)  }{ \sum_{\{\tilde\bsigma^\alpha\}}    \rme^{\frac{\beta   }{2n\rho}\sum_{\mu=1}^{M}\left(\sum_{\alpha=1}^{n}F_\mu(\tilde{\bsigma}^\alpha) \right)^2} } . \label{eq:P(F)-integer-n}
%
\end{eqnarray}
Formula (\ref{eq:P(c)-integer-n}) gives the distribution of B-clone sizes for the fast B-clone  and fast T-clone equilibration  regimes when $n=1$ and $\tilde\beta=n\beta$, respectively. This log-normal distribution can be interpreted as the (asymptotic) distribution of the ``size'' $c_i$ of an element  in a growth process, which changes its size at rate $w(c\!\rightarrow \!c^\prime)=\lambda \delta\left(c^\prime\!-\! (1\!+\!g)c\right)$, in which $\lambda$ is a (mean) growth rate and $g$ is a growth factor~\cite{ Goh2010}. The growth factor $g$ is  related to  the apoptosis parameter $\rho$ via  $\log^2(1\!+\!g)=\rho^{-1}$ and the noise parameter $\tilde\beta$ is related to the (rescaled) time $\lambda t$ , where $\lambda$ is a (mean) growth rate, via  $\lambda t={\tilde{\beta}}^{-1}$.    Furthermore, when the  initial size $c_i=c_0$ then  $c_0=\exp[(F/n\!-\!\sqrt{\rho}/\tilde{\beta})/\rho]$ and the distribution (\ref{eq:P(c)-integer-n}) represents an average over random initial conditions. Since $c$ is a clone size,  the growth  process picture  is consistent with the adaptive immune response:  those B-cells which ``survived''  interactions with the T cells, of which there are  $c_0$, serve to `seed' the B cell proliferation process. 

Average and variance of (\ref{eq:P(c)-integer-n}) are given by 
\begin{eqnarray}
\langle c \rangle&=&\rme^{{1}/{2\rho\tilde\beta}}\int\!\rmd F~P(F)\rme^{{F}/{n\rho}} 
\label{def:<c>}
\\
 \langle (c -\langle c \rangle)^2\rangle&=&\rme^{{1}/{2\rho\tilde\beta}}\Big( \rme^{{1}/{2\rho\tilde\beta}}\! -1    \Big)
 \int\!\rmd F~P(F)\rme^{{2F}/{n\rho}} 
 \label{def:Var-c}.
 \end{eqnarray}  
Let us next define the function  $\langle \rme^{{F}/{n\rho}} \rangle_\beta=\int\!\rmd F~P(F)\rme^{{F}/{n\rho}}$ and consider its properties. For $\beta\rightarrow0$ this function can be expanded around $\beta=0$ which gives $\langle \rme^{{F}/{n\rho}} \rangle_\beta= M^{-1}\sum_{\mu\leq M}\rme^{J_\mu \theta_\mu/\rho} \cosh^{n\vert\partial\mu\vert}(J_\mu/n\rho)+O(\beta)$, while in the opposite limit $\beta\rightarrow\infty$ it is dominated by  the ferromagnetic ground  state $\bsigma=(1,\ldots, 1)$  which gives us  $\langle \rme^{{F}/{n\rho}} \rangle_\infty=  M^{-1} \sum_{\mu\leq M}\rme^{(J_\mu/\rho)(\vert\partial\mu\vert + \theta_\mu)}$ (note that for $\theta_\mu\neq0$ the dynamics (\ref{eq:T-clone-dynamics}) can also ``select'' the  $(1,\ldots, 1)$ state as $N\rightarrow\infty$), so $\langle \rme^{{F}/{n\rho}} \rangle_\infty\geq\langle \rme^{{F}/{n\rho}} \rangle_0$.  Furthermore,  $\langle \rme^{{F}/{n\rho}} \rangle_\beta$ is monotonic non-decreasing function of $\beta$ and $J$ ($J_\mu=J\sum_{\nu\leq M} S_{\mu\nu} a_\mu$), which for $N<\infty$ is smooth everywhere except at  $\beta_c$ , when $N\rightarrow\infty$  with $\theta_\mu=0$ (see \ref{app:<exp(F)>} for details).  

From the above analysis of  $\langle \rme^{{F}/{n\rho}} \rangle_\beta$  we infer that  in the fast B-clone  equilibration regime ($n=1$) the average number of B-cells  $\langle c \rangle$  is a monotonic non-decreasing function of $\beta$ and $J$ (see Figure \ref{figure:PD}  for one of the possible behaviours). Combining the two plots in Figure \ref{figure:PD} then shows that $\langle c \rangle$ must be a monotonic non-decreasing (non-increasing) function of the fraction of helper T-cells $m_+$ (regulator T-cells  $m_-$) and Ag, which is consistent with what was observed in \emph{in vivo} ~\cite{Baumjohann2013, Vanderleyden2014}.  We note that the variance   $\langle (c -\langle c \rangle)^2\rangle$ has a similar behaviour in this regime (the function $\langle \rme^{2{F}/{n\rho}} \rangle_\beta=\int\!\rmd F~P(F)\rme^{2{F}/{n\rho}}$ used in (\ref{def:Var-c}) has the same behaviour as  the function   $\langle \rme^{{F}/{n\rho}} \rangle_\beta$).

The behaviour of the average  $\langle c\rangle $ in the case of fast T-clone equilibration with $n\in\mathbb{Z}^+$ is  not so clear, due to the relation $\tilde\beta\!=\!n\beta$. Here $\langle c \rangle\!=\!\rme^{{1}/{2\rho n\beta}} \langle \rme^{{F}/{n\rho}} \rangle_\beta$ diverges for $\beta\!\rightarrow\! 0$ (it is finite when $\beta\!\rightarrow\!\infty$), and the derivative $ \partial\langle c \rangle/\partial\beta =\rme^{{1}/{2\rho n\beta}} \big(\partial\langle \rme^{{F}/{n\rho}} \rangle_\beta/\partial\beta\! -\!  \langle \rme^{{F}/{n\rho}} \rangle_\beta / 2\rho n\beta^2\big)$ is  negative when $\beta\!\rightarrow\!0$ (in this limit $ \partial\langle \rme^{{F}/{n\rho}} \rangle_\beta/\partial\beta\!<\!\infty$,  and  $ \langle \rme^{{F}/{n\rho}} \rangle_\beta/2\rho n\beta^2= O({1}/{\beta})$). For $\beta\to\infty$  the derivative $ \partial \langle c \rangle/\partial\beta$ could be positive or negative; now $\langle \rme^{{F}/{n\rho}} \rangle_\beta/2\rho n\beta^2 =  O({1}/{\beta^2})$, and one expects the convergence to zero of $\partial \langle \rme^{{F}/{n\rho}} \rangle_\beta/\partial\beta$  to be strongly influenced by the topology of the effective T-clone network. However, if $\beta$ is fixed and we vary only $J$  then  behaviours summarised in the Figure \ref{figure:PD} are also observed (due to monotonicity of the average $\langle \rme^{{F}/{n\rho}} \rangle_\beta$  with respect to $J$ which is shown in the \ref{app:<exp(F)>}) in  the $n\in\mathbb{Z}^+$ fast T-clone equilibration regime.

%\emph{Analysis of  real $n$ scenario:} %latter - being the second of two mentioned 
The fast T-clone equilibration regime with $n\in\mathbb{R}^+$  is much more difficult to analyse, but in the low B-clone noise  $\tilde\beta\rightarrow\infty$ limit  we expect at least in some regimes  the same phase diagram as in the Figure \ref{figure:PD}.  To show this we first note that  the marginal distribution (\ref{eq:Pi(s)-real-n}) can be written as the integral 
\begin{eqnarray}
P_i(\sigma)&=&   \int\!\rmd\bb~\frac{\rme^{\tilde\beta \phi(\bb)}}{  \int\!\rmd\tilde{\bb}~\rme^{\tilde\beta \phi(\tilde{\bb})}} \frac{ \int\!\rmD\sigma_i\,\rme^{ \beta \sigma_i\sum_{\mu\in\partial i} b_\mu J_\mu \xi_i^{\mu} } \delta\!\left(\sigma-\sigma_i\right)}{ \int\!\rmD\tilde\sigma_i\,\rme^{ \beta \tilde\sigma_i \sum_{\mu\in\partial i} b_\mu J_\mu \xi_i^{\mu} } }\label{eq:Pi(s)-real-n-Laplace},
\end{eqnarray}
where
\begin{eqnarray}
\phi(\bb)&=& \!      -\!\frac{1}{2}\rho \! \sum_{\mu=1}^{M}\!\Big(b_\mu \!-\!\frac{J_\mu \theta_\mu}{\rho}\Big)^2 \!\!+\! \frac{1}{\beta}\! \sum_{i=1}^{N}    \log\!\int\! \rmD\sigma\, \rme^{ \beta\sigma\!\sum_{\mu\in\partial i} \!b_\mu J_\mu \xi_i^{\mu}    }\!. \label{def:phi}
\end{eqnarray}
 This  integral  can be computed exactly by the Laplace method~\cite{DeBruijn1970}, which gives us  %as $\tilde\beta\rightarrow\infty$,  
\begin{eqnarray}
P_i(\sigma)&=& \frac{ \int\!\rmD\sigma_i\,\rme^{ \beta \sigma_i\sum_{\mu\in\partial i} b^*_\mu J_\mu \xi_i^{\mu} } \delta\!\left(\sigma-\sigma_i\right)}{ \int\!\rmD\tilde\sigma_i\,\rme^{ \beta \tilde\sigma_i \sum_{\mu\in\partial i} b^*_\mu J_\mu \xi_i^{\mu} } }\label{eq:Pi(s)-large-n},
\end{eqnarray}
where $b_\mu^*$ is a solution of the following system of equations 
\begin{eqnarray}
b_\mu/J_\mu=\frac{1}{\rho}\sum_{i\in\partial \mu} \xi_i^{\mu}    \frac{ \int\!\rmD\sigma\,\rme^{ \beta \sigma\sum_{\nu\in\partial i} b_\nu J_\nu \xi_i^{\nu} } \sigma}{ \int\!\rmD\tilde\sigma\,\rme^{ \beta \tilde\sigma\sum_{\mu\in\partial i} b_\nu J_\nu \xi_i^{\nu} } }  +\theta_\mu\label{eq:large-n-extr}
\end{eqnarray}
corresponding  to a maximum of the function $\phi(\bb)$.  From the first moment 
\begin{eqnarray}
\langle\sigma_i\rangle&=& \frac{ \int\!\rmD\sigma_i\,\rme^{ \beta \sigma_i\sum_{\mu\in\partial i} b^*_\mu J_\mu \xi_i^{\mu} }    \sigma_i     }{ \int\!\rmD\tilde\sigma_i\,\rme^{ \beta \tilde\sigma_i \sum_{\mu\in\partial i} b^*_\mu J_\mu \xi_i^{\mu} } }\label{def:Pi(s)-large-n-1-moment}
\end{eqnarray}
of the  distribution (\ref{eq:Pi(s)-large-n}), and the extremum condition (\ref{eq:large-n-extr}), it follows that 
\begin{eqnarray}
b_\mu &=& J_\mu\Big(\frac{1}{\rho}\sum_{j\in\partial \mu} \xi_j^{\mu}\langle\sigma_j\rangle     +\theta_\mu\Big) \label{eq:b-large-n}
\end{eqnarray}
which gives us the equation 
\begin{eqnarray}
\langle\sigma_i\rangle&=& \frac{ \int\!\rmD\sigma\,\rme^{ 
\beta \sigma                        \sum_{\mu\in\partial i} J^2_\mu  \xi_i^{\mu} \left(      \frac{1}{\rho} \sum_{j\in\partial \mu}   \xi_j^{\mu}\langle\sigma_j\rangle +  \theta_\mu           \right)     }    \sigma
 }{ \int\!\rmD\tilde\sigma\,\rme^{ \beta \tilde\sigma   
 \sum_{\mu\in\partial i} J^2_\mu  \xi_i^{\mu} \left(      \frac{1}{\rho} \sum_{j\in\partial \mu}   \xi_j^{\mu}\langle\sigma_j\rangle +  \theta_\mu           \right)
  } }\label{eq:Pi(s)-large-n-1-moment}.
\end{eqnarray}
Solutions of this equation can be used to compute the marginal T-clone distribution (\ref{eq:Pi(s)-large-n}) via the equation 
\begin{eqnarray}
P_i(\sigma)&=& \frac{ \int\!\rmD\sigma_i\,\rme^{ \beta \sigma_i   
 \sum_{\mu\in\partial i} J^2_\mu  \xi_i^{\mu} \left(      \frac{1}{\rho} \sum_{j\in\partial \mu}   \xi_j^{\mu}\langle\sigma_j\rangle +  \theta_\mu           \right)
} \delta\!\left(\sigma-\sigma_i\right)}{ \int\!\rmD\tilde\sigma_i\,\rme^{ \beta \tilde\sigma_i
 \sum_{\mu\in\partial i} J^2_\mu  \xi_i^{\mu} \left(      \frac{1}{\rho} \sum_{j\in\partial \mu}   \xi_j^{\mu}\langle\sigma_j\rangle +  \theta_\mu           \right)
 }}\label{eq:Pi(s)-large-n-final}
\end{eqnarray}
and the distribution (\ref{def:P(F)-real-n}) via the equation 
\begin{eqnarray}
P_\mu(F)&=&  \prod_{i\in\partial \mu} \frac{ \int\!\rmD\sigma_i\,\rme^{ \beta \sigma_i   
 \sum_{\nu\in\partial i} J^2_\nu  \xi_i^{\nu} \left(      \frac{1}{\rho} \sum_{j\in\partial \nu}   \xi_j^{\nu}\langle\sigma_j\rangle +  \theta_\nu           \right)
} }{\int\!\rmD\tilde\sigma_i\,\rme^{ \beta \tilde\sigma_i
 \sum_{\nu\in\partial i} J^2_\nu  \xi_i^{\nu} \left(      \frac{1}{\rho} \sum_{j\in\partial \nu}   \xi_j^{\nu}\langle\sigma_j\rangle +  \theta_\nu           \right)
 }}\nonumber\\
 &&~~~~\times\delta\!\left(F-F_\mu (\bsigma)\right)\label{eq:P(F)-large-n}.
\end{eqnarray}
The average B-cell concentration  $\langle c\rangle=\int_{0}^{\infty}\!\rmd c~c P(c)$ can be also computed from the integral $\langle c\rangle=M^{-1} \sum_{\mu\leq M} \lim_{\tilde\beta\rightarrow\infty}\int\!\rmd\bb~ P_{\tilde{\beta}, \beta}(\bb) \rme^{b_\mu }$  which gives us the equation 
\begin{eqnarray}
\langle c\rangle&=& \frac{1}{M} \sum_{\mu=1}^M\rme^{J_\mu\left(\frac{1}{\rho}\sum_{j\in\partial \mu} \xi_j^{\mu}\langle\sigma_j\rangle     +\theta_\mu\right)} \label{eq:<c>-large-n}.
\end{eqnarray}
In the case of Ising variables $\sigma_i\in\{-1,1\}$ and $ \xi_i^{\mu}\in\{-1,1\}$, which also includes the binary case $\sigma_i\in\{0,1\}$, the above framework leads to
\begin{eqnarray}
\langle\sigma_i\rangle&=& \tanh\Big(\beta  \sum_{\mu\in\partial i} J^2_\mu  \xi_i^{\mu}  \big(      \frac{1}{\rho} \sum_{j\in\partial \mu}  \xi_j^{\mu} \langle\sigma_j\rangle +  \theta_\mu           \big) \Big)   
\label{eq:<s>-large-n-Ising}.
\end{eqnarray}
%
%We note that this  is a  mean-field equation  which can be obtained by minimising the ``distance'' $\sum_\bsigma Q(\bsigma)\log\frac{Q(\bsigma)}{P(\bsigma)}$  between the distributions $Q(\bsigma)=\prod_{i=1}^N\frac{1}{2}(1+\sigma_i m_i)$ and $P(\bsigma)$ which is the distribution (\ref{eq:P(s)-integer-n-Ising}) with $n=1$, but the equation (\ref{eq:<s>-large-n-Ising}) contains $\langle\sigma_i\rangle$ on the RHS which is at difference with the MF solution of  (\ref{eq:P(s)-integer-n-Ising}). Suppose that  the system  governed by $P(\bsigma)$ has a phase diagram as in the Figure  \ref{figure:PD},  the question then arises if the mean-field system (\ref{eq:<s>-large-n-Ising}) has the same behaviour?  

The simplest nontrivial clonal interaction structure  is a regular network topology, as in Figure \ref{figure:2d}, with uniform interactions\footnote{We expect that  having non-uniform interactions $J_\mu$ would lead only to  quantitative but not qualitative differences with the uniform $J_\mu\!=\!J$ case studied here, i.e. we expect the same phase diagram in both cases,  provided that non-uniform $J_\mu$ are well behaved.} $J_\mu\!=\!J$, $\xi_i^\mu\!=\!1$ for all $(i,\mu)$,  and no self-activation, i.e. $\theta_\mu\!=\!0$.  Here equation (\ref{eq:<s>-large-n-Ising})  simplifies considerably. From  (\ref{eq:b-large-n})  we infer that $b_\mu=\frac{J}{\rho}K \langle \sigma_i\rangle$, with $\vert\partial \mu\vert=K$,  and if we define  the average magnetization  $m=\langle \sigma_i\rangle$ we find that 
\begin{eqnarray}
m&=&\tanh\beta(J^2LK m/\rho)\label{eq:m-large-n-ferro},
\end{eqnarray}
with $\vert\partial i\vert=L$.  The average B cell concentration $\langle c \rangle=\rme^{\frac{J}{\rho} K m}$ follows from (\ref{eq:<c>-large-n}).   The distribution of fields  (\ref{eq:P(F)-large-n}) takes the following simple form, with $P(F)=P_\mu(F)$:
 \begin{eqnarray}
 P(F)&=&\sum_{\{\sigma_j\}} \frac{\rme^{\beta \frac{J^2}{\rho}LK m\sum_{j=1}^K \sigma_j}}{2^K\cosh^K\big(\beta\frac{J^2}{\rho}LK m\big)}\delta\big(F-J\sum_{j=1}^K \sigma_j\big),\label{eq:P(F)-large-n-ferro}
 \end{eqnarray}

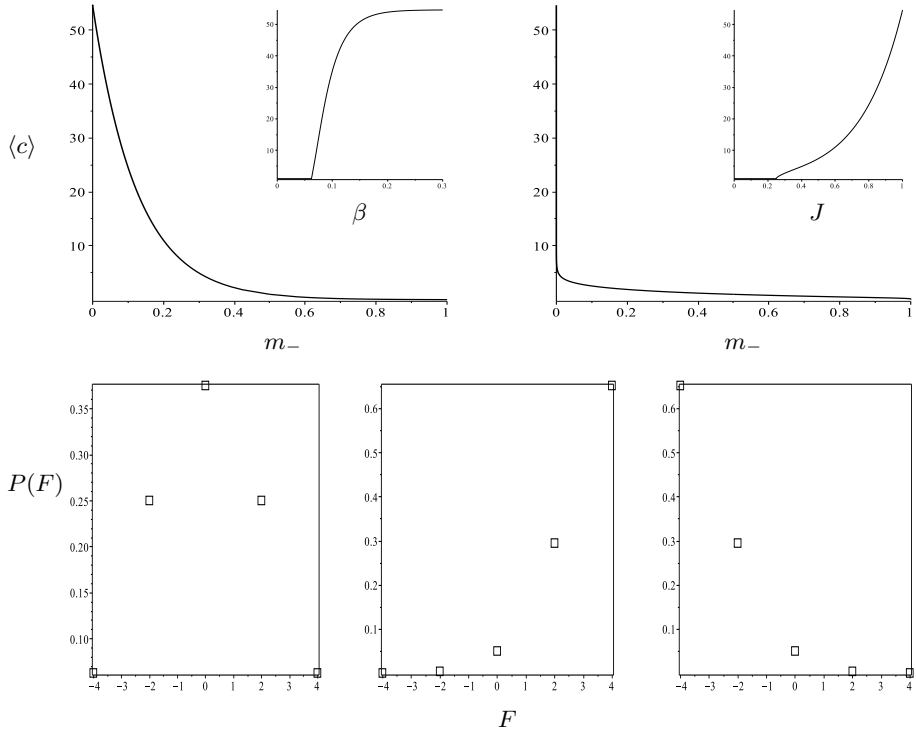
\begin{figure}[t]
%\vspace*{0mm} \hspace*{-9mm}
 \setlength{\unitlength}{0.45mm}
 \hspace*{5mm}
\begin{picture}(250,210)
 \put(8,110){\includegraphics[height=100\unitlength, width=120\unitlength]{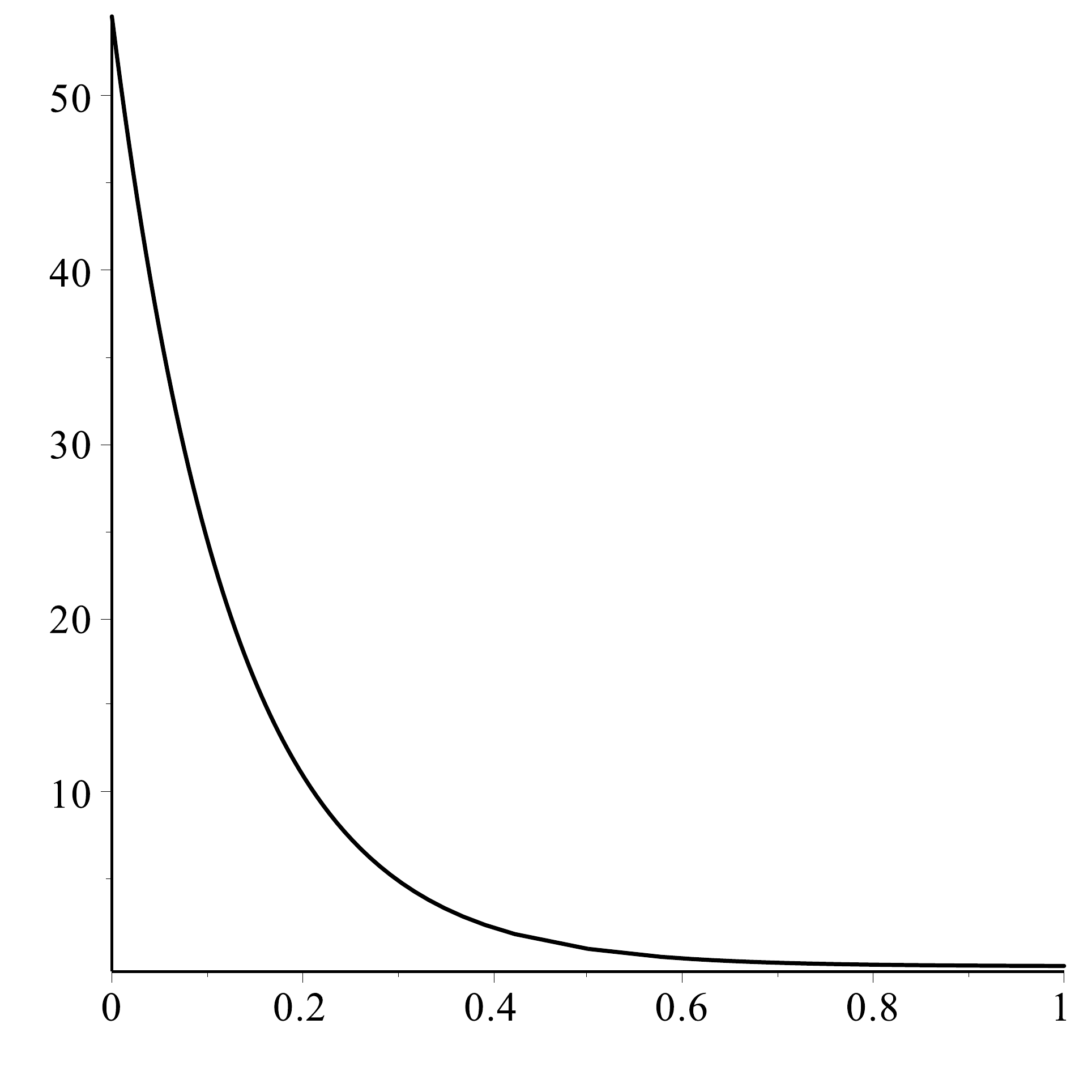}}
 %left inset
  \put(69,151){\includegraphics[width=57\unitlength]{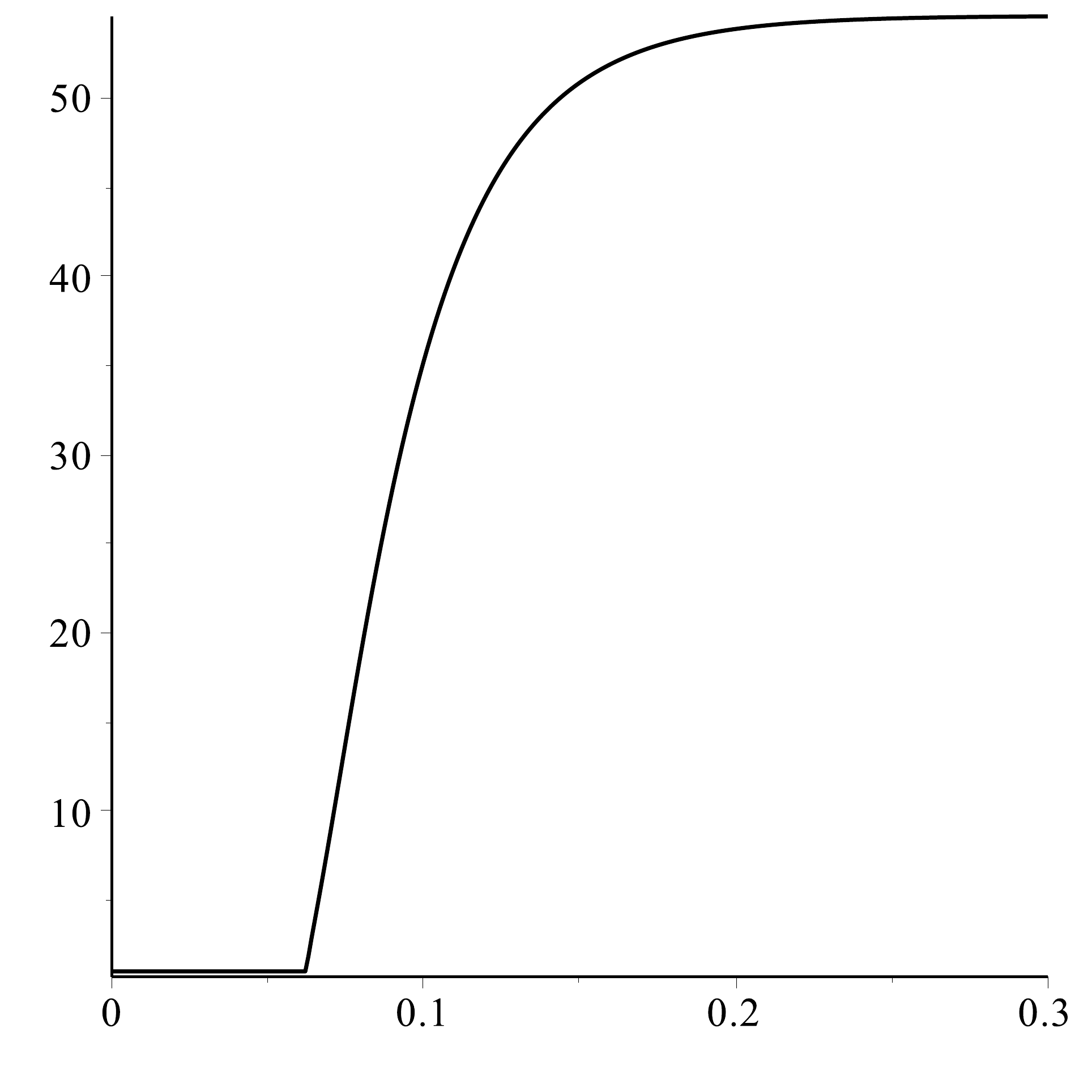}}%height=90\unitlength,
\put(97,145){\small{$\beta$}}
 \put(145,110){\includegraphics[height=100\unitlength, width=120\unitlength]{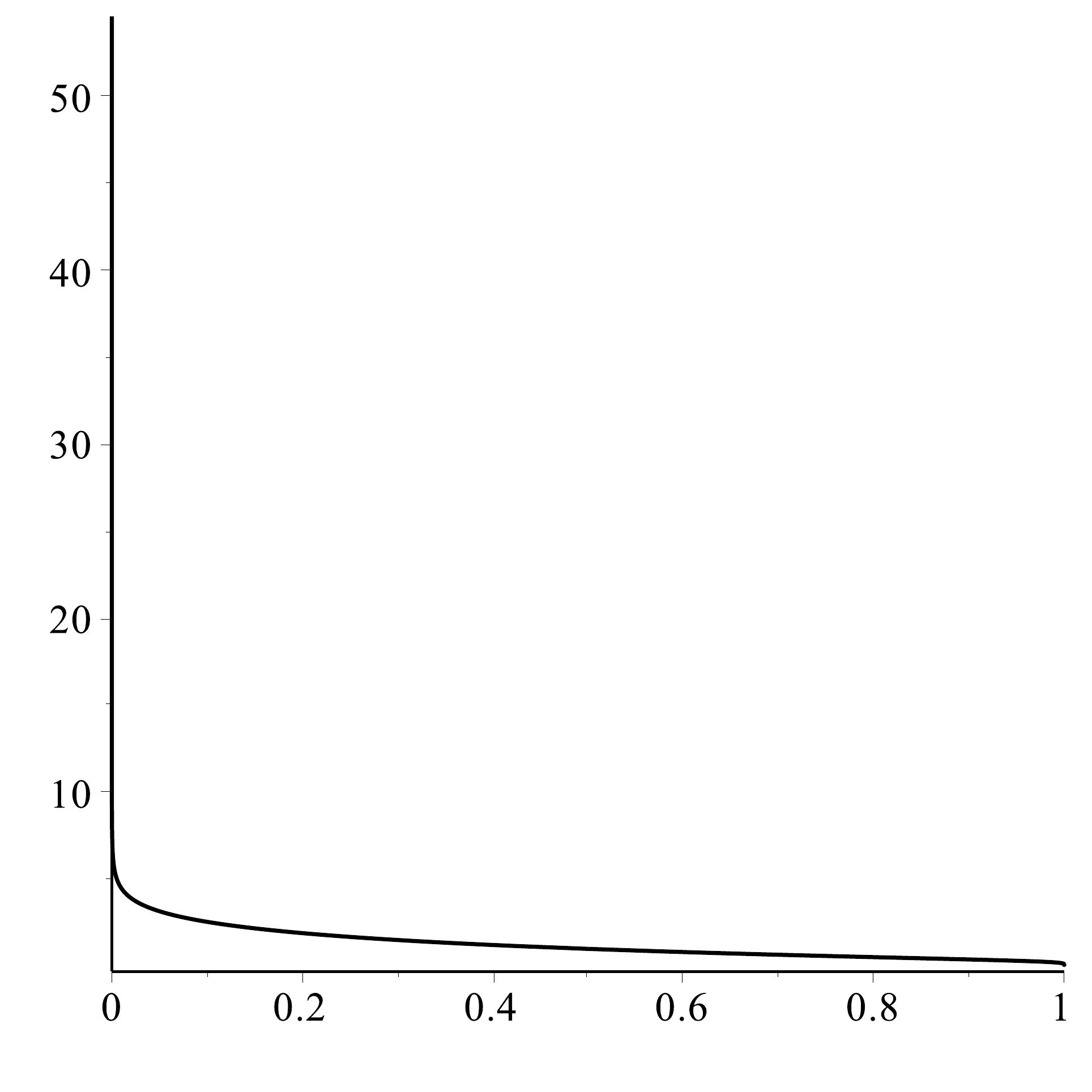}}
 %left inset
  \put(204,151){\includegraphics[width=57\unitlength]{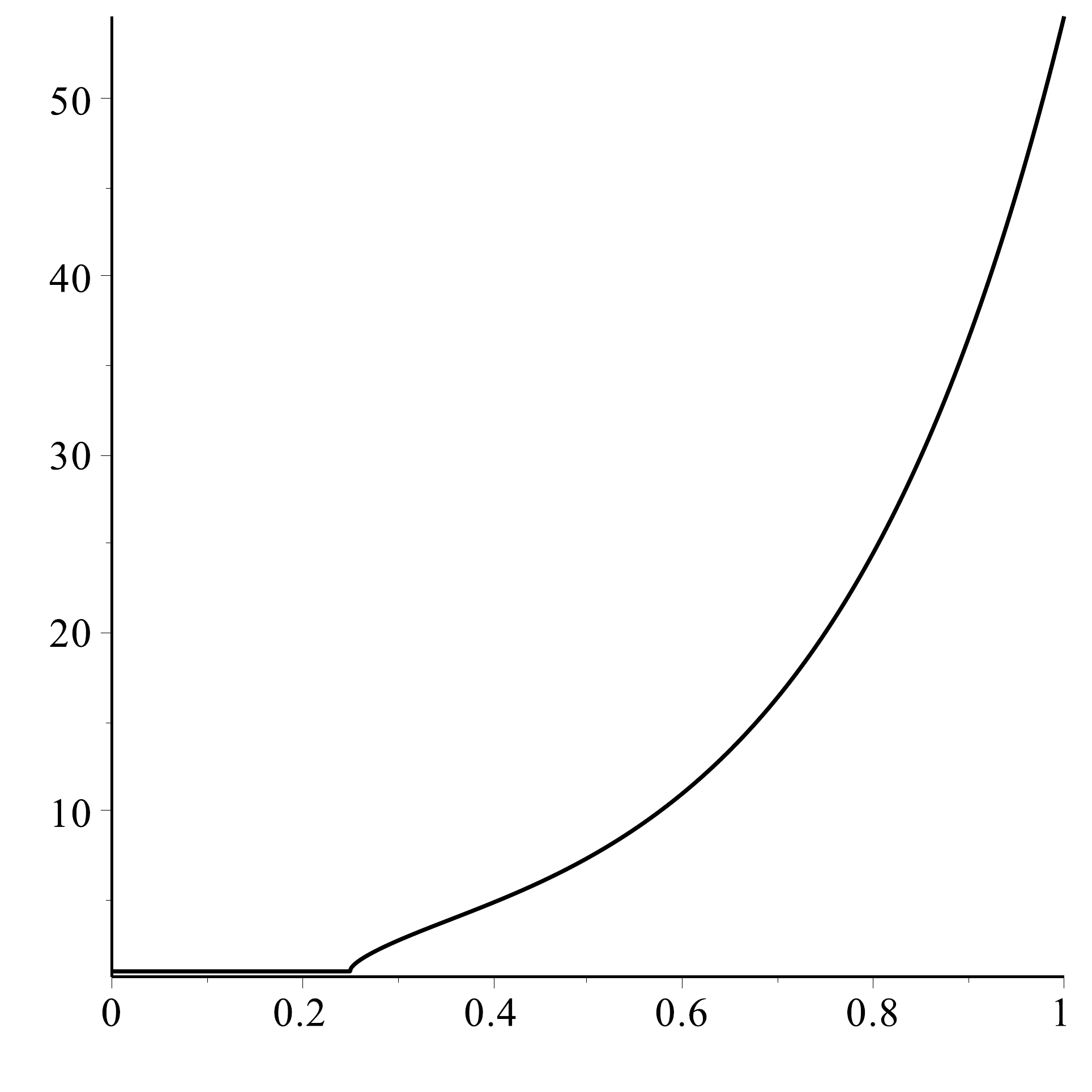}}%height=90\unitlength,
\put(232,145){\small{$J$}}
   \put(10,0){\includegraphics[height=100\unitlength, width=80\unitlength]{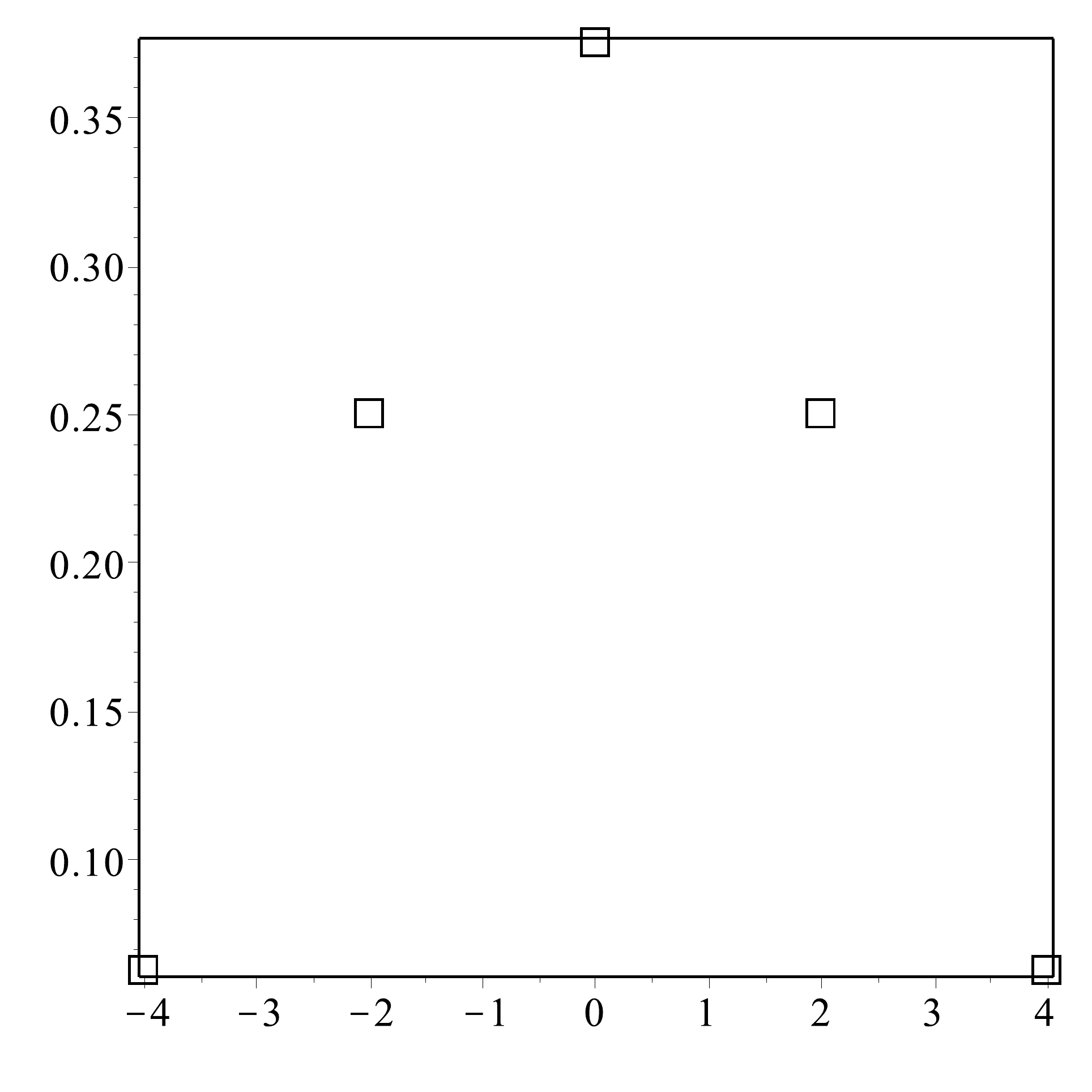}}
    \put(97,0){\includegraphics[height=100\unitlength, width=80\unitlength]{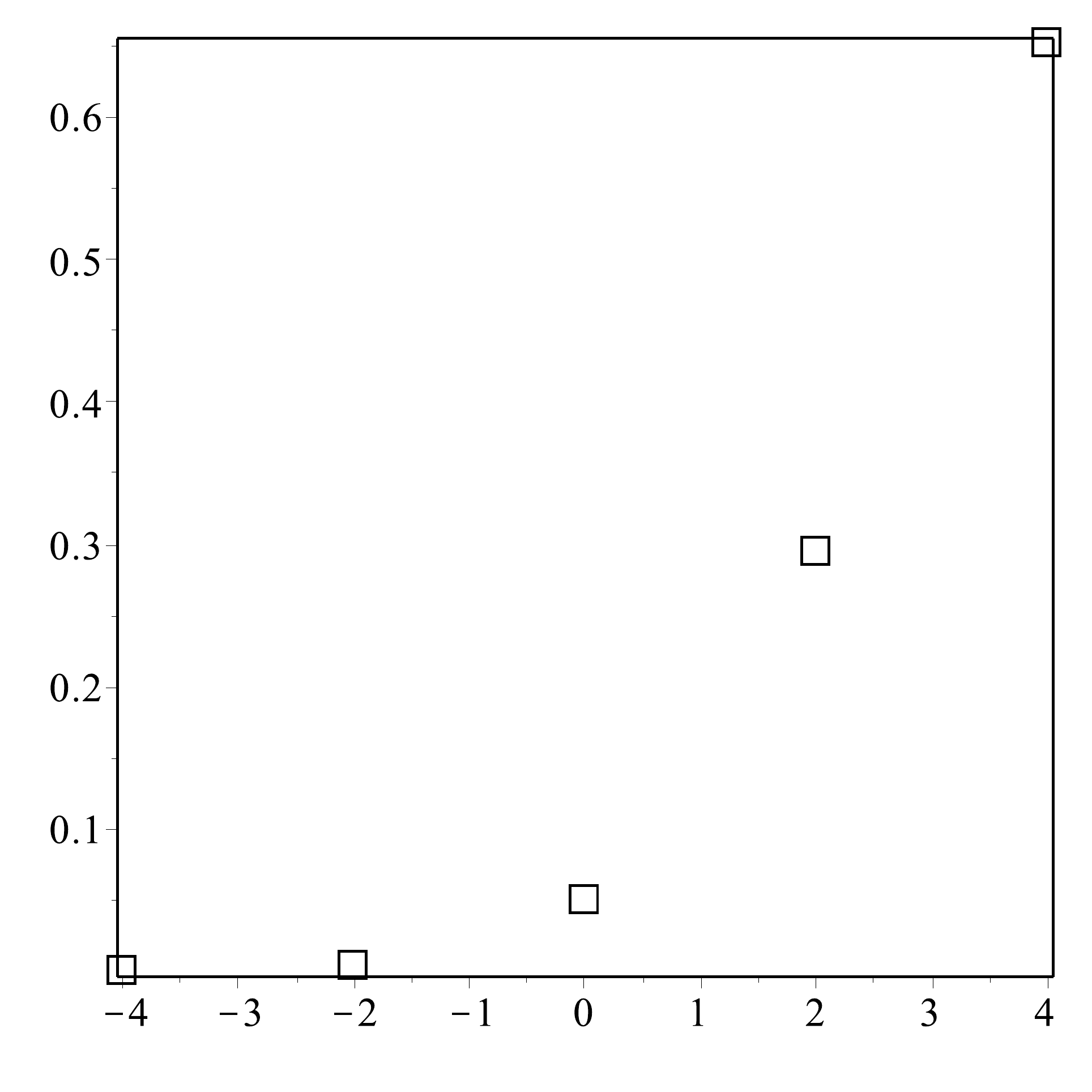}}
    \put(185,0){\includegraphics[height=100\unitlength, width=80\unitlength]{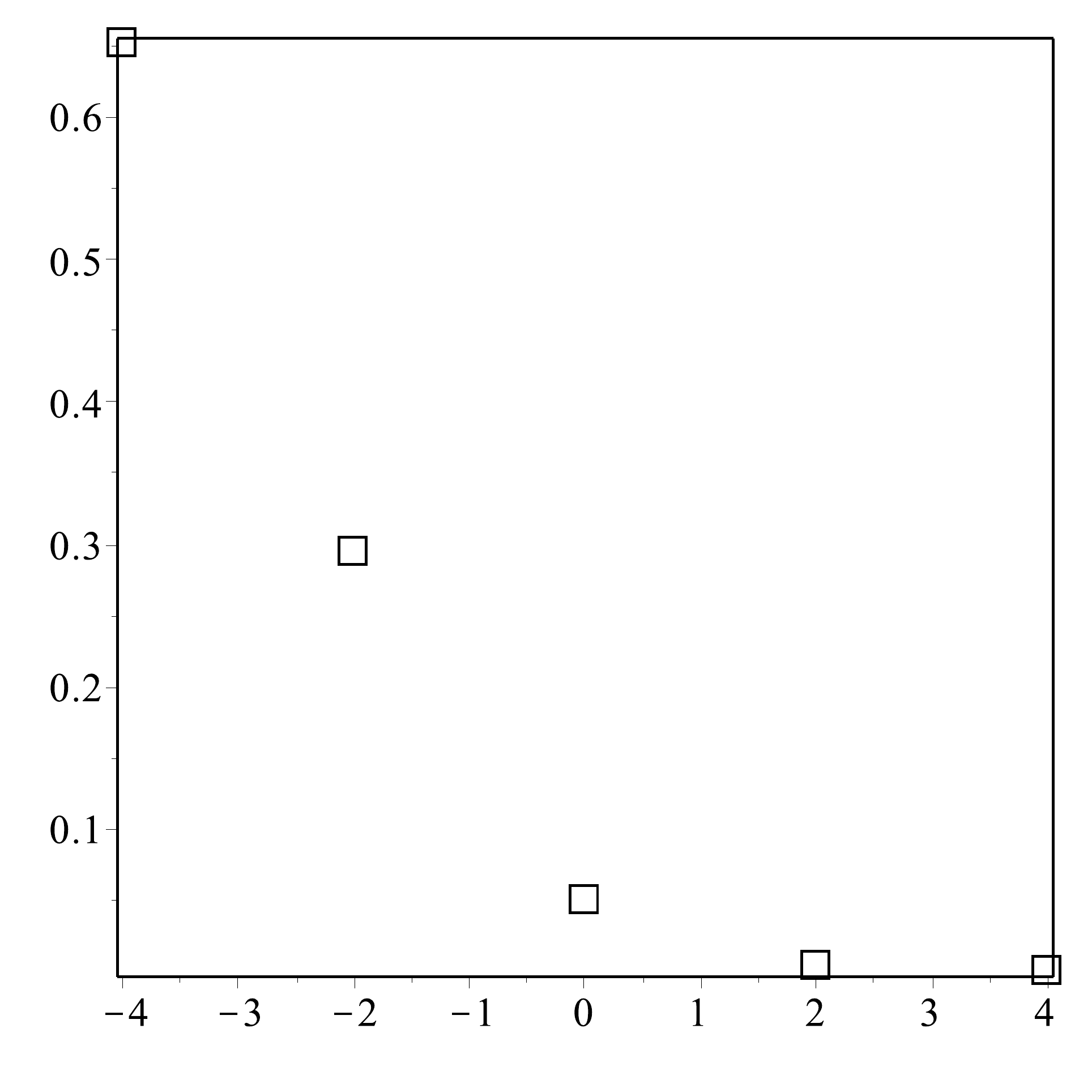}}
   \put(-5,65){\small{$P(F)$}} 
 \put(-5,165){\small{$\langle c\rangle$}}   
\put(140,-5){\small{$F$}}  % \put(196,-5){\small$F$}  
   \put(71,107){\small{$m_-$}}   \put(207,107){\small{$m_-$}}
  \end{picture}
 \vspace*{2mm}
\caption{Behaviour of  B- and T-clones in the  model with fast T-clone equilibration in the noiseless B-clone regime  ($\tilde\beta\rightarrow\infty$) . The system, here defined on a graph with  connectivity   $L=K=4$,     was studied  for the high $\beta<\beta_c$ ($\beta_c=0.0625$) and low $\beta>\beta_c$   T clone noise levels with $J=1$ and for the low $J<J_c$ ($J_c=0.25$) and high $J>J_c$  Ag levels with $\beta=1$.  Top left: The average B-clone size, $\langle c\rangle$, as a function of the  fraction of T-regulator cells , $m_-=\frac{1}{2}(1-m)$, for $J=1$ and $\beta\in\left[0, 0.5\right]$. Inset:  $\langle c\rangle$  as a function of $\beta$.   Top right:  $\langle c\rangle$ as a function of  $m_-$ for $\beta=1$ and $J\in\left[0, 1\right]$.  Inset:  $\langle c\rangle$  as a function of $J$.   Bottom:  The distribution $P(F)$  for $\beta=0.0525$  with $m_-=\frac{1}{2}$   (left),  for $\beta=0.0855$  with $m_-=0.1$ (centre) and for $\beta=0.0855$ with  $m_-=0.9$  (right).}  \label{figure:T-fast-large-n}
\end{figure}

 Equation (\ref{eq:m-large-n-ferro}) has  $m=0$ as its solution for any $\beta$, but this solution becomes unstable for $\beta>\beta_c$, where  $\beta_c=\frac{\rho}{J^2 LK}$,  and  two stable solutions $\vert m\vert\neq0$ emerge. Furthermore, the average number of B  cells, $\langle c \rangle$, is a monotonic non-increasing function of   the fraction of T regulator cells $m_{-}=\frac{1}{2}(1- m)$ (the representative case of $L=K=4$ and  $\rho=1$ is studied in the Figure \ref{figure:T-fast-large-n}).  The B-clones are receiving predominantly excitatory signals from the T clones, described by $P(F)$, when $m_{-}<\frac{1}{2}$, and predominantly inhibitory signals when $m_{-}>\frac{1}{2}$, as can be seen in Figure \ref{figure:T-fast-large-n}.  For $\beta<\beta_c$ the point $m=0$ is the only solution, and the B  cells  are no longer  controlled by the T cells: $\langle c \rangle=1$ for all $\beta$ and $J$.  For  $\beta>\beta_c$ and  $m_{-}<\frac{1}{2}$ the average number of B cells $\langle c \rangle$ is  increasing with $\beta$ and $J$ (see inset in Figure \ref{figure:T-fast-large-n}). Thus $\beta_c$  separates the  ``low-dose tolerance''  PM phase, where immune system is insensitive to Ag,  from the ``vigorous  immune response'' FM phase where it is very sensitive to Ag. 
%

%\emph{To obtain the  distributions of B-clones P(c) we have to specify networks:} 
So far we have been able to discuss the  behaviour of T- and B-clones by making only weak assumptions about network topologies and interactions.  This analysis is qualitative, and only covers the cases  where we can map our model onto  ferromagnetic Ising spin systems, which are quite well understood~\cite{Domb1972, Baxter1982}.  Unfortunately, one cannot construct such arguments for the case where $\sigma_i\in\{0,1\}$ with $\xi_i^\mu\in\{-1,1\}$, which is again equivalent to an Ising model but is no longer ferromagnetic; the map $\sigma:\{-1,1\}\rightarrow\{0,1\}$, where $\sigma=\frac{1+s}{2}$, here gives rise to site-dependent external fields, which could be positive or negative.  Moreover,  to make quantitive predictions about the observables such as fractions of T cells, $m_{\pm}$, and  concentrations of B cells, $\langle c\rangle$, and more informative observables  such as the B-clone distribution (\ref{eq:P(c)-integer-n}),  we have to be more specific about the   graph $\mathcal{G}$ and the interaction parameters $\{\xi_i^\mu,J_\mu\}$.  

In the absence of microscopic knowledge about interactivity in real immune systems, we will follow the ``Bayesian'' route and choose 
topological and interaction parameters randomly,  but subject to biological constraints such as  the average number of regulator T cells $p=\frac{1}{N}\sum_{i=1}^N\delta_{\xi_i;-1}$ or the number of T cells that a single B cell can interact with (for example in the graphs of  Figure  \ref{figure:2d} these numbers are $4$ and $8$)  which can be deduced from~\cite{Butler2013}.  
In real immune systems the interactions of lymphocytes with Ag occur  in a $3$-dimensional  volume, so it seems reasonable  to choose  a regular $3d$ lattice  for our graph  topology.  However, even if we choose a simpler $2d$ lattice, there are still many possibilities for how to construct the graph $\mathcal{G}$ (two of which are shown in Figure \ref{figure:2d}). Systems interacting on $d$-dimensional lattices, with  $d\!>\!1$, are hard to study analytically.  One therefore often uses locally tree-like~\cite{Mezard2009} random topologies,  an approximation which is expected to be good away from phase transitions ~\cite{Mozeika2015}.

\subsection{Bethe$€"-$Peierls approximation\label{section:BP}}
For systems interacting on trees, relevant local observables such as the distribution of fields (\ref{def:P(F)-real-n}) can be computed recursively. Let us first consider the distribution (\ref{eq:P(s)-integer-n-Ising}), which governs the (replicated) T-clone variables $\{\sigma_i^\alpha\}$ in both the fast B-clone ($n=1$) and fast T-clone ($n\in \mathbb{Z}^{+}$) equilibration regimes,  defined  on a factor-tree $\mathcal{T}_\mu(r)$ of radius $r$ which is centred at the factor-node $\mu$ (see Figure \ref{figure:factor-tree-integer-n}  in \ref{app:BP-integer-n}). The field $F_\mu(\{h_i\})=   J_\mu\big(\sum_{i\in \partial \mu}\xi_i^\mu h_i   + \theta_\mu \big) $, where $h_i=\sum_{\alpha=1}^n\sigma^\alpha_i$, on this factor node is governed  by the distribution\footnote{Derivation details of the first three equations in this section are provided in \ref{app:BP-integer-n}. }  
\begin{eqnarray}
P_\mu(F)
 &=&
\frac{\sum_{\{h_i\}} \left\{
 \prod_{i\in \partial \mu}  P_{i\mu} [h_i]\right\}\, \rme^{   \frac{\beta}{2n\rho}   F^2  } \,     }{
 \sum_{\{\tilde{h}_i\}}   \left\{ \prod_{i\in \partial \mu}   P_{i\mu} [\tilde{h}_i]\right\}  \rme^{ 
 \frac{\beta J_\mu^2}{2n\rho}  \left( \sum_{i\in \partial \mu}\xi_i^\mu \tilde{h}_i  + \theta_\mu   \right)^2           }}\nonumber\\
  &&~~~~~~~\times\delta\Big(F- J_\mu\big(\sum_{i\in \partial \mu}\xi_i^\mu h_i   + \theta_\mu \big)\Big)\label{eq:P(F)-using-P[h]},
\end{eqnarray}
where the cavity distribution $P_{\mu i} [h_i]$  can be computed recursively, starting from the variables  located in the boundary $\partial\mathcal{T}_\mu$ of the tree $\mathcal{T}_\mu$,   via the equation 
\begin{eqnarray}
 P_{\mu i} [h_i]&=&  \sum_{\{h_j  \}} \Bigg\{\!\prod_{\nu\in \partial i\setminus \mu }\prod_{j  \in \partial \nu \setminus i}\!   P_{\nu j} [h_j]\!\Bigg\}\! 
           \sum_{\sigma^1,\ldots,\sigma^n} \delta_{h_i; \sum_{\alpha=1}^n\sigma^\alpha}                  \nonumber
           \\
 &&\times  \rme^{        \frac{\beta}{2n\rho} \! \sum_{\nu\in \partial i\setminus \mu} \! J_\nu^2\left(\sum_{j\in \partial \nu}\xi_j^\nu h_j +\theta_\nu\right)^2 } \nonumber\\
 && \times\Bigg[
    \sum_{\tilde{h}_i}\!\sum_{\{\tilde{h}_j\}}\!      
    \Bigg\{\!\prod_{\nu\in \partial j\setminus \mu }\prod_{j\in \partial \nu \setminus i  }\!   P_{\nu j} [\tilde{h}_j]\!\Bigg\}  \sum_{\tilde{\sigma}^1,\ldots,\tilde{\sigma}^n} \delta_{\tilde{h}_i; \sum_{\alpha=1}^n\tilde{\sigma}^\alpha} \nonumber\\
    &&~~~~ \times \rme^{        \frac{\beta}{2n\rho} \! \sum_{\nu\in \partial i\setminus \mu}\!  J_\nu^2\left(\sum_{j\in \partial \nu}\xi_j^\nu \tilde{h}_j +\theta_\nu\right)^2 }\Bigg]^{-1}.
     \label{eq:P[h]} 
\end{eqnarray} %\sum_{i\in \partial \nu\setminus j}\xi_i^\nu h_i  + \xi_j^\nu h_j
Similarly the magnetization $ \langle\sigma_i\rangle=\frac{1}{n}\sum_h P_i(h)h$, where $P_i(h)=\sum_\bsigma P_i(\bsigma)\delta_{h; \sum_{\alpha=1}^n\sigma^\alpha}$, of the tree $\mathcal{T}_i$ can be computed  via the equation
\begin{eqnarray}
\langle\sigma_i\rangle
&=&\sum_h\frac{1}{n}\sum_{\{h_j\}} \Bigg\{  \prod_{\mu\in\partial i}  \prod_{j\in\partial\mu\setminus i} P_{\mu j}[h_j]  \Bigg\}  \sum_\bsigma \delta_{h; \sum_{\alpha=1}^n\sigma^\alpha}
\nonumber\\
&&~~\times \rme^{\frac{\beta   }{2n\rho}\sum_{\mu\in\partial i}J_\mu^2\left(\sum_{j\in\partial\mu\setminus i} \xi_j^\mu h_j +  \xi_i^\mu h  +\theta_\mu)\right)^2 }h\nonumber\\ 
&&~~\times\Bigg\{\sum_{\tilde{h}}\sum_{\{\tilde{h}_j\}}  \left\{  \prod_{\mu\in\partial i}  \prod_{j\in\partial\mu\setminus i} P_{\mu j}[\tilde{h}_j]  \right\}\sum_{\tilde{\bsigma}}\delta_{\tilde{h}; \sum_{\alpha=1}^n\tilde{\sigma}^\alpha}  
\label{eq:m-using-P[h]}
\\
&&~~~~~~~\times  \rme^{\frac{\beta   }{2n\rho}\sum_{\mu\in\partial i}J_\mu^2\left(\sum_{j\in\partial\mu\setminus i} \xi_j^\mu \tilde{h}_j +  \xi_i^\mu \tilde{h} +\theta_\mu)\right)^2 }  \Bigg\}^{-1}
\nonumber.
\end{eqnarray}

 We note that the equations derived in this section can be used to compute local observables, such as the distribution $P_{\mu}(F)$ or the local magnetization $\langle\sigma_i\rangle$, on locally-tree like graphs.  In these graphs a  ``ball''  of radius $r$ centred on any variable node $i$ (or factor node $\mu$ ) converges to the tree of radius $r$ centred at this node, $\mathcal{T}_i(r)$, when $N\rightarrow\infty$ (or when $M\rightarrow\infty$ with  $M/N<\infty$), for any $r$~\cite{Mezard2009}.   An observable associated with node $i$ in such graphs is usually approximated by the same observable computed  on $\mathcal{T}_i(\infty)$\footnote{This procedure is exact for ferromagnetic Ising models with uniform interactions~\cite{Dembo2010}.}.  Furthermore, local observables computed on random trees can be used to compute densities: for example, upon assuming that the density $P(F)=\lim_{M\rightarrow\infty}\frac{1}{M}\sum_{\mu=1}^M P_{\mu}(F)$ is self-averaging~\cite{Mezard2009} we can replace this average with the average over the trees  $\mathcal{T}_\mu(\infty)$ which for $n\in \mathbb{Z}^{+}$ gives us the equation 
\begin{eqnarray}
\hspace*{-15mm}
P(F)
&=& \sum_{K\geq1}P(K) \left\{\prod_{i=1}^K \int\left\{\rmd P_i\right\} W[\{P_i\}] \right\} \int \!\rmd J ~P(J) \int\! \rmd\theta~ P(\theta)\nonumber\\
\hspace*{-15mm}
  &&\times\Bigg\langle\!
\frac{\sum_{\{h_i\}} \left\{
 \prod_{i=1}^K  P_i[h_i\vert\xi_i]\right\}\! \rme^{   \frac{\beta}{2n\rho}   F^2  } \delta\!\left(F- J\left(\sum_{i=1}^K\xi_i h_i   + \theta \right)\right) }{
 \sum_{\{\tilde{h}_i\}}   \left\{ \prod_{i=1}^K   P_i [\tilde{h}_i\vert\xi_i]\right\}  \rme^{ 
 \frac{\beta J^2}{2n\rho}  \left( \sum_{i=1}^K\xi_i \tilde{h}_i  + \theta   \right)^2           }} \!\Bigg\rangle_{\!\{\xi_i\}}.\label{eq:P(F)-integer-n-pop-dyn}
\end{eqnarray}
The 
above equation was derived for the ensemble of random graphs with the prescribed distributions $P(L)$ and  $P(K)$ of the variable-node and factor-node connectivities,  respectively.  We have also assumed that all parameters $\{\xi_i\}$,  $\{J_\mu\}$ and $\{\theta_\mu\}$ are were drawn independently at random from the distributions $P(\xi)$,  $P(J)$ and $P(\theta)$ respectively.  The distribution $W[\{P\}]$ is usually approximated by the (empirical) density $W[\{P\}]=\frac{1}{\mathcal{N}}\sum_{i\leq \mathcal{N}}\prod_{h, \xi}\delta\left(P[h\vert\xi] \!-\! P_i[h\vert\xi]\right)$ which is obtained via a population dynamics algorithm~\cite{Mezard2001},  which at each step replaces a member $i$ of the population $\mathcal{P}=\{P_j[h\vert\xi]\}$ with  the new value  
%$W[\{P\}]\equiv W[\{P[h\vert\xi]\}]$ %\lim_{\mathcal{N}\rightarrow\infty}
%
\begin{eqnarray}
  P_i[h\vert\xi]  &=&   
 \sum_{\{h_j^a\}}   \Bigg\{ \prod_{a=1}^{L-1}\prod_{j=1}^{K-1}   P _{i_{(a,j)}}[h_j^a\vert\xi_j^a]\Bigg\}  \sum_{\sigma^1,\ldots,\sigma^n}\delta_{h; \sum_{\alpha=1}^n\sigma^\alpha} \label{eq:P[h]-pop-dyn}\\
 &&\hspace*{10mm}\times \rme^{        \frac{\beta }{2n\rho}\! \sum_{a=1}^{L-1}\! J_a^2 (   \sum_{j=1}^{K-1} \! \xi_j^a h_j^a    +\theta_a+\xi h )^2  }  \nonumber\\
 &&\times \Bigg[ 
  \sum_{\tilde{h}} \sum_{\{\tilde{h}_j^a\}}    \Bigg\{\prod_{a=1}^{L-1}\prod_{j=1}^{K-1}   P_{i_{(a,j)}} [\tilde{h}_j^a\vert\xi_j^a]\Bigg\}   \sum_{\tilde{\sigma}^1,\ldots,\tilde{\sigma}^n}\delta_{\tilde h; \sum_{\alpha=1}^n\tilde{\sigma}^\alpha}\nonumber\\
  &&\hspace*{10mm}\times \rme^{        \frac{\beta }{2n\rho} \sum_{a=1}^{L-1} J_a^2 (   \sum_{j=1}^{K-1}  \xi_j^a\tilde{h}_j^a    +\theta_a+\xi\tilde{h} )^2  } \Bigg]^{-1}\nonumber
  \end{eqnarray}
 which is computed using the $(L\!-\!1)\!\times\!(K\!-\!1)$  distributions $P _{i_{(a,j)}}[h\vert\xi]$. Here  $i_{(a,j)}\in\{1,\ldots,\mathcal{N}\}$ is drawn randomly and uniformly  from the population $\mathcal{P}$, and $L$ and $K$ are random integers drawn from $Q(L)=LP(L)/\sum_{\tilde{L}\geq1} \tilde{L} P(\tilde{L})$ and $Q(K)=KP(K)/\sum_{\tilde{K}\geq1} \tilde{K} P(\tilde{K})$,  respectively. The parameters $\{\xi_i^a\}$,  $\{J_a\}$ and $\{\theta_a\}$ are also random numbers, drawn from the distributions $P(\xi)$,  $P(J)$ and $P(\theta)$, respectively.
 
Finally,  we note that Bethe$€"-$Peierls approximation also can be  used to study the $n\in \mathbb{R}^{+}$ case (see \ref{app:BP-real-n}). 

\subsection{Analysis of homogeneous systems on random regular factor-graphs\label{section:ferro-n-integer}}
\begin{figure}[!t]
\setlength{\unitlength}{1mm}
\begin{center}
\hspace*{4mm}
\begin{picture}(130,53)
\put(0,0){\includegraphics[width=120\unitlength]{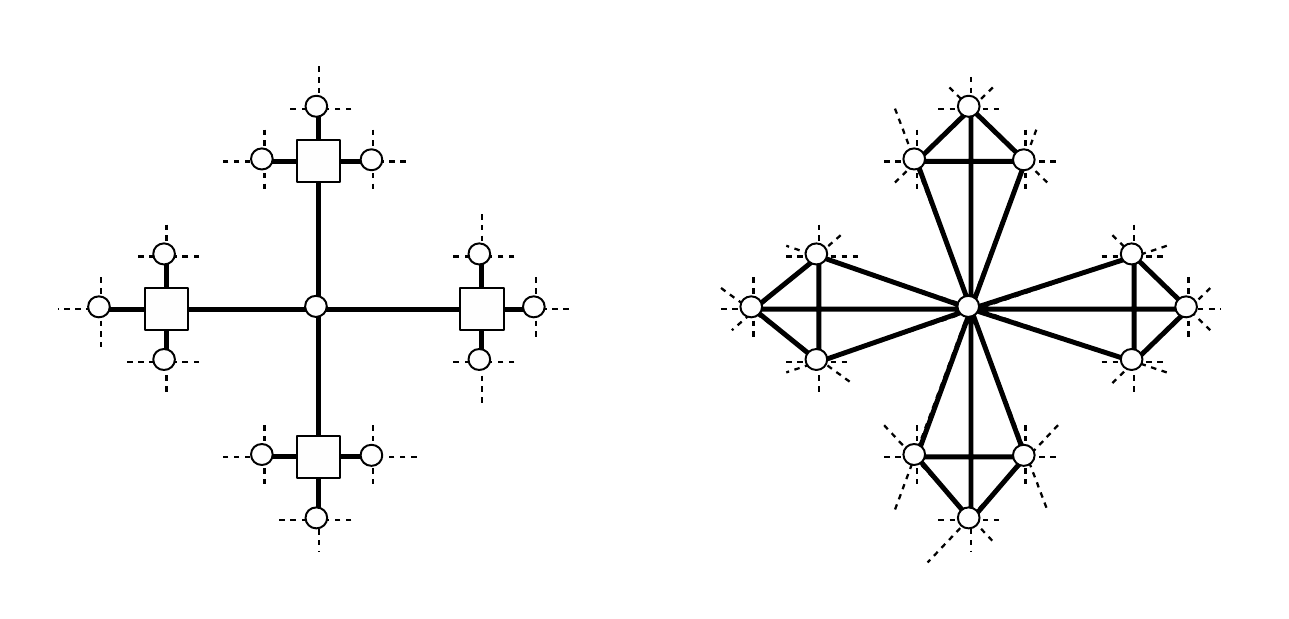}}
\end{picture}
\end{center}
\vspace*{-7mm}
\caption{An immune system with  the interaction  topology of  a random  $(K, L)$-regular factor-graph (left)  gives rise to an effective system of T-clones  interacting on a random $(K, L)$-regular clique graph (right). %T-clones and B-clones are represented by circles and squares respectively
 \label{figure:KL4} }
\end{figure}

The simplest non-trivial interaction topologies  that allows us to obtain more quantitative results are models defined on random regular  factor-graphs.  The ensemble of these graphs contains all graphs with fixed connectivities of variable-nodes, $\vert \partial i \vert=L$, and  factor-nodes, $\vert \partial \mu \vert=K$. This regularity imposes the constraint $L/K=M/N$ on the ratio  of factor-nodes (B-clones) to variable nodes (T-clones).  We also assume that $J_\mu=J$ and  $\theta_\mu=\theta$, and consider the  case of $\sigma_i\in\{-1,1\}$ with $n\in \mathbb{Z}^{+}$.   For $N\rightarrow\infty$, random regular factor-graphs are locally tree-like~\cite{Mezard2009}, and we therefore expect that the ``tree approximation'', described in the previous section will be exact on such graphs.  Furthermore,  the  (local) topology of  the T-clones system (\ref{eq:P(s)-integer-n-Ising}) in this connectivity regime  resembles a variant of  the Bethe lattice~\cite{Baxter1982}, where each node is connected to exactly $L$ fully connected graphs (cliques) of size $K$  (see Figure \ref{figure:KL4}).  From now on we will call such a graph a random $(K, L)$-regular clique graph. 

Let us consider the system (\ref{eq:P(s)-integer-n-Ising}) on one of such graphs. Since the variable nodes and factor nodes are now all equivalent,  the cavity distribution $P_{i\mu} [h] $ in equation (\ref{eq:P[h]}) is the same for all $i$ and $\mu$, i.e. $P_{i\mu} [h] =  P[h]$, which gives us the recursive equation 
%
%\begin{eqnarray}
%  P[h]  &=&   
% \sum_{\{h_j^a\}}   \left\{ \prod_{a=1}^{L-1}\prod_{j=1}^{K-1}   P [h_j^a]\right\}  \rme^{        \frac{\beta J^2}{2n\rho} \sum_{a=1}^{L-1}  (   \sum_{j=1}^{K-1}  h_j^a    +h )^2  }  {{n}\choose{\frac{n+h}{2}}} \label{eq:P[h]-ferro}\\
% &&~~~\times \Bigg\{ 
%  \sum_{\tilde{h}} \sum_{\{\tilde{h}_j^a\}}    \left\{\prod_{a=1}^{L-1}\prod_{j=1}^{K-1}   P [\tilde{h}_j^a]\right\}\nonumber\\
%  &&~~~~~~~~~\times \rme^{        \frac{\beta J^2}{2n\rho} \sum_{a=1}^{L-1}  (   \sum_{j=1}^{K-1}  \tilde{h}_j^a    +\tilde{h} )^2  } {{n}\choose{\frac{n+\tilde{h}}{2}}}\Bigg\}^{-1}\nonumber,
%  \end{eqnarray}
%
\begin{eqnarray}
  P[h]&=&\frac{\Big[\sum_{\{h_j\}}\left\{ \prod_{j=1}^{K-1}P [ h_j]\right\}\rme^{ \frac{\beta J^2}{2n\rho} \left(\sum_{j=1}^{K-1} h_j +h\right)^2}\Big]^{L-1}   {{n}\choose{\frac{n+h}{2}} }    } {
 \sum_{\tilde{h}}   \Big[\sum_{\{\tilde{h}_j\}} \left\{ \prod_{j=1}^{K-1}P [ \tilde{h}_j]\right\}\rme^{ \frac{\beta J^2}{2n\rho} \left(\sum_{j=1}^{K-1} \tilde{h}_j +\tilde{h} \right)^2}\Big]^{L-1}                                      {{n}\choose{\frac{n+\tilde{h}}{2}} }    }\label{eq:P[h]-ferro} ,
\end{eqnarray}
where  $h\in\{-n,-n+1, \ldots, n-1, n\}$, and where the binomial coefficient ${{n}\choose{\frac{n+h}{2}}}$ results from the computation of $\sum_{\sigma^1,\ldots,\sigma^n} \delta_{h; \sum_{\alpha=1}^n\sigma^\alpha}$. The solution of this equation  can be used to compute the field distribution (\ref{eq:P(F)-using-P[h]}),  via the equation
\begin{eqnarray}
  P(F)  &=&    \frac{  \sum_{\{h_j\}}  \left\{ \prod_{j=1}^{K}   P [h_j]\right\} \rme^{        \frac{\beta}{2n\rho}F^2  }   \delta\big(F-   J\sum_{j=1}^{K} h_j\big)
 }{
 \sum_{\{\tilde{h}_j\}}   \left\{ \prod_{j=1}^{K}   P [\tilde{h}_j]\right\} \rme^{        \frac{\beta J^2}{2n\rho}( \sum_{j=1}^{K} \tilde{h}_j)^2  }
  }\label{eq:P(F)-ferro},
\end{eqnarray}
% the density of fields (\ref{eq:P(F)-integer-n})
The same is true for the average magnetization $m=\frac{1}{n}\sum_h P(h) h$, where $P(h)$ is  the RHS of  the equation (\ref{eq:P[h]-ferro})  with $L=L+1$.  The latter result is obtained by comparing the equations  (\ref{eq:m-integer-n}) and (\ref{eq:P[h]}). 

\begin{figure}[t]
%\vspace*{0mm} \hspace*{-9mm}
 \setlength{\unitlength}{0.42mm}
 \hspace*{2mm}
\begin{picture}(310,102)
\put(0,0){\includegraphics[height=100\unitlength,width=100\unitlength]{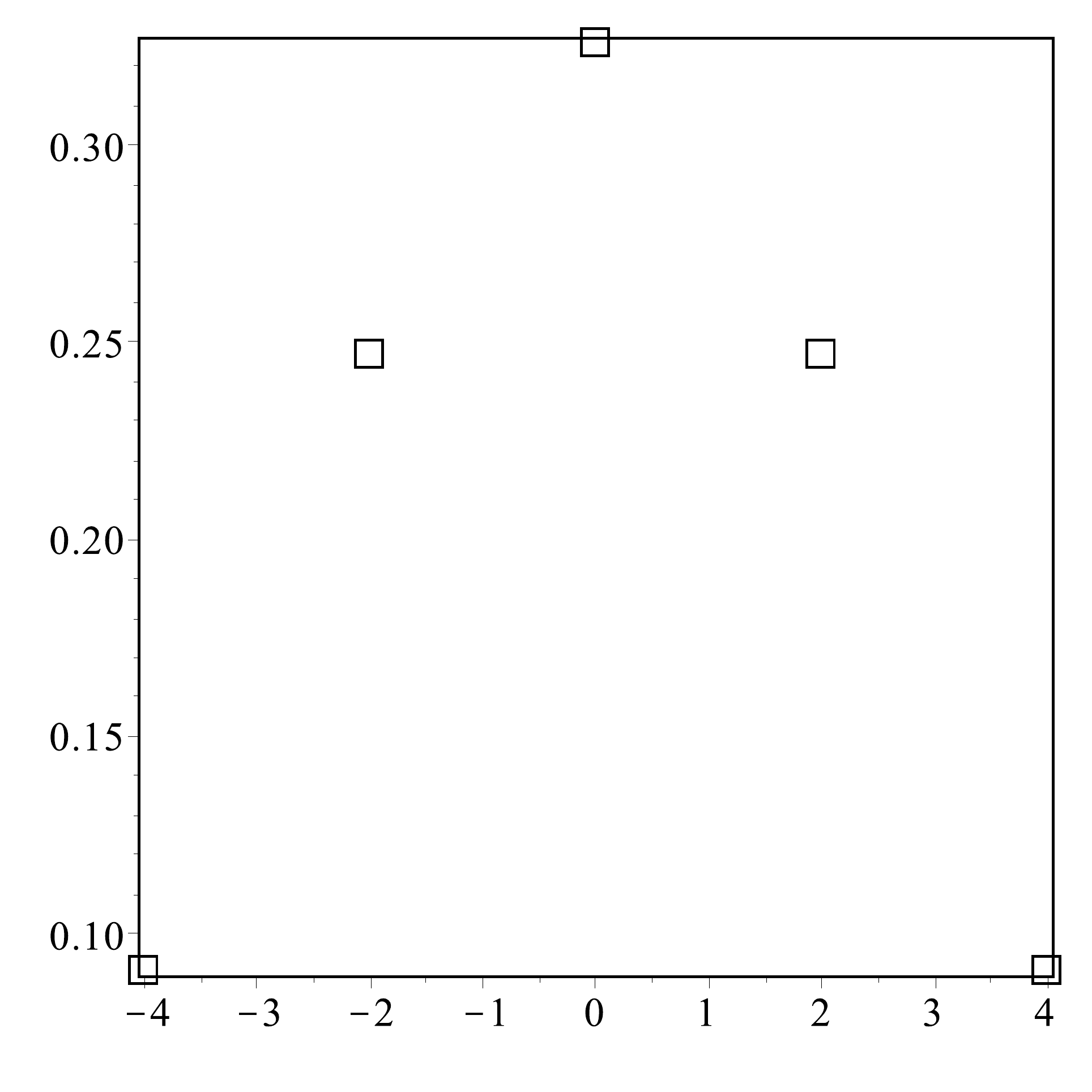}}%height=100\unitlength,
   
   \put(100,0){\includegraphics[height=100\unitlength,width=100\unitlength]{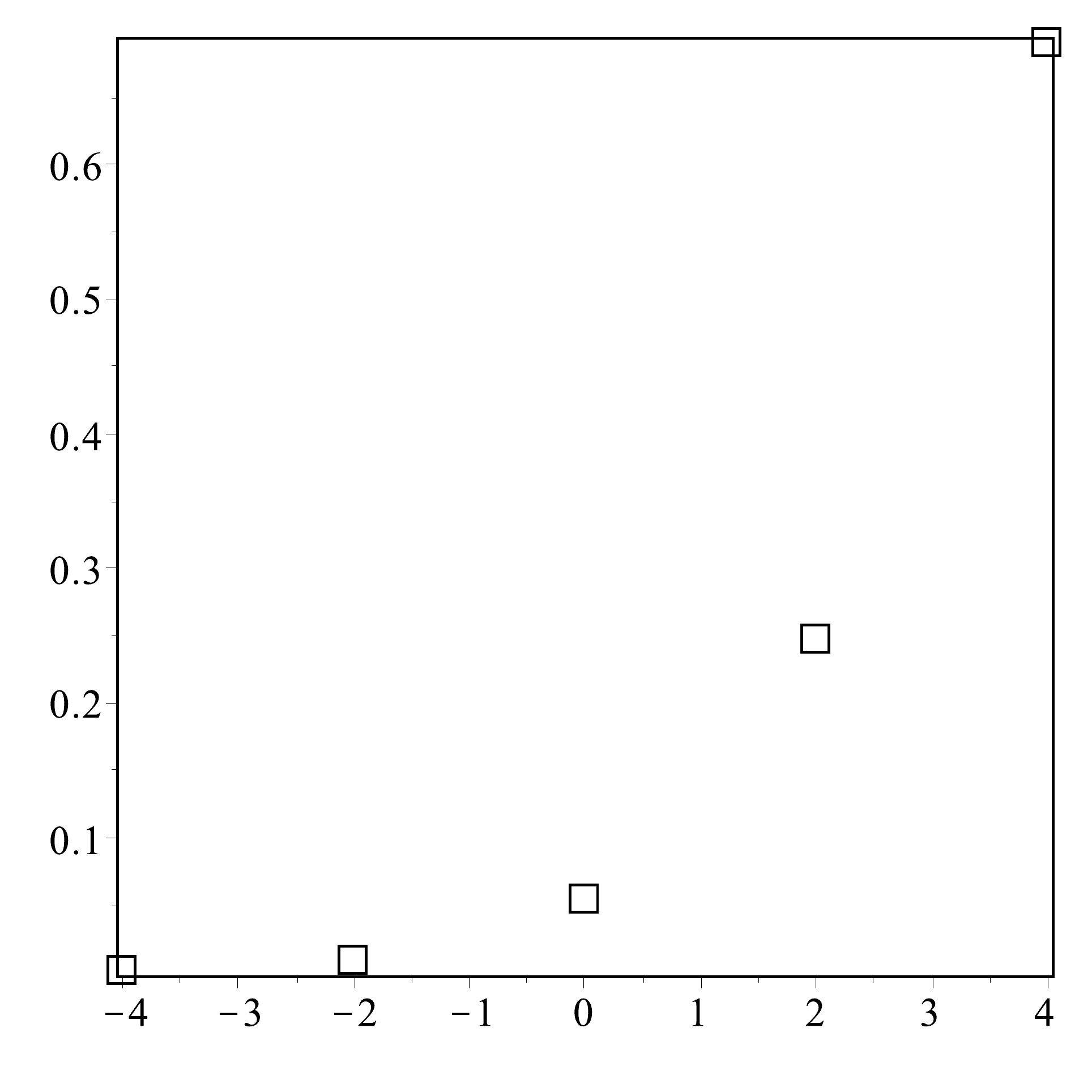}}
   
    \put(200,0){\includegraphics[height=100\unitlength,width=100\unitlength]{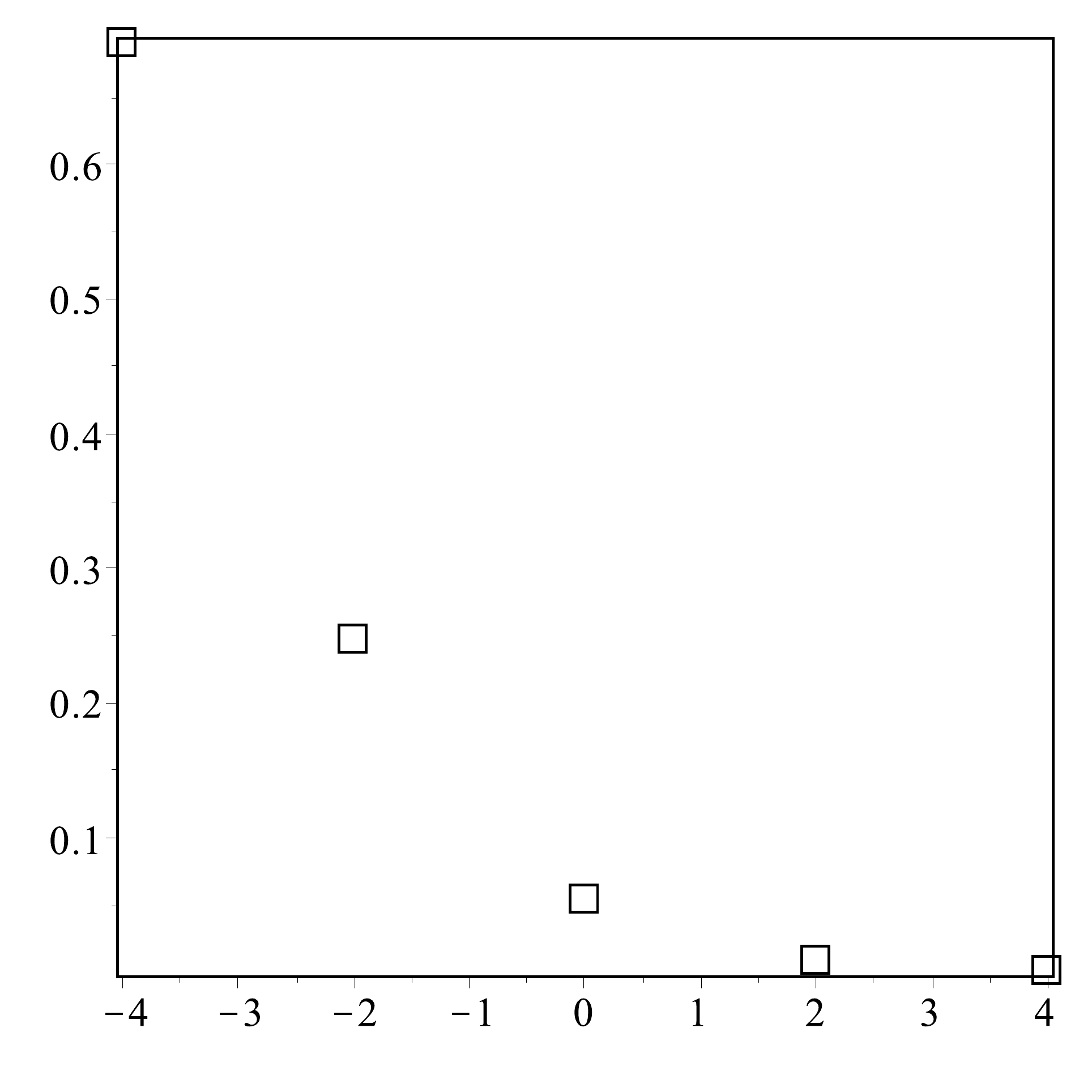}}
   
  \put(-15,65){\small{$P(F)$}} 
 
\put(150,-2){\small{$F$}}   
  
  \end{picture}
 \vspace*{-2mm}
\caption{The distribution $P(F)$ of the T-clone activity $F$ in the immune system model with fast B-clone equilibration,   defined on a random regular factor-graph with connectivity $L\!=\!K\!=\!4$. We show data for the high $\beta\!<\!\beta_c$ and low $\beta\!>\!\beta_c$   T-clone noise regimes (note: $\beta_c\!\approx\!0.0929$), for  the values $\beta=0.0639$, with $m_-=\frac{1}{2}$ (left), $\beta=0.1219$ with  $m_-=0.1$ (centre) and  $\beta=0.1219$ with $m_-=0.9$ (right).} \label{figure:P(F)-B-fast}
\end{figure}

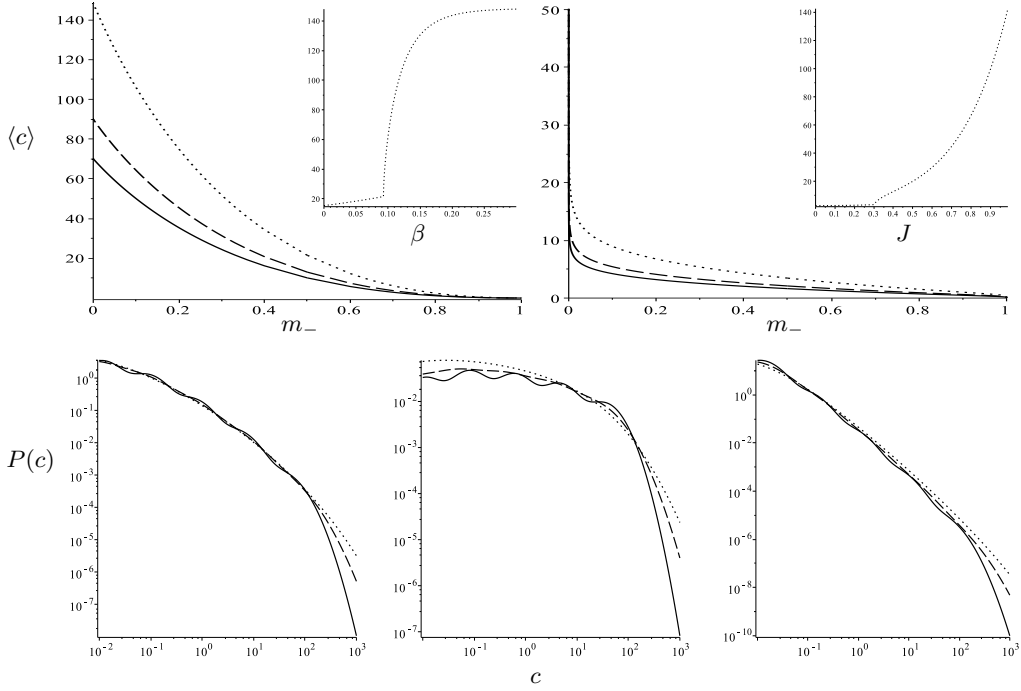
\begin{figure}[t]
%\vspace*{0mm} \hspace*{-9mm}

 \setlength{\unitlength}{0.43mm}
\begin{picture}(310,210)

 \put(-5,165){\small{$\langle c\rangle$}}
 \put(8,110){\includegraphics[height=100\unitlength,width=150\unitlength]{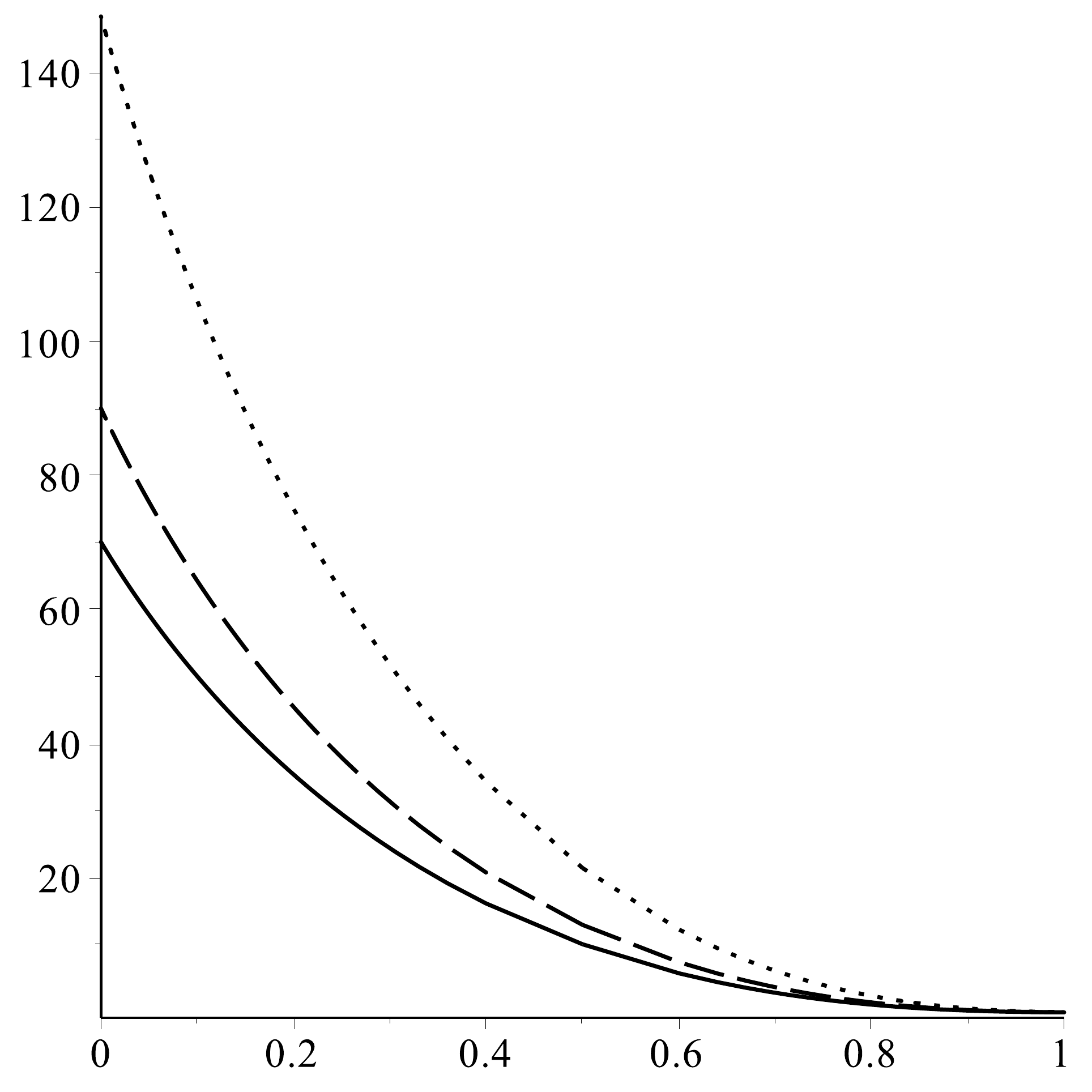}} 
  \put(80,107){\small{$m_-$}} 
 \put(87,141){\includegraphics[width=67\unitlength]{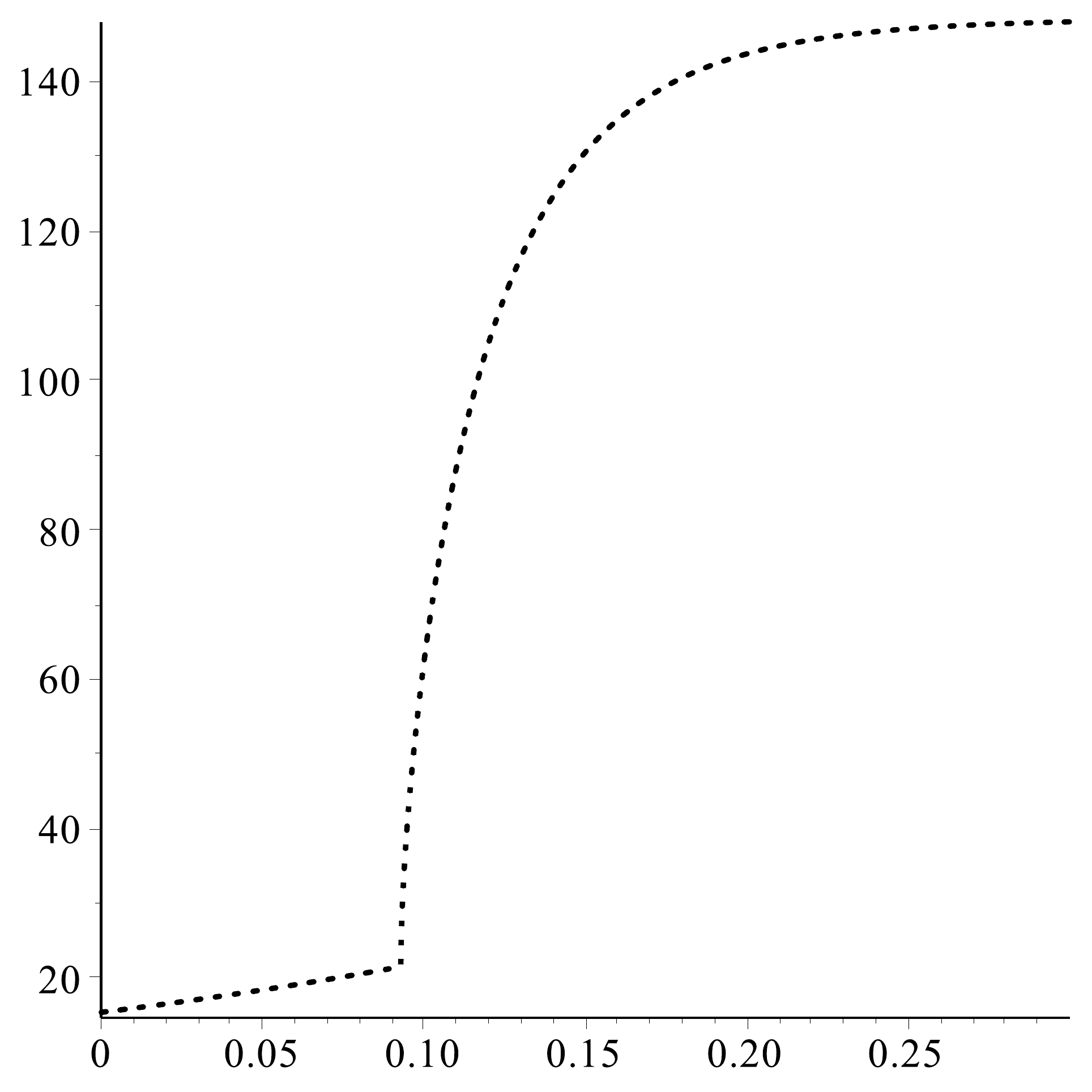}}
\put(120,135){\small{$\beta$}}

 \put(158,110){\includegraphics[height=100\unitlength,width=150\unitlength]{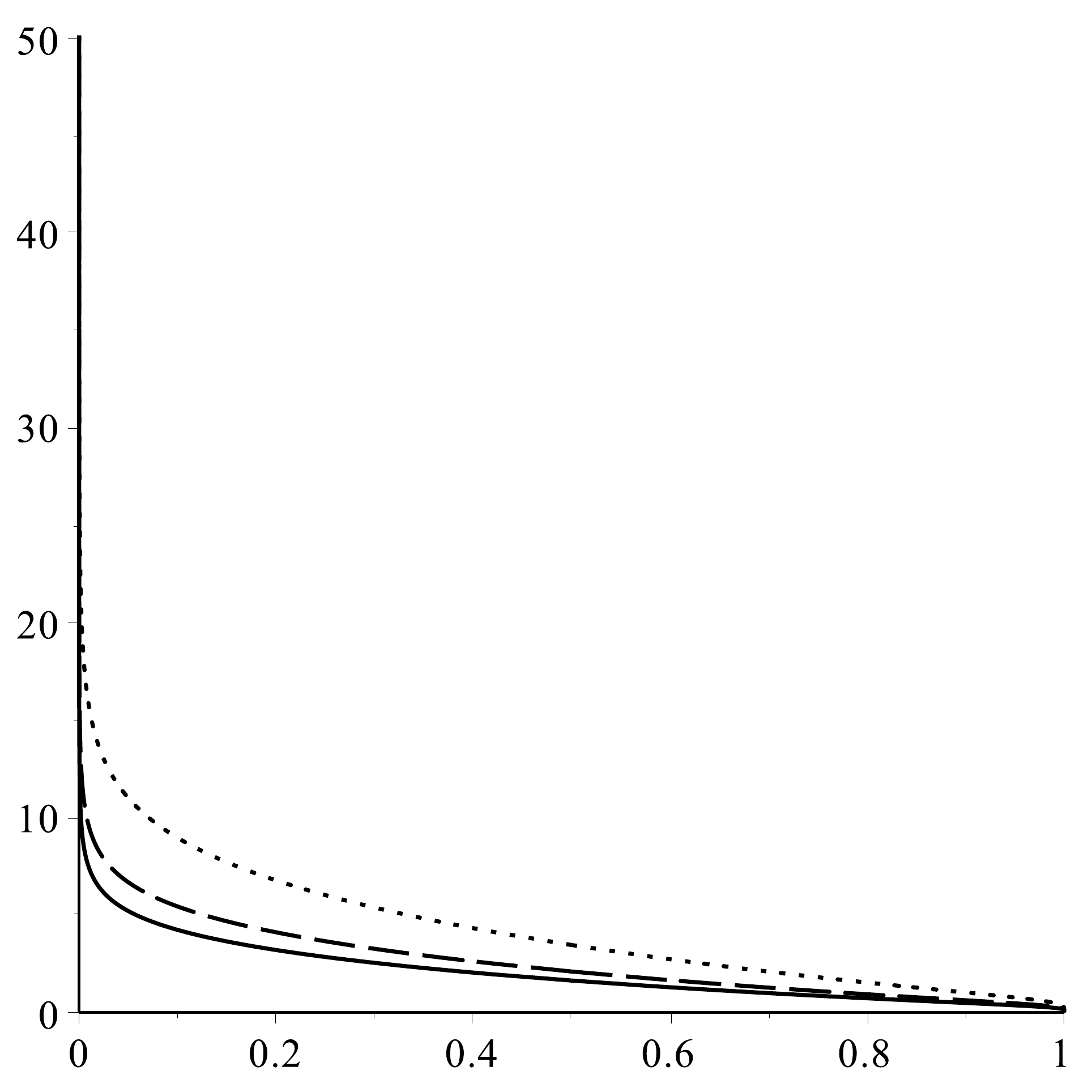}}
  \put(230,107){\small{$m_-$}}
  \put(239,141){\includegraphics[width=67\unitlength]{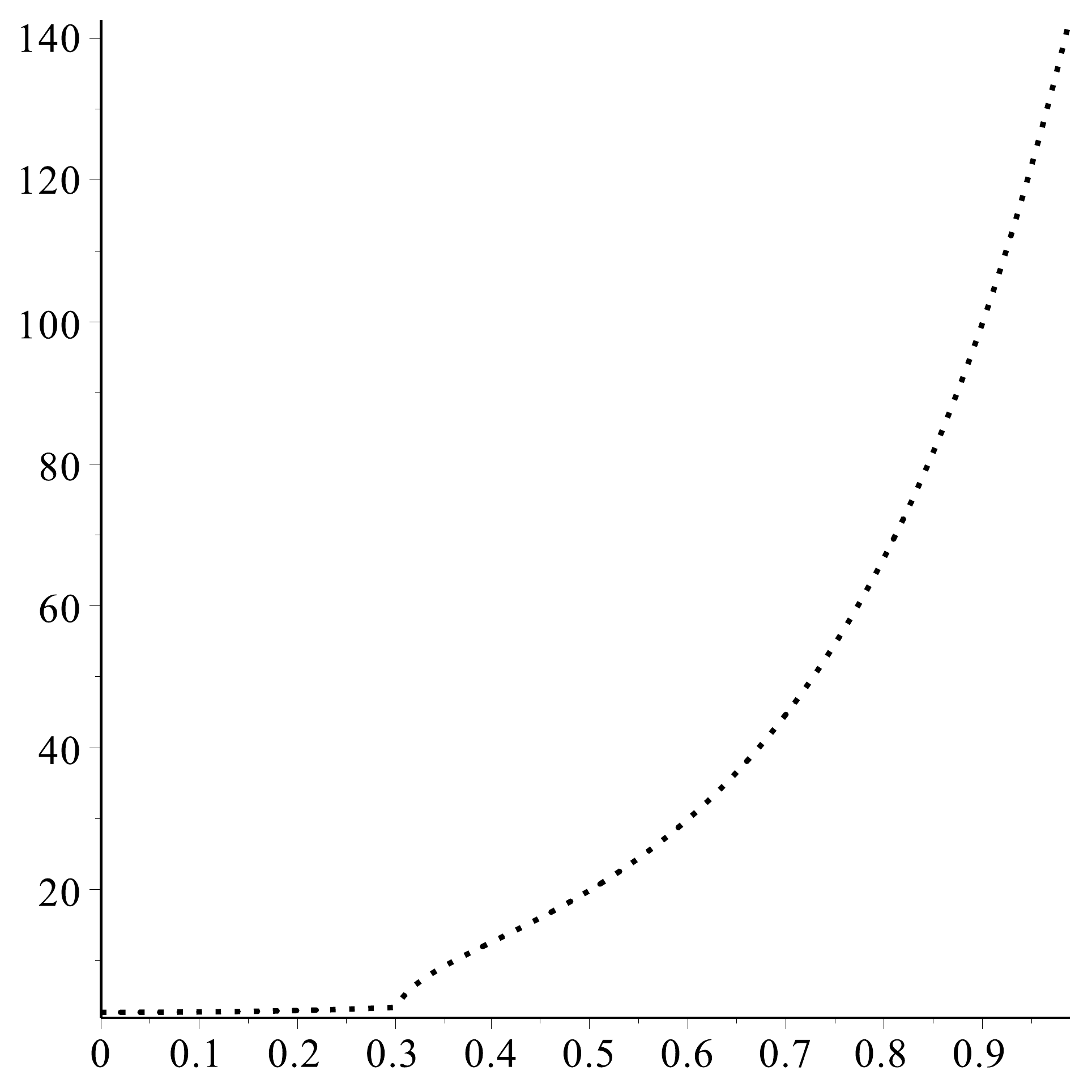}}
\put(270,135){\small{$J$}}

  \put(8,0){\includegraphics[height=100\unitlength,width=100\unitlength]{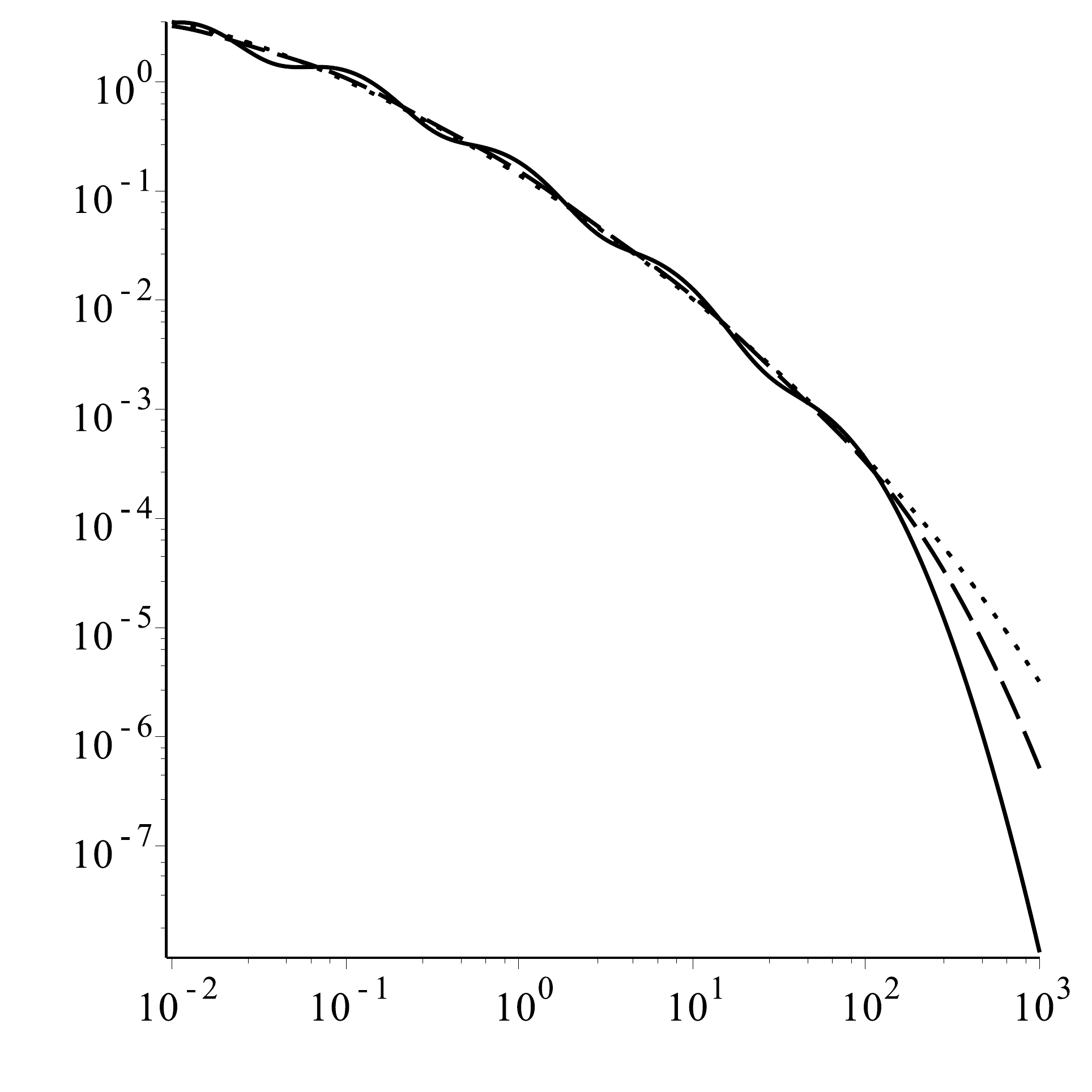}}
  \put(108,0){\includegraphics[height=100\unitlength,width=100\unitlength]{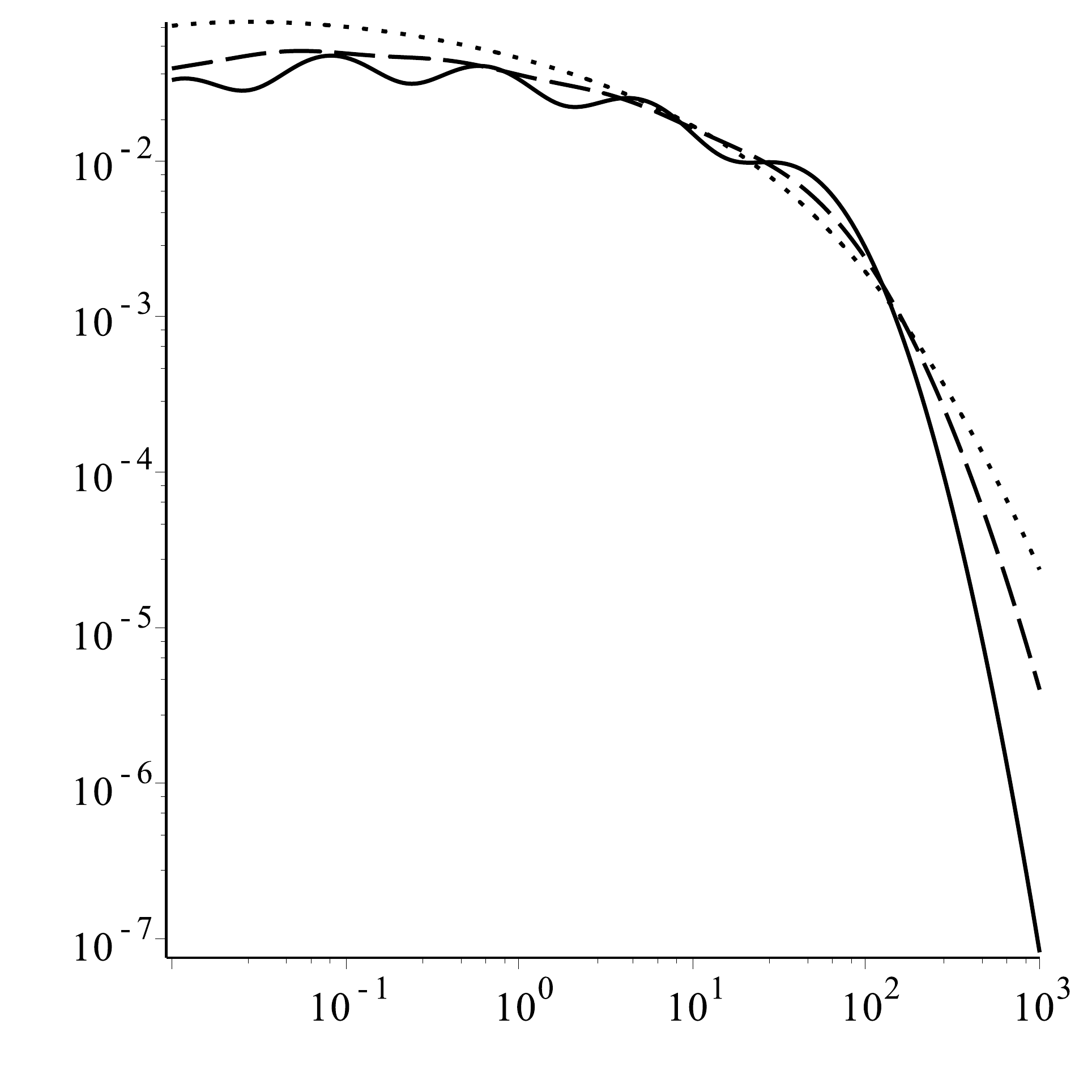}}
   \put(210,0){\includegraphics[height=100\unitlength,width=100\unitlength]{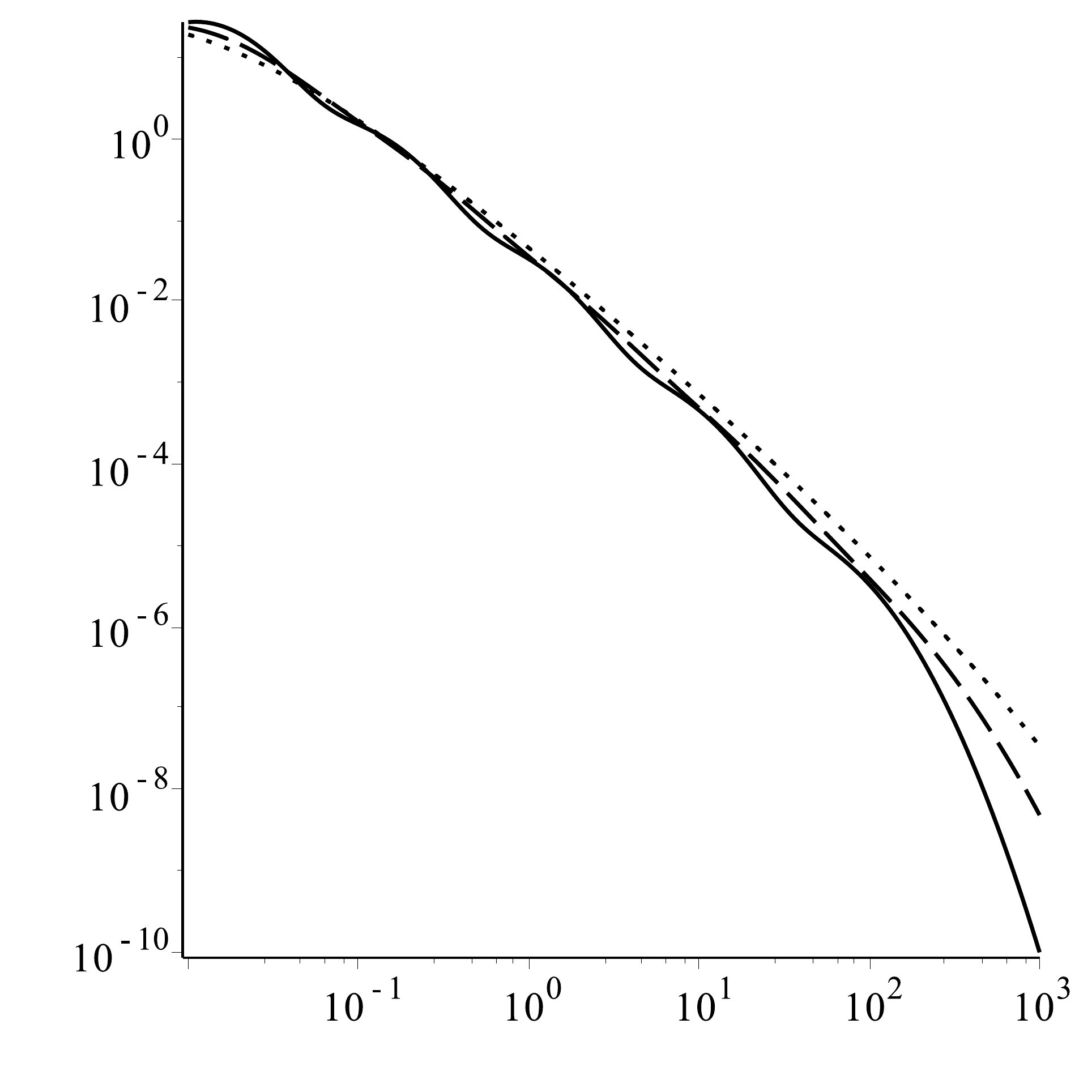}}
   \put(-5,65){\small{$P(c)$}} \put(157,-2){\small{$c$}}

  \end{picture}
 \vspace*{-3mm}
\caption{Behaviour of B-clone concentrations in the  immune system with fast B-clone equilibration. The system, interacting on a random regular factor-graph with   connectivity $L\!=\!K\!=\!4$,   was studied for B-clone noise parameters $\tilde\beta\in\{ 0.5, 1.0, 2.0\}$, represented by the dotted, dashed and solid lines respectively, in the high T-clone noise ($\beta<\beta_c$) and low T-clone noise ($\beta>\beta_c$)  regimes with $J=1$ and in the low Ag ($J<J_c$) and high Ag ($J>J_c$)  regimes with $\beta=1$. Note  that $\beta_c\approx0.0929$ and $J_c\approx0.3048$. Top left: average B-clone size, $\langle c\rangle$, as a function of the  fraction $m_-$ of  regulator T cells for $J=1$ and $\beta\in[0, 1]$.  Inset:  $\langle c\rangle$  as a function of $\beta$ for $\tilde\beta=0.5$. Top right: $\langle c\rangle$, as a function of $m_-$  for $\beta=1$ and $J\in[0, 1]$.  Inset:  $\langle c\rangle$  as a function of $J$ for $\tilde\beta=0.5$.  Bottom left:  distribution $P(c)$ of the B-clone size $c$ for $\beta=0.0639$ ($m_-=\frac{1}{2}$).     Bottom centre:  $P(c)$ for $\beta=0.1219$ and $m_-=0.1$ (excess of helpers). Bottom right: $P(c)$   for $\beta=0.1219$ and  $m_-=0.9$ (excess of  regulators). }
\label{figure:P(c)-B-fast}
\end{figure}

Let us first consider the case of fast B-clone equilibration. This regime can be studied using equations  (\ref{eq:P[h]-ferro}),  (\ref{eq:P(F)-ferro}) and  (\ref{eq:P(c)-integer-n}), with $n=1$. For $n=1$ the T-clone system is in the paramagnetic (PM) $m=0$ phase when $\beta<\beta_c$ (see \ref{app:n=1}).  Here the distribution of fields $P(F)$ is symmetric, see Figures \ref{figure:P(F)-B-fast}, 
which gives rise to the quadratic behaviour in the distribution of  B-clones $P(c)$ seen in Figure \ref{figure:P(c)-B-fast}. 
 As we increase $\beta$, the system will enter the ferromagnetic (FM)  $m\neq0$ phase when $\beta>\beta_c$.  Here the distribution $P(F)$ is no longer symmetric, see Figure \ref{figure:P(F)-B-fast}, and the distribution of  B-clones is seen to exhibit ``power law'' behaviour when $\beta\rightarrow\infty$. We note that in this regime the  distribution (\ref{eq:P(c)-integer-n}) is dominated by the probability $P(F)$ evaluated at the field values  $F=\pm JK$, corresponding to the magnetization $m>0$ or $m<0$. In particular when $\beta\rightarrow\infty$,  we have 
\begin{eqnarray}
P(c) \propto c^{-1+\tilde{\beta} F} \rme^{-\frac{1}{2}\rho \tilde{\beta}    \log^2\left(c\right)}.\label{eq:power-law}%& \rightarrow &\frac{  \rme^{-\frac{1}{2}\rho \tilde{\beta}\left(\log\left(c\right)- \frac{F}{\rho}\right)^2}  }{c\sqrt{2\pi/\rho \tilde{\beta}}}
\end{eqnarray}
 We note that power law behaviour is also present  in the empirical B-clone distributions~\cite{Desponds2016}.  Moreover, in the FM phase the average concentration of B cells (\ref{def:<c>}) can be controlled by the fraction $m_-$ of regulating  T-clones; see Figures \ref{figure:P(c)-B-fast}.  Comparing with the  results of  fast T-clone  equilibration, summarised in Figure \ref{figure:T-fast-large-n},  we note that here $\beta_c$ is larger and hence a larger amount of Ag (i.e. a larger value of $J$) is needed by the immune system to mount a vigorous immune response (see  Figure  \ref{figure:P(c)-B-fast}), i.e. the system  is less sensitive\footnote{This aspect of the model can be used to distinguish between the equilibration regimes when comparing model  with experimental data.}.  Also the behaviour of B-clones in the ``low-dose tolerance'' PM regime is different:  the average number of B-cells, $\langle c\rangle$, is increasing with $\beta$ and $J$ in Figure  \ref{figure:P(c)-B-fast},  but in  Figure \ref{figure:T-fast-large-n} it is a  constant.

In the regime of fast T-clone equilibration with $n\in\mathbb{Z}^+$ ($n>2$)  we solve the recurrence equation (\ref{eq:P[h]-ferro}) using the initial condition
\begin{eqnarray}
  P_0[h]&=& \Big(  \frac{1\!+\!m_0}{2}  \Big)^{\!\frac{n+h}{2}}   \Big(  \frac{1\!-\!m_0}{2}   \Big)^{\!\frac{n - h}{2}}   {{n}\choose{\frac{n+h}{2}} } \label{def:P0[h]} ,
\end{eqnarray}
where  $m_0\in[-1,1]$ is such that  $m_0=n^{-1}\sum_h P_0[h]h$.  For small $\beta$ the solution $P[h]$ is symmetric for any $m_0$, and $\sum_h P[h]h =0$, corresponding to the PM phase $m=0$. For large $\beta$ the solution $P[h]$ is no longer symmetric, and $\sum_h P[h]h >0$ (or $\sum_h P[h]h <0$) for  $m_0>0$ (or $m_0<0$), which corresponds to the FM phase. The transition from PM to FM happens at  $\beta_c$ which is, in addition to $L$ and $K$, also a function of the parameter $n$.  The T-clones apparently continue to behave  as in the fast B-clone equilibration regime, see Figures \ref{figure:P(F)-T-fast}, but the behaviour of the B-clones is different.

We observe that if we fix $J$ and vary $\beta$ then the   average concentration  of B cells $\langle c\rangle$ is no longer a monotonic function of $m_-$:  $\langle c\rangle$ increases with $m_-$ on the interval $[0, M_-)$ and decreases on the interval  $(M_-, 1]$, where $M_-$ denotes the location of the maximum. See Figure \ref{figure:P(c)-T-fast}.     For $n\rightarrow\infty$, the interval $[0, M_-)$ shrinks and in this limit we expect to recover the exact equations (\ref{eq:m-large-n-ferro}) and (\ref{eq:P(F)-large-n-ferro}) of the (equivalent) $\tilde\beta\rightarrow\infty$ limit.  However, this limit does not commute with the  $\beta\rightarrow0$ limit for which  the average $\langle c\rangle$ is diverging. This is very different  from the fast B-clone equilibration result where the average concentration of B-cells, $\langle c\rangle$, behaves  monotonically  and does not have any singularities (see Figure \ref{figure:P(c)-B-fast}) .  There is no qualitative difference in the behaviour of $\langle c\rangle$ for the different equilibration regimes when we fix $\beta$  and vary $J$,   but  the system in the  fast T-clone equilibration regime has a lower $\beta_c$  than in the  fast B-clone equilibration regime and hence it is more ``sensitive''.  The ``power law'' behaviour  of the distribution $P(c)$ is less pronounced, see Figure \ref{figure:P(c)-T-fast}, than in Figure \ref{figure:P(c)-B-fast} of the fast B-clone case.  We ascribe this to the dependence  on $\beta$ of the log-normal part of the distribution (\ref{eq:P(c)-integer-n})  via the relation $\tilde\beta=n\beta$. 

Finally, we note that equation (\ref{eq:P[h]-ferro}), which thus far we were able to analyse only numerically, can be  studied further analytically when the connectivity   $L$ or  $K$  is large.  Also, this analysis allows us to probe the extreme regime of large B-clone numbers $M\gg N$ or large T-clone numbers ($N\gg M$).
\begin{figure}[t]
%\vspace*{0mm} \hspace*{-9mm}
 \setlength{\unitlength}{0.43mm}
\begin{picture}(310,102)
\put(0,0){\includegraphics[height=100\unitlength,width=100\unitlength]{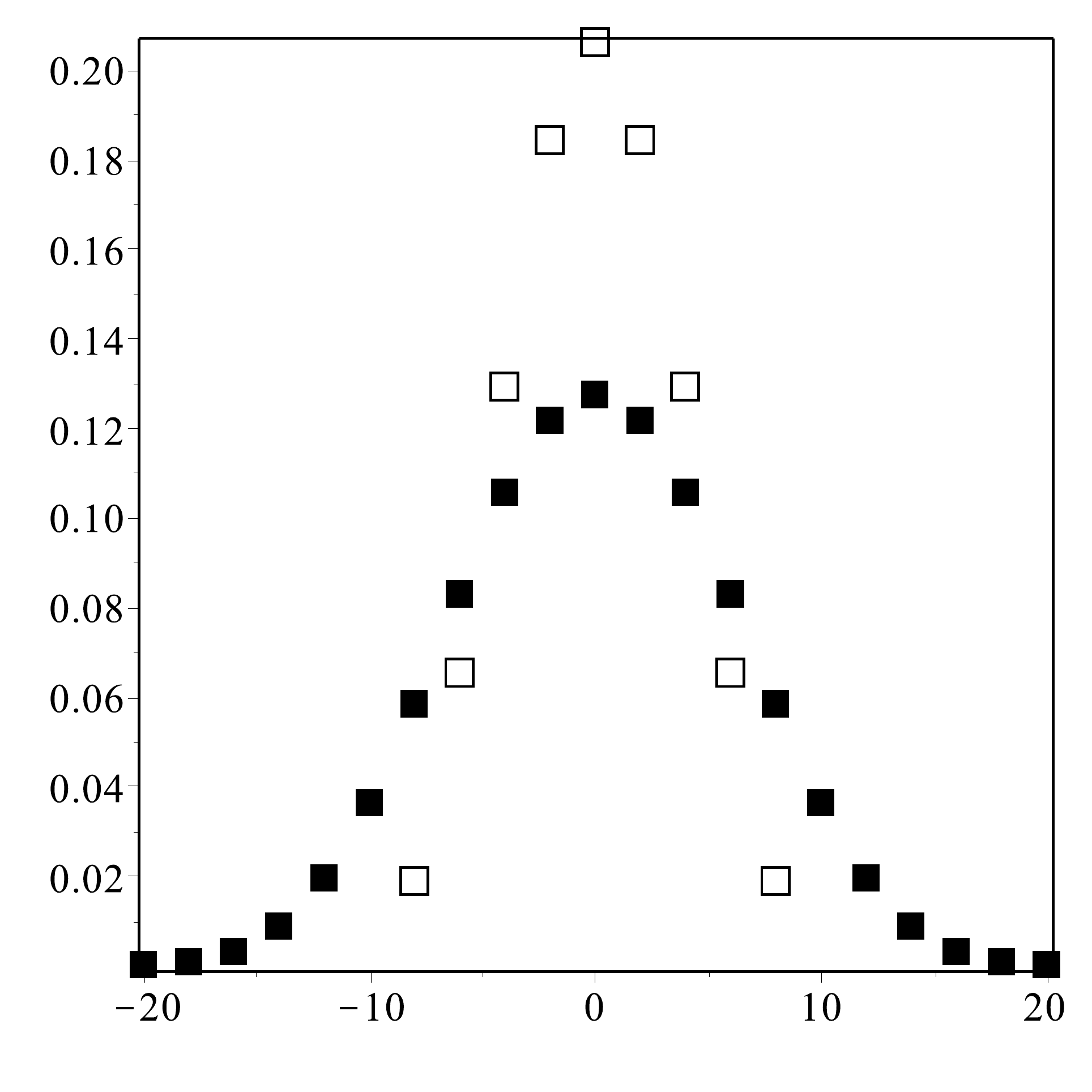}}%height=100\unitlength,
   
   \put(100,0){\includegraphics[height=100\unitlength,width=100\unitlength]{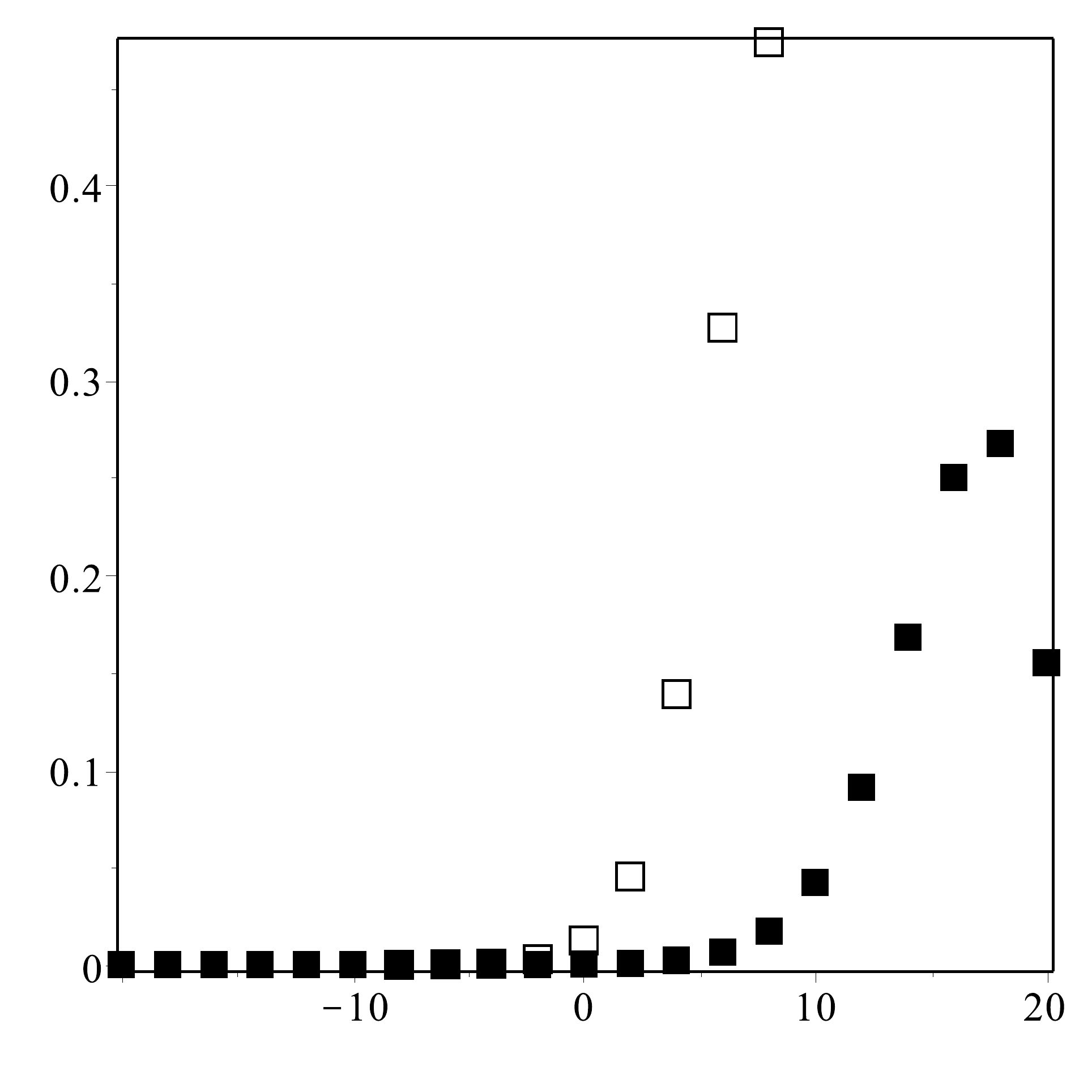}}
   
    \put(200,0){\includegraphics[height=100\unitlength,width=100\unitlength]{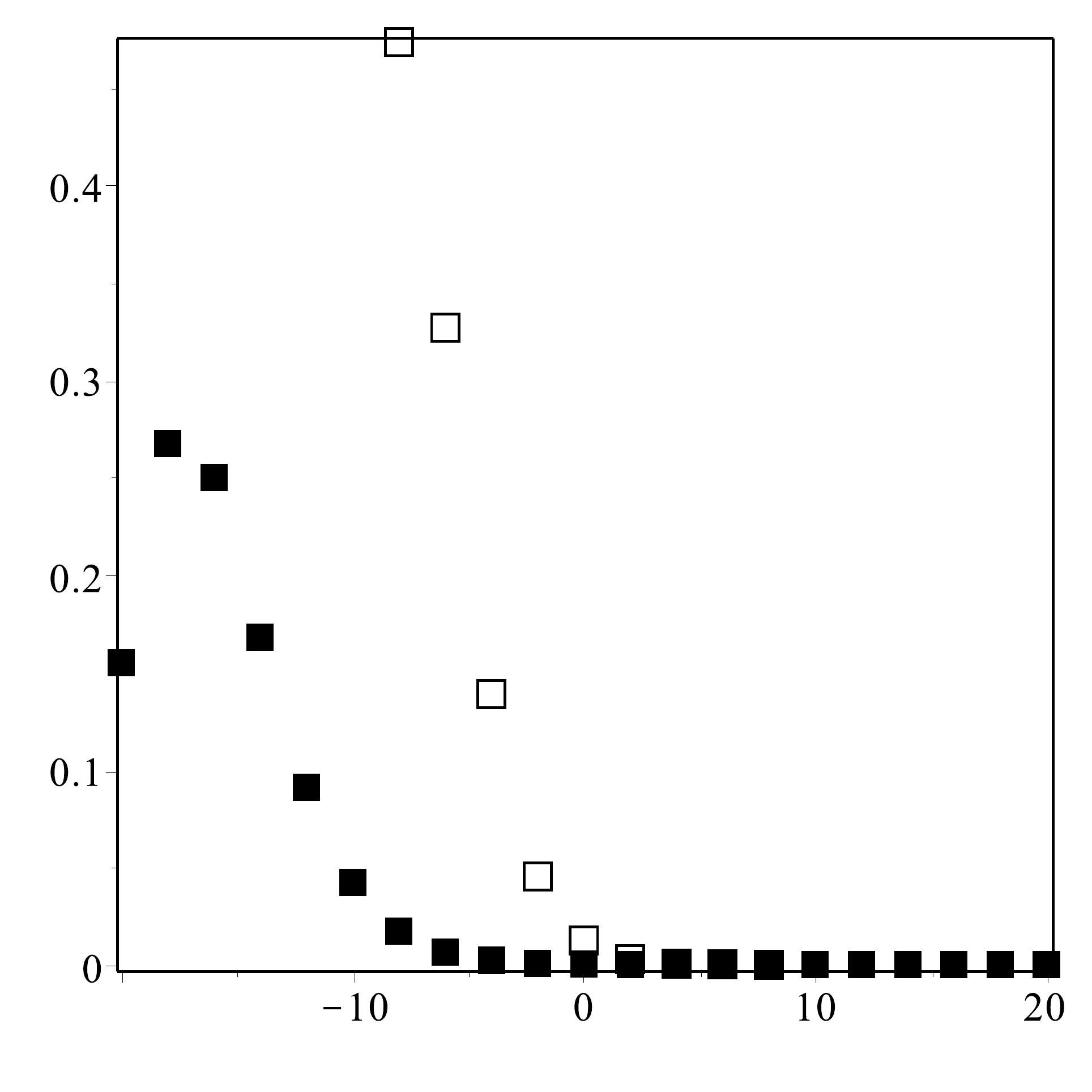}}
   
  \put(-15,65){\small{$P(F)$}} 
 
\put(150,-2){\small{$F$}}   
  
  \end{picture}
% \vspace*{-5mm}
\caption{The distribution $P(F)$ of  T-clone activity $F$ in the immune system model  with fast T-clone equilibration,  defined on a random regular factor-graph with connectivity   $L\!=\!K\!=\!4$.  The system was studied  in the high $\beta<\beta_c$ (left) and low $\beta>\beta_c$  (centre and right) T-clone noise regimes, for B-clone noise parameters  $\tilde\beta=2\beta$ ($\beta_c\approx0.077$) and $\tilde\beta=5\beta$ ($\beta_c\approx0.069$), represented by empty and filled boxes, respectively.  The T-clone noise parameters for  the case of $\tilde\beta=2\beta$  (empty boxes) were $\beta=0.076$ with $m_-=\frac{1}{2}$ (left),  $\beta=0.1015$  with  $m_-=0.1$ (centre) and  $\beta=0.1015$  with  $m_-=0.9$ (right). The T-clone noise parameters for  the case $\tilde\beta=5\beta$  (filled boxes) were $\beta=0.068$ with $m_-=\frac{1}{2}$ (left),  $\beta=0.0915$  with  $m_-=0.1$ (centre) and  $\beta=0.0915$  with  $m_-=0.9$ (right).  } \label{figure:P(F)-T-fast}
\end{figure}
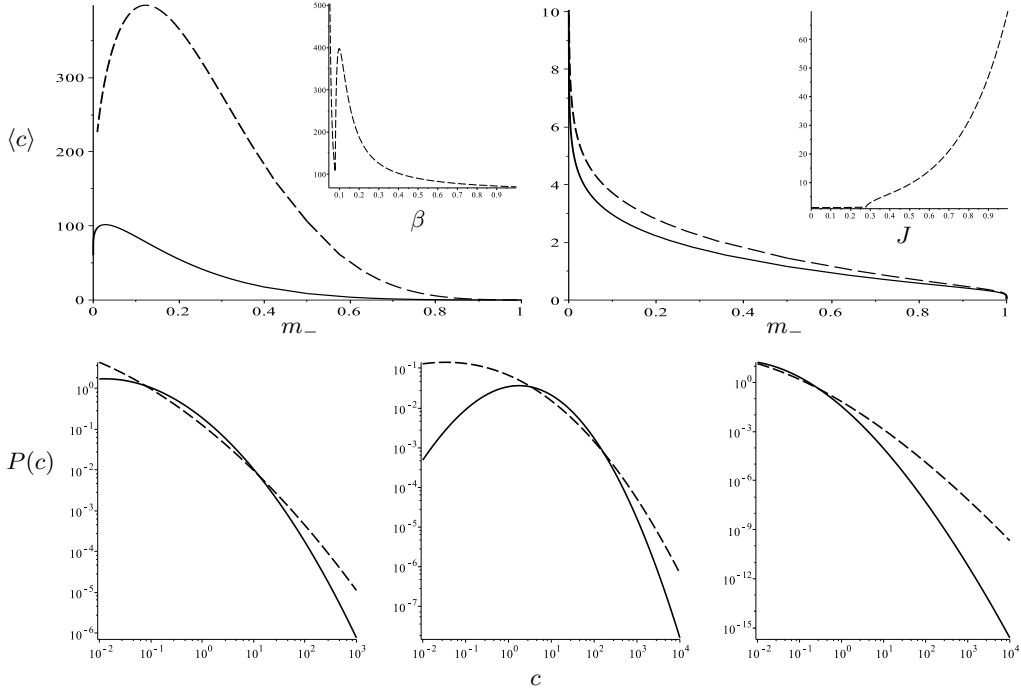
\begin{figure}[t]
%\vspace*{0mm} \hspace*{-9mm}

 \setlength{\unitlength}{0.43mm}
\begin{picture}(310,210)

 \put(-5,165){\small{$\langle c\rangle$}} 
 
\put(8,110){\includegraphics[height=100\unitlength,width=150\unitlength]{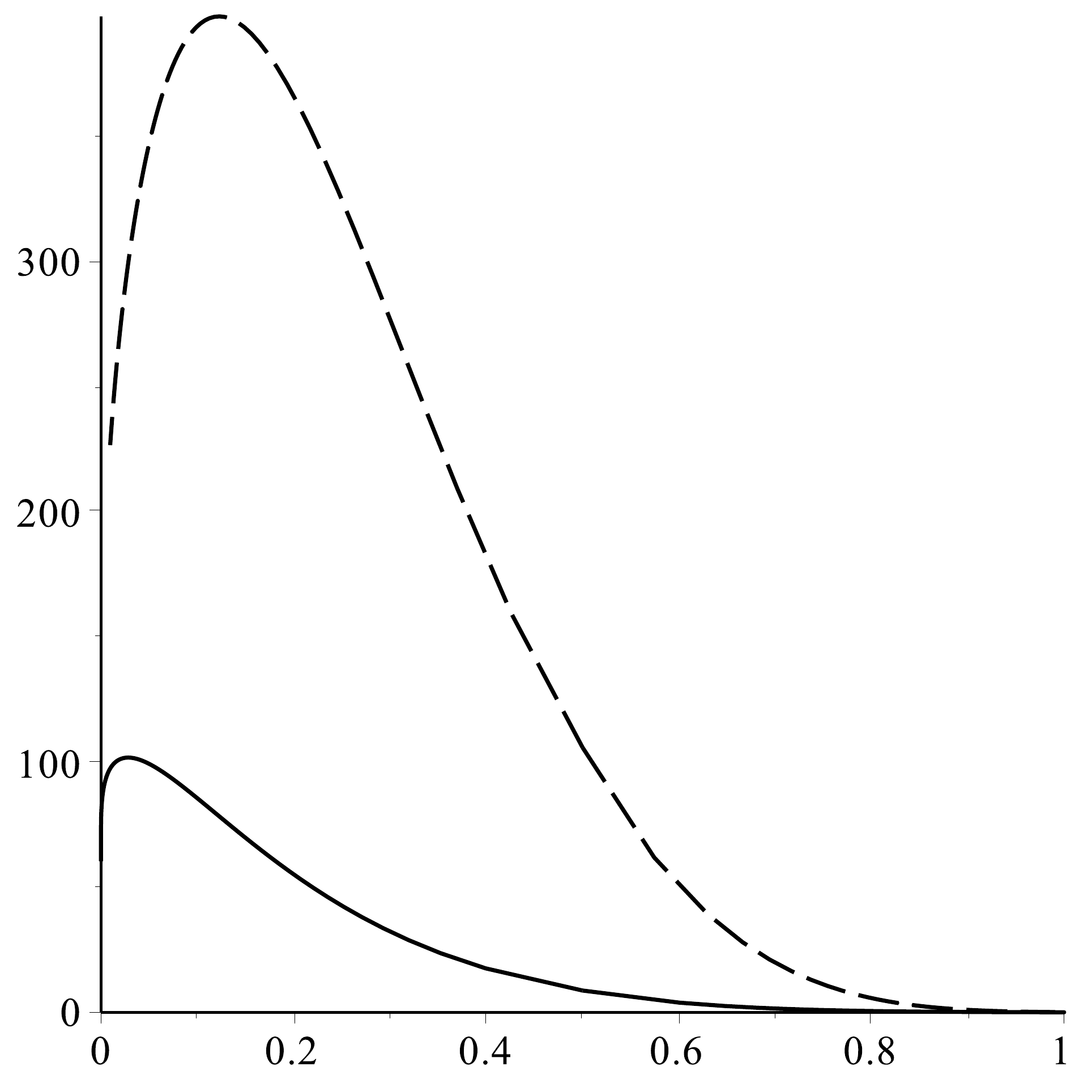}}
\put(80,107){\small{$m_-$}}
 \put(87,145){\includegraphics[width=67\unitlength]{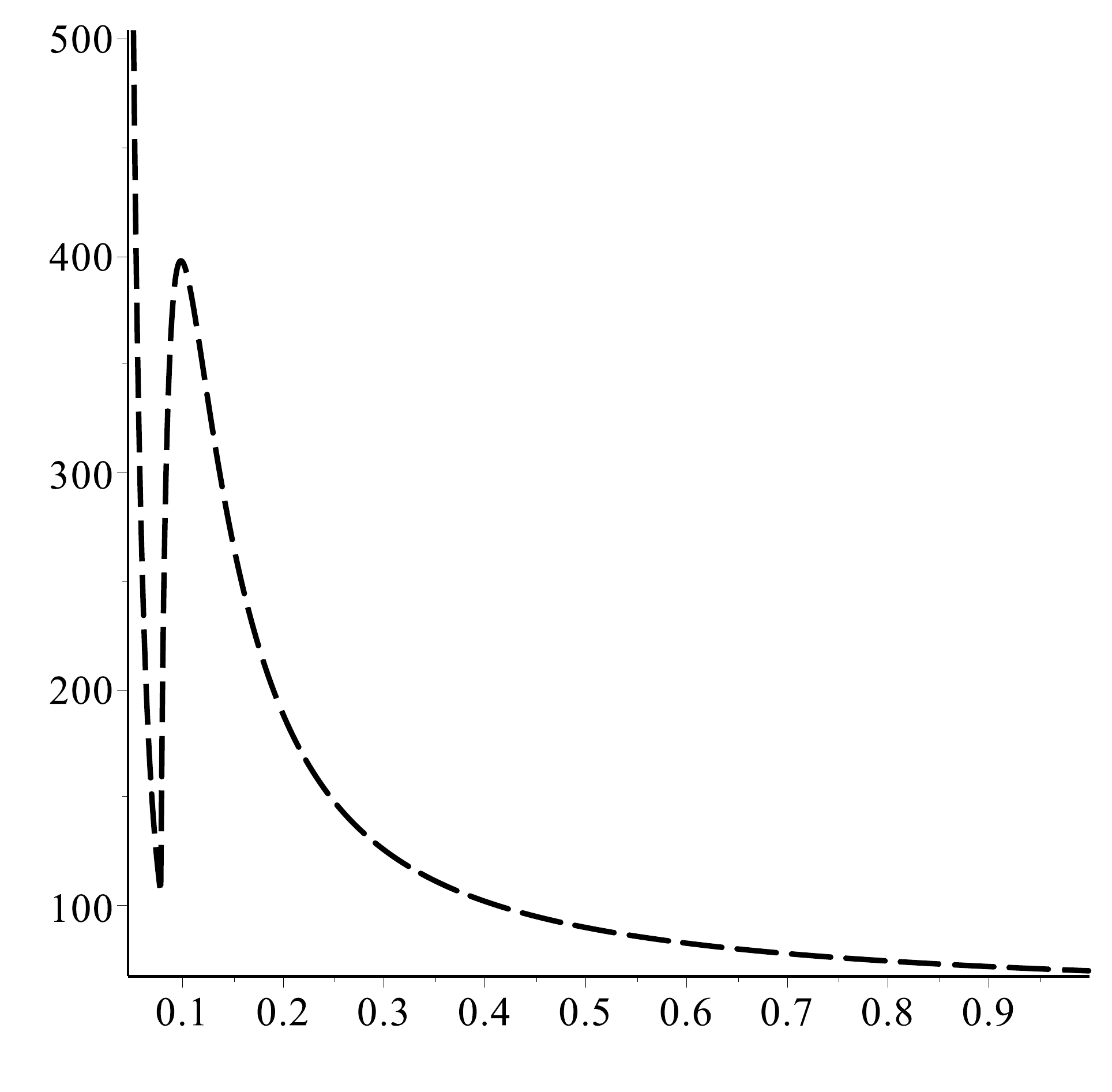}}
\put(120,140){\small{$\beta$}}

 \put(158,110){\includegraphics[height=100\unitlength,width=150\unitlength]{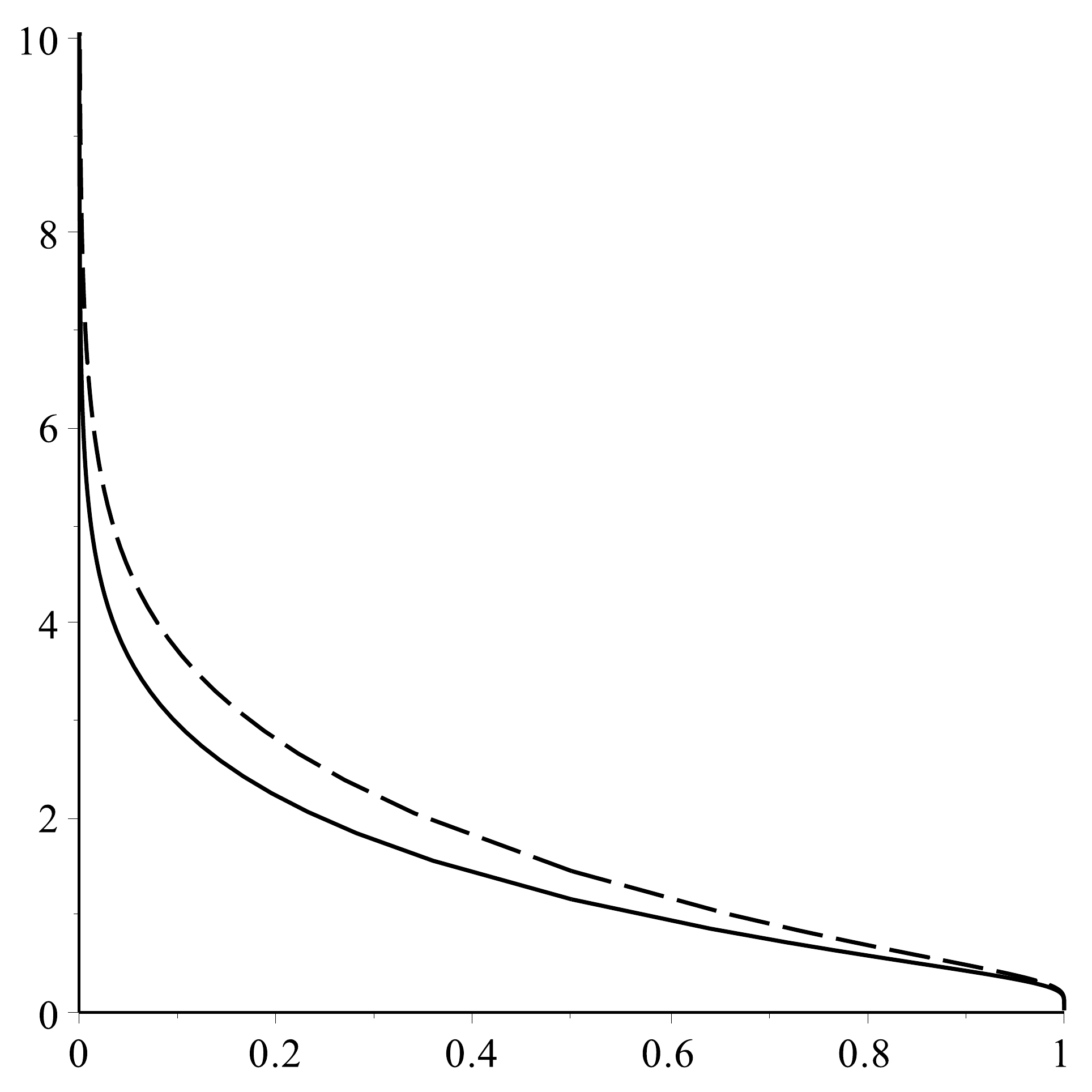}}
  \put(230,107){\small{$m_-$}}
  \put(239,141){\includegraphics[width=67\unitlength]{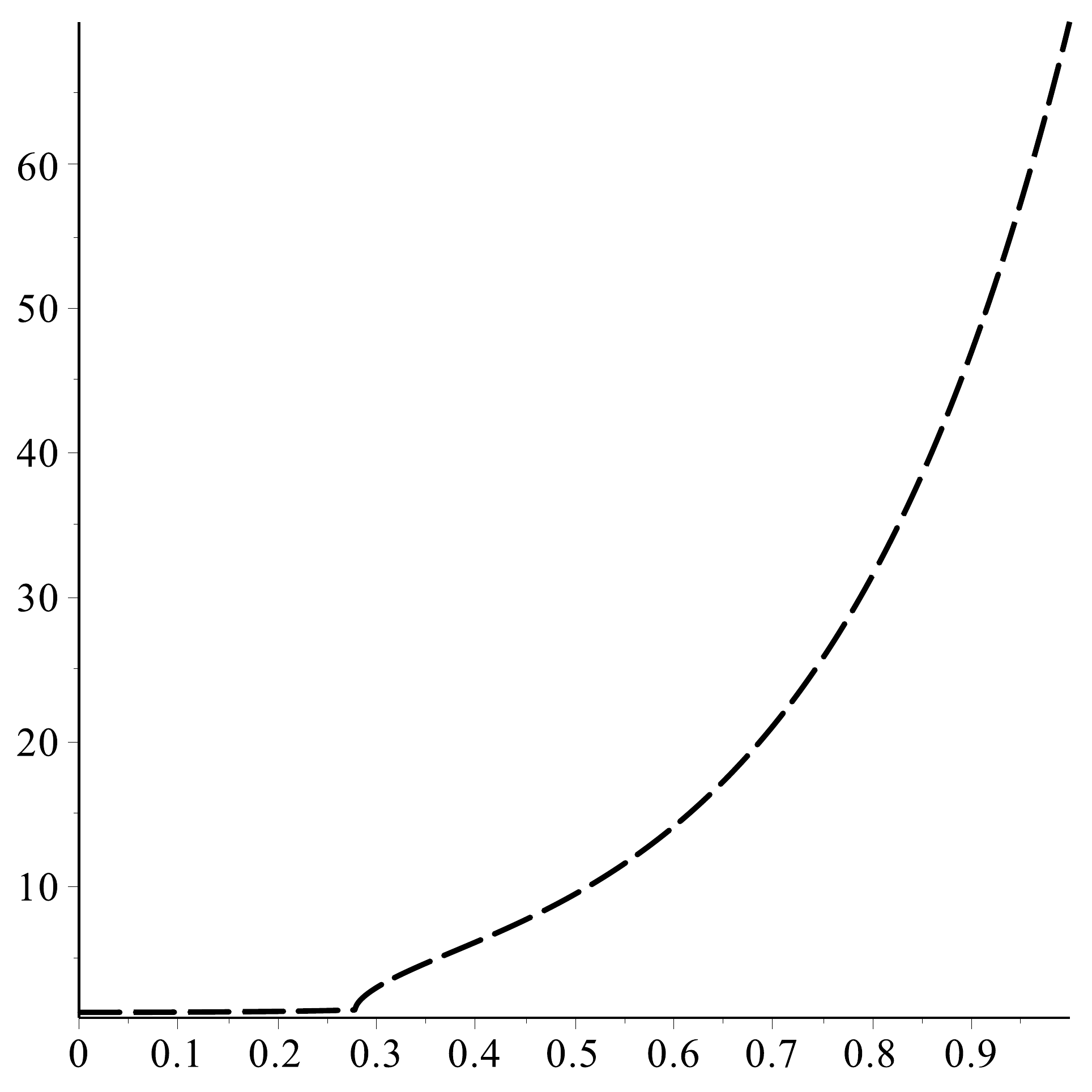}}
\put(270,135){\small{$J$}}

    \put(8,0){\includegraphics[height=100\unitlength,width=100\unitlength]{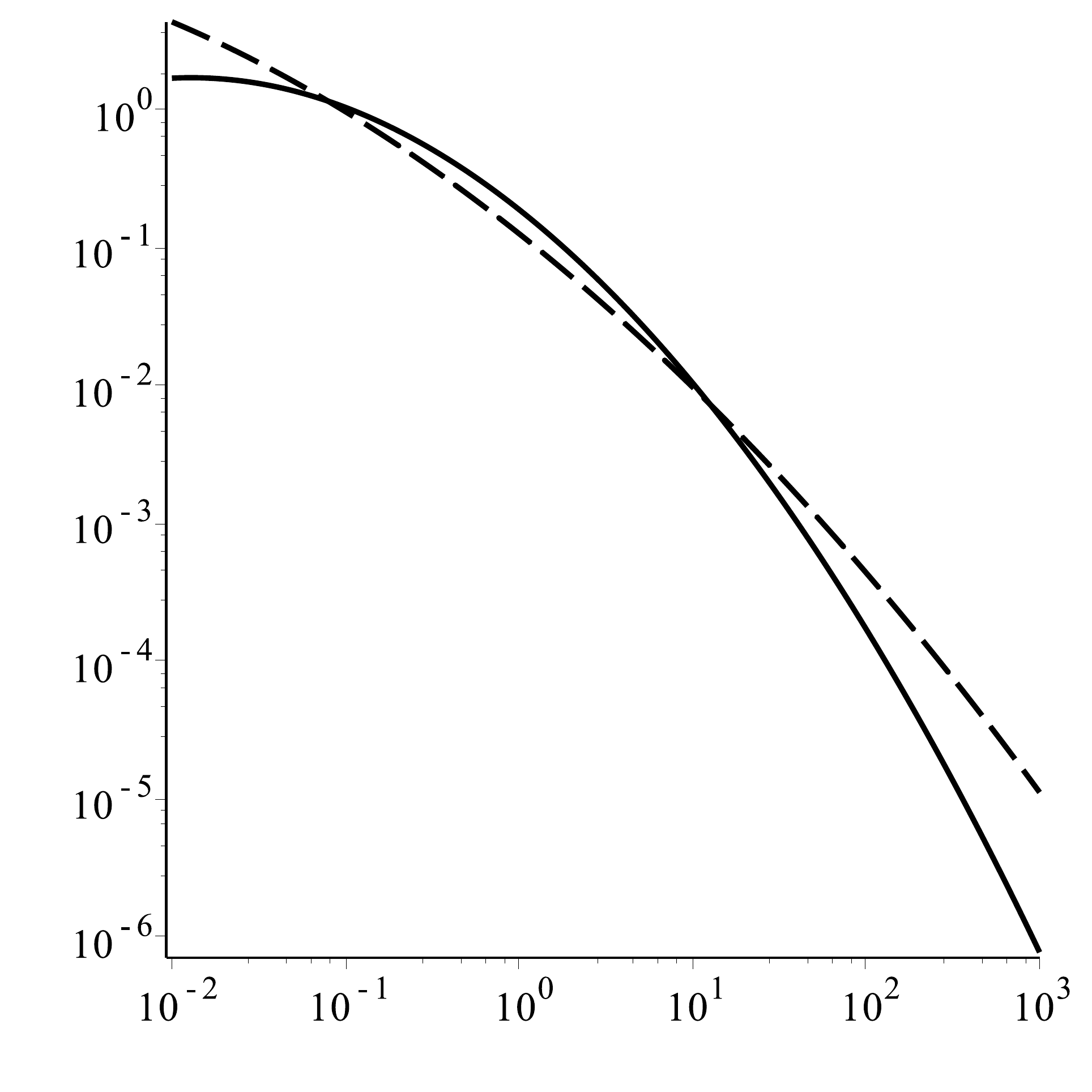}}
  \put(108,0){\includegraphics[height=100\unitlength,width=100\unitlength]{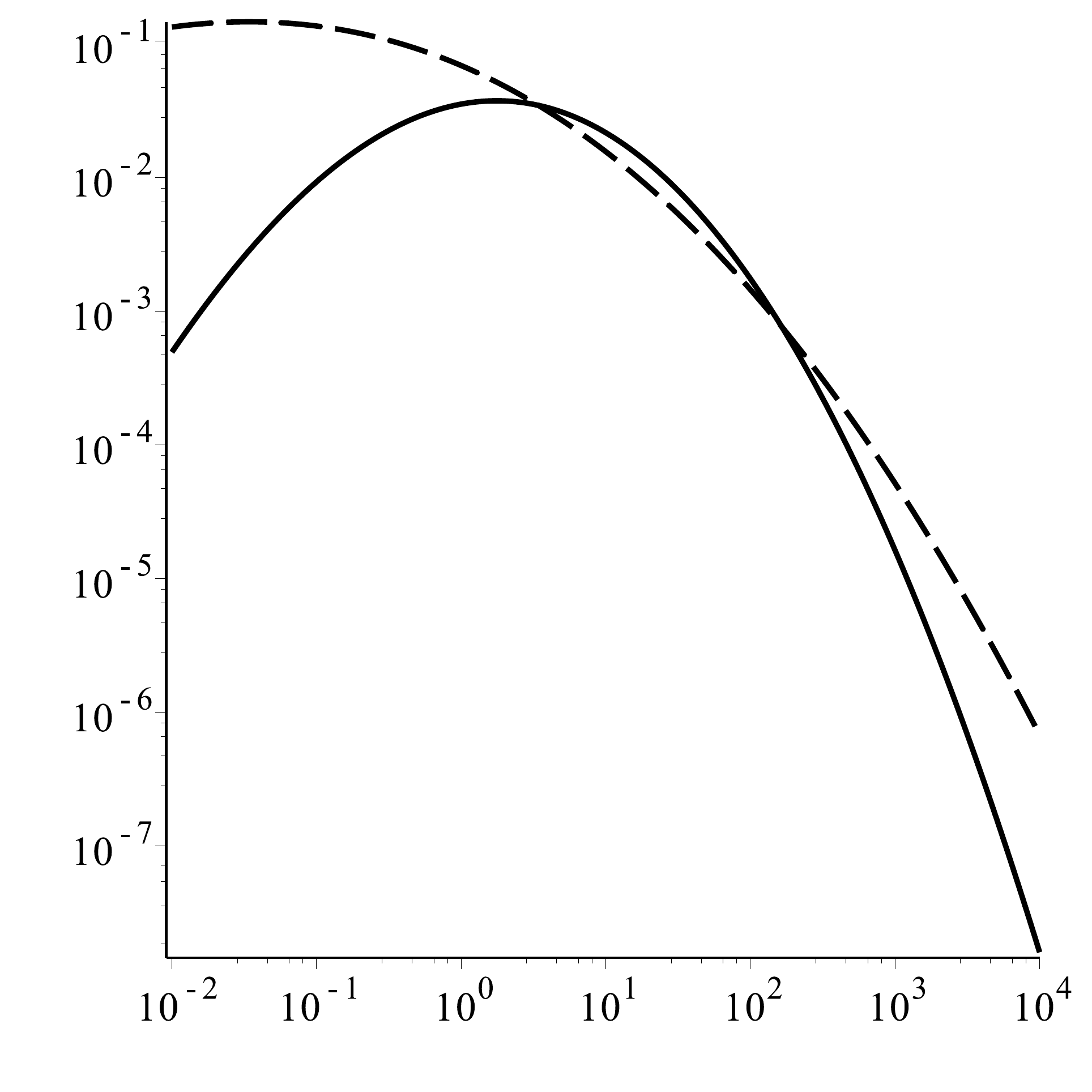}}
   \put(210,0){\includegraphics[height=100\unitlength,width=100\unitlength]{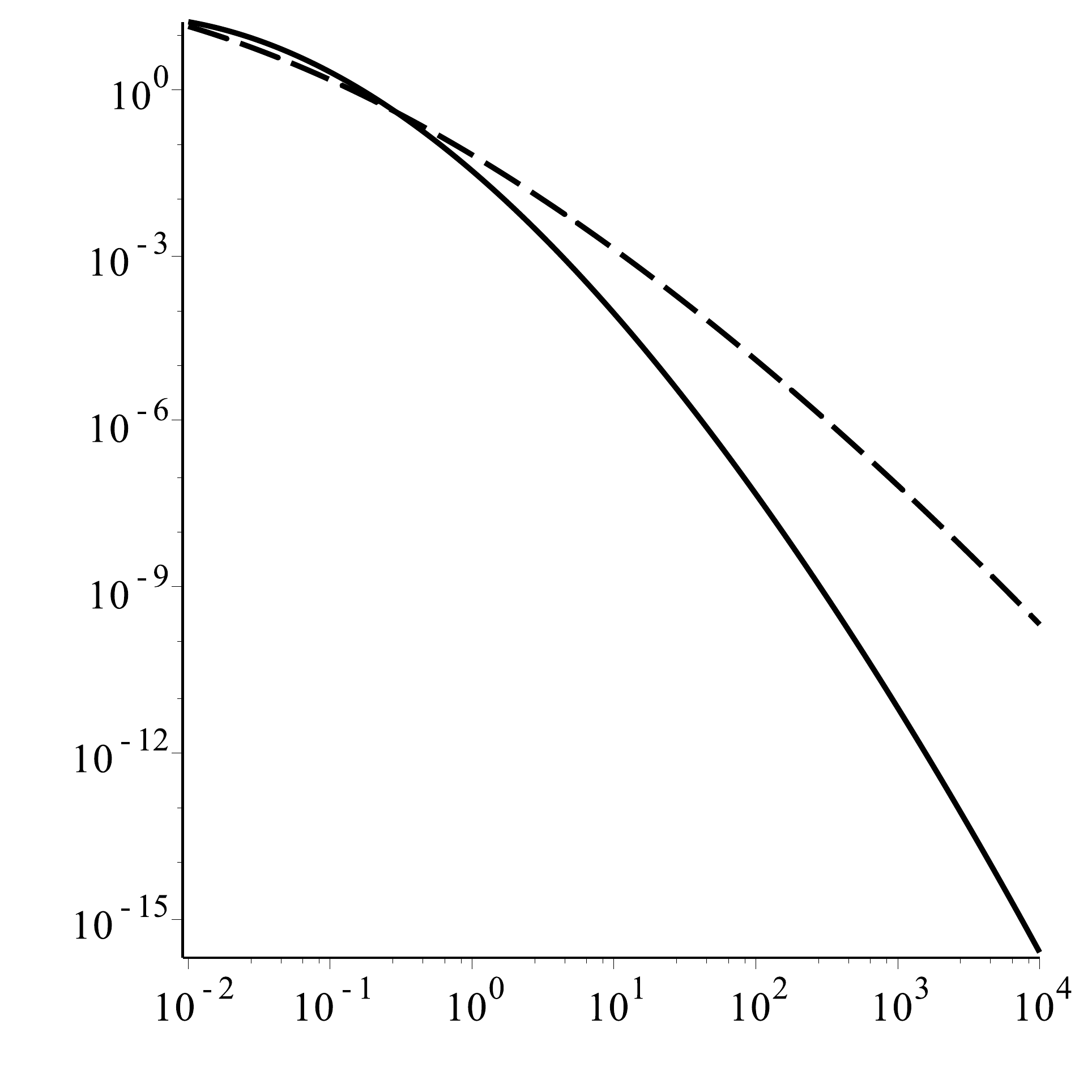}}
   \put(-5,65){\small{$P(c)$}} \put(157,-2){\small{$c$}}

 \end{picture}
 \vspace*{-4mm}
\caption{Behaviour of  B-clones in the immune model with fast T-clone equilibration, defined on a random regular factor-graph with  connectivity   $L\!=\!K\!=\!4$. The system    was studied  in the high $\beta<\beta_c$  and low $\beta>\beta_c$   T-clone noise regimes with $J=1$ and in the low  $J<J_c$ and  high $J>J_c$ Ag regimes with $\beta=1$, for the B-clone noise parameters  $\tilde\beta=2\beta$ and $\tilde\beta=5\beta$ , represented by dashed and solid lines respectively.  Note that $\beta_c\approx0.077$ for $\tilde\beta=2\beta$ and   $\beta_c\approx0.069$ for $\tilde\beta=5\beta$ when  $J=1$. For $\beta=1$: $J_c=\sqrt{\beta_c}$ .  Top left: The average B clone size, $\langle c\rangle$, as a function of the  fraction of T regulator cells, $m_-$, for $J=1$ and $\beta\in[\beta_c, 1]$. Inset:  $\langle c\rangle$  as a function of $\beta$ for $\tilde\beta=2\beta$. Top right:  $\langle c\rangle$, as a function of  $m_-$  for $\beta=1$ and $J\in[J_c, 1]$. Inset:  $\langle c\rangle$  as a function of $J$ for $\tilde\beta=2\beta$. Bottom left: The distribution $P(c)$ of the B-clone size $c$ for $\beta=0.076$ (dashed line) and  $\beta=0.068$ (solid line) with $m_-=\frac{1}{2}$. Bottom centre:  $P(c)$ for $\beta=0.1015$ (dashed line) and $\beta=0.0915$ (solid line), with $m_-=0.1$. Bottom right:  $P(c)$ for $\beta=0.1015$ (dashed line) and $\beta=0.0915$ (solid line), with  $m_-=0.9$.} \label{figure:P(c)-T-fast}
\end{figure}

\subsection{Large $L$ and finite $K$ (or $M/N\rightarrow\infty$) regime\label{section:large-L}}
To start we note that equation (\ref{eq:P[h]-ferro}) can also be written in the form 
\begin{eqnarray}
  P[h]&=& \sum_{\{\mathcal{N}\}}\frac{(L\!-\!1)!\Big\{\prod_{\{h_j\} } \Big( \prod_{j=1}^{K-1}P [ h_j]\Big)^{\mathcal{N}[\{h_j\}]} \Big\}      }{\Big\{  \prod_{\{h_j\}} \mathcal{N}[\{h_j\}]!\Big\}}\nonumber\\
&&~~~~~~~~~~~~~~~~\times  \rme^{\frac{\beta J^2}{2n\rho}\sum_{\{h_j\} } \mathcal{N}[\{h_j\}]\left(h+         \sum_{j=1}^{K-1} h_j \right)^2}  {{n}\choose{\frac{n+h}{2}}}   \nonumber\\
&&\times\Bigg[
\sum_{\tilde{h}}
 \sum_{\{\mathcal{N}\}}\frac{(L\!-\!1)!\Big\{\prod_{\{\tilde{h}_j\} } \Big( \prod_{j=1}^{K-1}P [\tilde{h}_j]\Big)^{\mathcal{N}[\{\tilde{h}_j\}]} \Big\}      }{\Big\{  \prod_{\{\tilde{h}_j\}} \mathcal{N}[\{\tilde{h}_j\}]!\Big\}}\nonumber\\
&&~~~~~~~~~~~~~~~\times
  \rme^{\frac{\beta J^2}{2n\rho}\sum_{\{\tilde{h}_j\} } \mathcal{N}[\{\tilde{h}_j\}]\left(\tilde{h}+         \sum_{j=1}^{K-1} \tilde{h}_j \right)^2} 
{{n}\choose{\frac{n+\tilde{h}}{2}}}\Bigg]^{-1}
  \label{eq:P[h]_calc4},
 % 
  % \label{eq:}\\  
\end{eqnarray}
where $\sum_{\{h_j\} }\mathcal{N}[\{h_j\}]=L\!-\!1$ and  $\mathcal{N}[\{h_j\}]\in\{0,\ldots,L\!-\!1\}$. Using $\mathcal{N}!=\rho_\mathcal{N} \mathcal{N}^\mathcal{N}\rme^{-\mathcal{N}}$ with $\rho_0=1$ and  $\rho_\mathcal{N}=\sqrt{2\pi \mathcal{N}}\rme^{\frac{\theta_\mathcal{N}}{12\mathcal{N}}}$, where $\vert\theta_\mathcal{N}\vert<1$ (see  \cite{Robbins1955}  ),  gives  us 
\begin{eqnarray}
\frac{(L\!-\!1)!     }{  \prod_{\{h_j\}} \mathcal{N}[\{h_j\}]!}&=&\frac{\rho_{L-1} (L\!-\!1)^{(L\!-\!1)}\rme^{- (L-1)  }  }{ \prod_{\{h_j\}} \rho_{\mathcal{N}[\{h_j\}]} \mathcal{N}^{\mathcal{N}[\{h_j\}]}[\{h_j\}]\rme^{-\mathcal{N}[\{h_j\}]}   }\nonumber\\
&=&\rme^{-(L-1)\sum_{\{h_j\}}\frac{\mathcal{N}[\{h_j\}]}{L-1} \log\frac{\mathcal{N}[\{h_j\}]}{L-1}+\mathcal{R}_{L-1}(\{\mathcal{N}\})}\label{eq:Stirling}, 
\end{eqnarray}
where $\mathcal{R}_{L-1}(\{\mathcal{N}\})\!=\!\log\rho_{L-1} \!-\! \sum_{\{h_j\}}\log\rho_{\mathcal{N}[\{h_j\}]}$ (it is easy to show that $\mathcal{R}_{L-1}(\{\mathcal{N}\})\!=\!\Omega(\{\frac{\mathcal{N}}{L-1}\})\!+\!O(\log(L\!-\!1))$).  Let us define the functional  
\begin{eqnarray}
D\left( \frac{\mathcal{N}}{L-1}\vert\vert P^{K-1}\right)=\sum_{\{h_j\} }\frac{\mathcal{N}[\{h_j\}]}{L\!-\!1}\log\Big(\frac{\mathcal{N}[\{h_j\}] }{(L\!-\!1) \prod_{j=1}^{K-1}P [ h_j]  }\Big),
\end{eqnarray} 
which  is the Kullback$-$Leibler  (KL) ``distance''~\cite{Cover2012} between the distributions $\mathcal{N}[\{h_j\}]/(L\!-\!1)$ and $\prod_{j=1}^{K-1}P [ h_j]$, and  use the identity (\ref{eq:Stirling}) to compute  the numerator in the equation (\ref{eq:P[h]_calc4}). This computation is greatly simplified if we assume that  $J^2=1/(L\!-\!1)$  and take  the limit $L\rightarrow\infty$, which gives us 
\begin{eqnarray}
  &&\hspace*{-20mm}\sum_{\{\mathcal{N}\}}\rme^{-(L-1)D\left( \frac{\mathcal{N}}{L-1}\vert\vert P^{K-1}\right)      +  \mathcal{R}_{L-1}(\{\mathcal{N}\})             +\sum_{\{h_j\} } \frac{\mathcal{N}[\{h_j\}]}{L-1}\frac{\beta }{2n\rho}\left(h+         \sum_{j=1}^{K-1} h_j \right)^2}  \nonumber\\ 
&=&\sum_{\{\mathcal{N}\}} \rme^{-(L-1)D\left( \frac{\mathcal{N}}{L-1}\vert\vert P^{K-1}\right)      +  \mathcal{R}_{L-1}(\{\mathcal{N}\})             +\sum_{\{h_j\} } \frac{\mathcal{N}[\{h_j\}]}{L-1}\frac{\beta }{2n\rho}\left(h+         \sum_{j=1}^{K-1} h_j \right)^2}  
\nonumber\\[-2mm]
&& \hspace*{20mm}
\times \Big[
  \sum_{\{\mathcal{N}\}}\rme^{-(L-1)D\left( \frac{\mathcal{N}}{L-1}\vert\vert P^{K-1}\right)      +  \mathcal{R}_{L-1}(\{\mathcal{N}\})              }                         \Big]^{-1} \nonumber\\[1mm]
&=& \rme^{\sum_{\{h_j\} } \left\{\prod_{j=1}^{K-1}P [ h_j]\right\} \frac{\beta }{2n\rho}\left(h+         \sum_{j=1}^{K-1} h_j \right)^2}=\rme^{ \frac{\beta }{2n\rho}\left(h^2+  2h(K-1)\langle\tilde{h}\rangle\right)},
\end{eqnarray}
where $\langle\tilde{h}\rangle=\sum_{\tilde{h}}P [ \tilde{h}]\tilde{h}$.
Using the above result to compute the distribution (\ref{eq:P[h]_calc4}) gives
\begin{eqnarray}
  P[h]&=&\frac{\rme^{ \frac{\beta }{2n\rho}\left(h^2+  2h(K-1)\langle\tilde{h}\rangle\right)}  {{n}\choose{\frac{n+h}{2}}  }   }{
\sum_{\hat{h}} \rme^{ \frac{\beta }{2n\rho}\left(\hat{h}^2+  2\hat{h}(K-1)\langle\tilde{h}\rangle\right)}
{{n}\choose{\frac{n+\hat{h}}{2}}}
  }\label{eq:P[h]_large_L}\\
\langle h\rangle&=&\frac{\sum_{h}\rme^{ \frac{\beta }{2n\rho}\left(h^2+  2h(K-1)\langle\tilde{h}\rangle\right)}  {{n}\choose{\frac{n+h}{2}} }    h     }{
\sum_{\hat{h}} \rme^{ \frac{\beta }{2n\rho}\left(\hat{h}^2+  2\hat{h}(K-1)\langle\tilde{h}\rangle\right)}
{{n}\choose{\frac{n+\hat{h}}{2}}}
  }\label{eq:<h>_large_L}  
   \end{eqnarray}
We note that in this limit the cavity distribution $P[h]$ and the physical distribution $P(h)$ are the same, and  the average $\langle h\rangle/n$ equals the average magnetization $m$.  

Let us next define the average 
\begin{eqnarray}
\langle g(h)\rangle_x&=&\frac{\sum_{h}\rme^{ \frac{\beta }{2n\rho}\left(h^2+  2h(K-1)x\right)}  {{n}\choose{\frac{n+h}{2}} }  g(h)  }{
\sum_{\hat{h}} \rme^{ \frac{\beta }{2n\rho}\left(\hat{h}^2+  2\hat{h}(K-1)x\right)}
{{n}\choose{\frac{n+\hat{h}}{2}}}          }\label{eq:<>_x}  
   \end{eqnarray}
It allows us to write the RHS of equation (\ref{eq:<h>_large_L}) as $\langle h\rangle_x$, where $x=\langle\tilde{h}\rangle$. The function $\langle h\rangle_x$ has the following properties for all $x\geq 0$:
\begin{enumerate}[i)]
  \item $\langle h\rangle_{-x}=-\langle h\rangle_x$
  \item $\frac{\partial }{\partial x}\langle h\rangle_x=\frac{\beta }{n\rho}(K\!-\!1)\big\langle \left(h\! -\! \langle h\rangle_x\right)^2\big\rangle_x\geq0$
  \item $\frac{\partial^2 }{\partial x^2}\langle h\rangle_x=\big(\frac{\beta }{n\rho}\big)^2(K\!-\!1)^2\big\langle \left(h\! -\! \langle h\rangle_x\right)^3\big\rangle_x\leq0$
\end{enumerate}
From i) it follows that $\langle\tilde{h}\rangle=0$ is a fixed point of the recursion (\ref{eq:<h>_large_L}). According to ii) this point becomes unstable when $\frac{\partial }{\partial x}\langle h\rangle_x\vert_{x=0}= \frac{\beta }{n\rho}(K\!-\!1)\langle h^2\rangle_0\geq1$. The derivative $\frac{\partial }{\partial x}\langle h\rangle_x\vert_{x=0}$ is a monotonic nondecreasing function of $\beta$, by the inequality 
\begin{eqnarray}
\frac{\partial^2 }{\partial \beta\partial x}\langle h\rangle_x\vert_{x=0}= \frac{(K\!-\!1)}{n\rho}\langle h^2\rangle_0  + \frac{\beta (K\!-\!1)}{n\rho}\left[   \langle h^4\rangle_0  \!-\! \langle h^2\rangle_0^2\right] \geq0\label{eq:ineq}  
   \end{eqnarray}
Furthermore, for $\beta$ large 
\begin{eqnarray}
\langle h^2\rangle_0&=&  \frac{\sum_{h}\rme^{ \frac{\beta }{2n\rho}h^2}  {{n}\choose{\frac{n+h}{2}} }  h^2  }{
\sum_{\hat{h}} \rme^{ \frac{\beta }{2n\rho}\hat{h}^2}
{{n}\choose{\frac{n+\hat{h}}{2}} }        }=\frac{\sum_{h}\rme^{ -\frac{\beta }{2n\rho}\left(n^2-h^2\right)}  {{n}\choose{\frac{n+h}{2}}}   h^2  }{
\sum_{\hat{h}} \rme^{- \frac{\beta }{2n\rho}\left(n^2-\hat{h}^2\right)}
{{n}\choose{\frac{n+\hat{h}}{2}} }       }\nonumber\\
&=& \frac{  {{n}\choose{0}}   (-n)^2  +{{n}\choose{n}} n^2 }{  {{n}\choose{0}}    + {{n}\choose{n}}           }=n^2\label{eq:h^2_beta_large}  
   \end{eqnarray}
and hence $\frac{\partial }{\partial x}\langle h\rangle_x\vert_{x=0}\in[0,   (\beta n/\rho)(K\!-\!1)]$ for $\beta\in[0, \infty)$. Thus there exists a finite value $\beta_c$ where the trivial solution $\langle\tilde{h}\rangle=0$ of the recursion (\ref{eq:<h>_large_L}) becomes unstable.  According to i) and iii) the function $\langle h\rangle_x$ is concave (respectively convex) on the interval  $x\in(0,\infty)$ (respectively $x\in(-\infty,0)$), and hence this function crosses the diagonal $x$ only once on this interval when $\frac{\partial }{\partial x}\langle h\rangle_x\vert_{x=0}>1$. This intersection corresponds to the stable solution $|\langle\tilde{h}\rangle\vert\neq0$ of equation (\ref{eq:<h>_large_L}). 
To prove iii)  we use the identity ${{n}\choose{\frac{n+h}{2}}}=\sum_{\{\sigma^\alpha\}}\delta_{h;\sum_{\alpha=1}^n\sigma^\alpha}$ in equation (\ref{eq:<>_x}), which gives 
\begin{eqnarray}
\langle g(h)\rangle_x
% 
% &=&\frac{\sum_{h}\rme^{ \frac{\beta }{2n\rho}\left(h^2+  2h(K-1)x\right)}  \binom{n}{\frac{h+n}{2}}   h  }{
% \sum_{\hat{h}} \rme^{ \frac{\beta }{2n\rho}\left(\hat{h}^2+  2\hat{h}(K-1)x\right)}
% \binom{n}{\frac{\hat{h}+n}{2}}}\nonumber\\
% %
% &=&\frac{\sum_{h}\rme^{ \frac{\beta }{2n\rho}\left(h^2+  2h(K-1)x\right)}  \sum_{\sigma^1,\ldots,\sigma^n}\delta_{h;\sum_{\alpha=1}^n\sigma^\alpha} h  }{           % \sum_{\sigma^1,\ldots,\sigma^n}\delta_{h;\sum_{\alpha=1}^n\sigma^\alpha}
% \sum_{\hat{h}} \rme^{ \frac{\beta }{2n\rho}\left(\hat{h}^2+  2\hat{h}(K-1)x\right)}
% \binom{n}{\frac{\hat{h}+n}{2}}}\nonumber\\
%
&=& \frac{  \sum_{\{\sigma^\alpha\}} \rme^{\frac{\beta }{n\rho}\sum_{\alpha<\gamma}\sigma^\alpha\sigma^\gamma +  \frac{\beta }{n\rho}(K-1)x\sum_{\alpha=1}^n\sigma^\alpha} g(\sum_{\alpha=1}^n\sigma^\alpha)  }{           
\sum_{\{\tilde{\sigma}^\alpha\}} \rme^{\frac{\beta }{n\rho}\sum_{\alpha<\gamma}\tilde{\sigma}^\alpha\tilde{\sigma}^\gamma +  \frac{\beta }{n\rho}(K-1)x\sum_{\alpha=1}^n\tilde{\sigma}^\alpha}
}
.\label{eq:spin-average_large_L} 
   \end{eqnarray}
From this we infer that the average $\langle h\rangle_x$ equals the  average magnetization  in ferromagnetic Ising system with interactions and external field given by $1/n\rho$ and  $(K\!-\!1)x$, respectively.  By the  the Griffiths--Hurst--Sherman (GHS) theorem~\cite{Griffiths1970} the average magnetization in such systems is a concave function of a positive external field, and hence inequality iii) is true. 

Furthermore, upon using (\ref{eq:spin-average_large_L}) to compute the average $\langle h^2\rangle_0$ we obtain the identity  $(\beta /n\rho)(K\!-\!1)\langle h^2\rangle_0=(\beta /\rho)(K\!-\!1)+(2\beta /n\rho)(K\!-\!1) \sum_{\alpha<\gamma}\langle\sigma^\alpha\sigma^\gamma\rangle_0$. The correlation function  $\langle\sigma^\alpha\sigma^\gamma\rangle_0\geq0$, by the Griffiths-Kelly-Sherman (GKS) theorem~\cite{Domb1972}, and $\langle\sigma^\alpha\sigma^\gamma\rangle_0\leq1$, since $\sigma^\alpha\in\{-1,1\}$ for all $\alpha$. From this follow the inequalities  $(\beta/\rho)(K\!-\!1)\leq(\beta /n\rho)(K\!-\!1)\langle h^2\rangle_0 < (n\beta /\rho)(K\!-\!1)$, which gives us the following lower and upper bounds on $\beta_c$: $(\rho /n(K\!-\!1))  <    \beta_c     \leq    \rho /(K\!-\!1)$. 

Next we define the function $m(h)=h/n$ and the average $m=\sum_h P[h] m(h)$.  Using the recursive equation (\ref{eq:<h>_large_L}) we obtain 
\begin{eqnarray}
m&=&\frac{\sum_{h}\rme^{ n\frac{\beta }{2\rho}\left(m^2(h)+  2m(h)(K-1)m\right)} {{n}\choose{n\frac{1+m(h)}{2}} }m(h)     }{
\sum_{\tilde{h}} \rme^{ n\frac{\beta }{2\rho}\left(m^2(\tilde{h})+  2m(\tilde{h})(K-1) m\right)}
{{n}\choose{n\frac{1+m(\tilde{h})}{2}}}
  }\label{eq:m_large_L_1}  
   \end{eqnarray}
for this average. If in addition we define the function $\phi_n(m(h))=  (\beta /2\rho)[m(h)^2+  2m(h)(K\!-\!1)m]+ \frac{1}{n}\log{{n}\choose{n\frac{1+m(h)}{2}}}$, then for $n\rightarrow\infty$ we obtain the equation 
\begin{eqnarray}
m&=&\frac{\sum_{h}\rme^{ n\, \phi_n(m(h))}  m(h)     }{
\sum_{\tilde{h}} \rme^{ n\,\phi_n(m(\tilde{h}))}
}\nonumber\\
&=& \frac{\sum_{h}\rme^{ -n\, \left(\sup_{M\in[-1,1]}\phi_\infty(M) - \phi_n(m(h)) \right)   }  m(h)     }{
\sum_{\tilde{h}} \rme^{ -n\,\left(\sup_{M\in[-1,1]}\phi_\infty(M) - \phi_n(m(\tilde{h})) \right)    }
}\nonumber\\
&=&\argsup_{M\in[-1,1]}  \left\{ \frac{\beta }{2\rho}[M^2+  2 M(K-1)m] + \mathcal{S}\left(\frac{1}{2}(1\!+\!M)\right)\right\}, \label{eq:m_large_L_2}
   \end{eqnarray}
where $\mathcal{S}(u)$ is the Shannon entropy of a binary variable $s\in\{0,1\}$ with average $\langle s\rangle =u$~\cite{Cover2012}. From the above it follows that $m$ satisfies the equation 
\begin{eqnarray}
m&=&\tanh(\beta K m/\rho) \label{eq:MF_large_L_n}
   \end{eqnarray}
which recovers the $\tilde\beta\!\rightarrow\!\infty$ equation  (\ref{eq:m-large-n-ferro}), in the limit $L\rightarrow\infty$ with $J^2=1/(L\!-\!1)$. 

Finally, we note that the field distribution (\ref{eq:P(F)-ferro}) converges to $\delta\left(F\right)$ in the limit  $L\rightarrow\infty$ with $J^2=1/(L\!-\!1)$, and  although the T-clones  are responding to Ag in this regime (one can set $J^2=\tilde{J}^2/(L\!-\!1)$), they are unable to control the B-clones. 

\subsection{Finite $L$ and large $K$ (or $M/N\rightarrow0$) regime\label{section:large-K}}
The starting point of our analysis in this parameter regime is to note that  the equation (\ref{eq:P[h]-ferro}) can be written in the following form:
\begin{eqnarray}
  P[h]  &=&\frac{ \Big[ \int\! \rmD z\,\rme^{  z J\sqrt{\frac{\beta}{n\rho}}h} \Big(\! \sum_{\tilde{h}}  P [\tilde{h}]\, \rme^{ z J\sqrt{\frac{\beta}{n\rho}} \tilde{h}}                         \Big)^{K-1}\Big]^{L-1}\!\!  {{n}\choose{\frac{n+h}{2}} }    }{
\sum_{\hat{h}}\!
  \Big[ \int\! \rmD z\,\rme^{  z J\sqrt{\frac{\beta}{n\rho}}\hat{h}} \Big( \!\sum_{\tilde{h}}  P [\tilde{h}]\, \rme^{ z J\sqrt{\frac{\beta}{n\rho}} \tilde{h}}                         \Big)^{K-1}\Big]^{L-1}\!\!
  {{n}\choose{\frac{n+\hat{h}}{2}}}
  }\label{eq:P[h]-ferro-gauss},
 % 
  % \label{eq:}\\  
\end{eqnarray}
where $\rmD z \equiv \rme^{-\frac{1}{2}z^2}\rmd z/\sqrt{2\pi}$. Using the multinomial theorem we can write
\begin{eqnarray}
 &&\hspace*{-10mm} \Big( \sum_{h}  P [ h]\, \rme^{ z J\sqrt{\frac{\beta}{n\rho}} h }                  \Big)^{K-1} 
 =\sum_{\{\mathcal{N}[h]\}}\frac{(K-1)!}{\left\{\prod_h \mathcal{N}[h]!\right\}}\left\{\prod_h P^{\mathcal{N}[h]} [ h]\rme^{ z J\sqrt{\frac{\beta}{n\rho}} \mathcal{N}[h]h } \right\}\nonumber\\
&&\hspace*{15mm} =\sum_{\{\mathcal{N}[h]\}}\!\frac{(K\!-\!1)!}{\left\{\prod_h \mathcal{N}[h]!\right\}}\left\{\prod_h P^{\mathcal{N}[h]} [ h]\right\}\rme^{ z J\sqrt{\frac{\beta}{n\rho}} \sum_h\mathcal{N}[h]h }\label{eq:multi},
\end{eqnarray}
where $\sum_h\mathcal{N}[h]=K\!-\!1$ and $\mathcal{N}[h]\in\{0,1,\ldots,K\!-\!1\}$. Using the above in equation (\ref{eq:P[h]-ferro-gauss}) gives us 
\begin{eqnarray}
\hspace*{-12mm}
  P[h]&=&\Bigg[ \sum_{\{\mathcal{N}[\tilde{h}]\}}\frac{(K\!-\!1)!}{\left\{\prod_{\tilde{h}} \mathcal{N}[\tilde{h}]!\right\}}\Bigg\{\prod_{\tilde{h}} P^{\mathcal{N}[\tilde{h}]} [\tilde{h}]\Bigg\} \rme^{\frac{\beta J^2}{2n\rho}\left(h+\sum_{\tilde{h}}\mathcal{N}[\tilde{h}]\tilde{h} \right)^2}\Bigg]^{L-1}
 \! {{n}\choose{\frac{n+h}{2}}  } \nonumber\\  
  \hspace*{-12mm}
&&\times\Bigg\{\sum_{\hat{h}}
\Bigg[ \sum_{\{\mathcal{N}[\tilde{h}]\}}\frac{(K-1)!}{\left\{\prod_{\tilde{h}} \mathcal{N}[\tilde{h}]!\right\}}\Bigg\{\prod_{\tilde{h}} P^{\mathcal{N}[\tilde{h}]} [\tilde{h}]\Bigg\}
 \rme^{\frac{\beta J^2}{2n\rho}\left(\hat{h}+\sum_{\tilde{h}}\mathcal{N}[\tilde{h}]\tilde{h} \right)^2}\Bigg]^{L-1}
\nonumber\\[-2mm]
\hspace*{-12mm}
&&\hspace*{70mm}\times 
 {{n}\choose{\frac{n+\hat{h}}{2}}}\Bigg\}^{-1}
  \label{eq:P[h]_calc2},
 % 
  % \label{eq:}\\  
\end{eqnarray}
Let us consider the multinomial distribution in equation (\ref{eq:P[h]_calc2}).  Using the formula $K!\!=\!\rho_K K^K\rme^{-K}$, with $\rho_0\!=\!1$ and  $\rho_{K>0}\!=\!\sqrt{2\pi K}\rme^{\frac{\theta_K}{12K}}$, where $\vert\theta_K\vert<1$,  we obtain 
\begin{eqnarray}
 &&\hspace*{-15mm} \frac{(K\!-\!1)!}{\left\{\prod_{\tilde{h}} \mathcal{N}[\tilde{h}]!\right\}}\left\{\prod_{\tilde{h}} P^{\mathcal{N}[\tilde{h}]} [\tilde{h}]\right\}
 =\frac{ \rho_{K-1} (K-1)^{K-1}\rme^{-(K-1)}     }{   \prod_{\tilde{h}} 
 \rho_{ \mathcal{N}[\tilde{h}]}\,  \mathcal{N}^{ \mathcal{N}[\tilde{h}]} [\tilde{h}]\,  \rme^{- \mathcal{N}[\tilde{h}]}   }\,
 \rme^{ \sum_{\tilde{h}} \mathcal{N}[\tilde{h}]\log P [\tilde{h}] }\nonumber\\
  &&\hspace*{25mm} = \rme^{ -(K-1)\sum_{\tilde{h}}       \frac{     \mathcal{N}[\tilde{h}]}{K-1}     \log\left(  \frac{     \mathcal{N}[\tilde{h}]}{(K-1)P [\tilde{h}]}\right)   -\sum_{\tilde{h}}\log  \rho_{ \mathcal{N}[\tilde{h}]}  + \log\rho_{K-1} }\nonumber\\
  &&\hspace*{25mm} = \rme^{ -(K-1)D\left(  \frac{     \mathcal{N}}{K-1}  \vert\vert P\right) \, - \,R\left(\frac{\mathcal{N}}{K-1}\right)\, +\mbox{const.}}\label{eq:}, 
\end{eqnarray}
The functional $D\left(  \frac{     \mathcal{N}}{K-1}  \vert\vert P\right)=\sum_{\tilde{h}}       \frac{     \mathcal{N}[\tilde{h}]}{K-1}     \log  \left(\frac{     \mathcal{N}[\tilde{h}]}{(K-1)P [\tilde{h}]}\right)$ is the KL  distance between the distributions $\mathcal{N}[h]/(K\!-\!1)$ and $P [h]$.  We have also defined the remainder functional  
\begin{eqnarray}
R\left(\frac{\mathcal{N}}{K-1}\right)=  \frac{1}{2}\sum_{\tilde{h}}\log \left( \frac{ \mathcal{N}[\tilde{h}]}{K-1}  \right)+ \sum_{\tilde{h}}\frac{\theta_{\mathcal{N}[\tilde{h}]}/(K-1)}{12\mathcal{N}[\tilde{h}]/(K-1)}
\end{eqnarray}
Let us next consider the average 
\begin{eqnarray}
 &&\hspace*{-15mm} \frac{ \sum_{\{\mathcal{N}[\tilde{h}]\}}\frac{(K-1)!}{\left\{\prod_{\tilde{h}} \mathcal{N}[\tilde{h}]!\right\}}\left\{\prod_{\tilde{h}} P^{\mathcal{N}[\tilde{h}]} [\tilde{h}]\right\}
  \rme^{\frac{\beta J^2}{2n\rho}\left(h+\sum_{\tilde{h}}\mathcal{N}[\tilde{h}]\tilde{h} \right)^2} }
  {\sum_{\{\mathcal{N}[\tilde{h}]\}}\frac{(K-1)!}{\left\{\prod_{\tilde{h}} \mathcal{N}[\tilde{h}]!\right\}}\left\{\prod_{\tilde{h}} P^{\mathcal{N}[\tilde{h}]} [\tilde{h}]\right\}}\nonumber\\
  &=& \frac{ \sum_{\{\mathcal{N}[\tilde{h}]\}}  
  \rme^{ -(K-1)D\left(  \frac{     \mathcal{N}}{K-1}  \vert\vert P\right) \, - \,R\left(\frac{\mathcal{N}}{K-1}\right)}\,
  \rme^{\frac{\beta J^2}{2n\rho}\left(h+(K-1)\sum_{\tilde{h}}\frac{\mathcal{N}[\tilde{h}]}{K-1}\tilde{h} \right)^2} }
  {\sum_{\{\mathcal{N}[\tilde{h}]\}}
   \rme^{ -(K-1)D\left(  \frac{     \mathcal{N}}{K-1}  \vert\vert P\right) \, - \,R\left(\frac{\mathcal{N}}{K-1}\right)}
 }
  \end{eqnarray}
For $K$ large,  with $J=1/(K\!-\!1)$,  the above average is dominated by the summands $\mathcal{N}[h]/(K\!-\!1)=P[h]$, and is equal to $ \exp[\frac{\beta }{2n\rho}\left(\frac{h}{K-1}+\sum_{\tilde{h}}P[\tilde{h}] \tilde{h}\right)^2]$. This allows us to compute the distribution (\ref{eq:P[h]_calc2}) for large $K\!-\!1$, which satisfies the equation
\begin{eqnarray}
   P[h]
   % &=&\frac{ \rme^{\frac{\beta }{2n\rho}(L-1)\left(\frac{h}{K-1}+\sum_{\tilde{h}}P[\tilde{h}] \tilde{h}\right)^2}\binom{n}{\frac{h+n}{2}}     }{
% \sum_{\hat{h}}\rme^{\frac{\beta }{2n\rho}(L-1)\left(\frac{\hat{h}}{K-1}+\sum_{\tilde{h}}P[\tilde{h}] \tilde{h}\right)^2}
%  \binom{n}{\frac{\hat{h}+n}{2}}
%   }\nonumber\\
  %
  &=&\frac{ \rme^{\frac{\beta }{2n\rho}(L-1)\left(  \frac{h^2}{(K-1)^2}+2\frac{h}{(K-1)}\sum_{\tilde{h}}P[\tilde{h}] \tilde{h}        \right)}{{n}\choose{\frac{h+n}{2}} }    }{
\sum_{\hat{h}}\rme^{\frac{\beta }{2n\rho}(L-1)\left(  \frac{\hat{h}^2}{(K-1)^2}+2\frac{\hat{h}}{(K-1)}\sum_{\tilde{h}}P[\tilde{h}] \tilde{h}        \right)}
 {{n}\choose{\frac{n+\hat{h}}{2}}}
  }.\label{eq:P[h]_large_K}
\end{eqnarray}
Clearly $\lim_{K\to\infty} P[h]=     {{n}\choose{\frac{\hat{h}+n}{2}} }   / \sum_{\hat{h}}    {{n}\choose{\frac{\hat{h}+n}{2}} } $, so this limit is equivalent to the infinite temperature regime $\beta=0$, and the T-clones are unable to control the B-clones.

\section{Discussion\label{section:discuss}}

%Summary of  mathematical  results, open problems,  biological interpretation
In this paper we study lymphocyte network models of the adaptive immune system.  We derive dynamic equation for the B  cell clones (B-clones), which depend on the T cell clones (T-clones) and antigen, and  assume (following \cite{Agliari2011,Agliari2013}) that the Hamiltonian of  this dynamic equation  also governs the dynamics of the T-clones.  Furthermore,  we propose that the dynamics of B-  and T-clones is subject to different thermal noise environments, and that they may evolve on different characteristic  timescales.  We compute the stationary distribution of the process in the limit of  infinite (adiabatic) separation of timescales, which corresponds to an equilibration scenario in which the  fast (or slow) B-clone variables are interacting with the slow (or fast) T-clone variables.  From the stationary distribution we obtain  the average density of B-clone sizes, and  the thermal averages of other macroscopic observables. 

To simplify our analysis we consider the  scenario where T-clones can be modelled by either  Ising  spin variables $\{-1,1\}$ and binary variables $\{0,1\}$.  The former definition describes activated helper T-cells, modelled by $+1$, and activated regulator T-cells, modelled by $-1$.  In the latter case we only have helper T-clones, which are either inactive,  modelled by $0$, or active, modelled by $+1$. We show that in the fast B-clone equilibration regime and in the fast T-clone equilibration regime, with the ratio of B clone noise to T-clone noise parameter $n=\tilde\beta/\beta\in \mathbb{Z}^+$,  the behaviour   of T-clones is governed by an equilibrium distribution of  $n\times N$ interacting Ising spins,  in a effectively ferromagnetic model  with  inverse  temperature $\beta$ and  interactions $J_\mu^2/\rho$.  As a consequence,  there are network topologies for which there exists the critical noise $\beta_c$ such that for $\beta> \beta_c$ the fraction of helper (regulator)  T cells is a monotonic  function of  the  Ag concentration.  Furthermore,  we show that  the average B-clone size (or concentration), $\langle c\rangle$, is a monotonic  increasing function of the fraction of helper  T-clones $m_+$ (i.e. a decreasing function of the fracion $m_-$ of regulators). This result  is consistent with experimental data~\cite{Baumjohann2013, Vanderleyden2014}.    Unfortunately, at present we are unable to carry out  such  a topology-independent  analysis  for the fast T-clone equilibration regime with non-integer ratios $n\in \mathbb{R}^+$.  

Obtaining distributions $P(c)$ of B-clone sizes requires  a more detailed knowledge of the topology of the lymphocyte networks.  Assuming this topology to be locally tree-like, we use the  Bethe-Peierls  (BP) approximation to derive equations for $P(c)$ and other observables in models on random networks. We solve these equations for the case when the model is homogeneous ($J_\mu=J$ for all $\mu$) and defined on a random regular graph, where each B-clone is connected to exactly $K$ T-clones and each T-clone is connected to exactly $L$ B-clones;  here the ratio of B- to T-clones $\alpha=M/N$ equals $L/K$. We study this model in the fast B-clone equilibration regime and the fast T-clone equilibration regime, with $n\in \mathbb{Z}^+$ and $\tilde\beta\rightarrow\infty$.  We find that for fast B-clone equilibration 
the distribution  $P(c)$ in an ``overregulated''    ($m_-=0.9$) immune response regime is different from the distribution in a ``normal''   ($m_-=0.1$) regime:  it behaves as a ``power law'' when $m_-=0.9$. The overregulated regime corresponds to the ``branch'' of the phase diagram (see Figure \ref{figure:PD}) where the ``signal'' from the T cells to the B cell clones is predominantly suppressive (see Figure  \ref{figure:P(F)-B-fast}). We envisage that in the  real immune system such a situation can occur when the B cell clones are \emph{self-reactive}, which is possible in  the immune response to tumours~\cite{Zou2006}. 

We also study the regimes when $\alpha\rightarrow\infty$ (`large'' B-clone number limit) and $\alpha\rightarrow0$ (``large'' T-clone number limit), with rescaled interactions $J^2=\frac{1}{L-1}$ and $J=\frac{1}{K-1}$ respectively  (for technical reasons). We find that in both of these regimes the B-clones are operating independently of the T-clones. In the regime $\alpha\rightarrow\infty$ we find that the response of T-clones to Ag is following the same phase transition  pattern as in the Figure  \ref{figure:PD}.  Also, in this regime the ``cavity'' distribution equation (\ref{eq:P-cavity-spin-integer-n}) simplifies significantly, and much can be learned about this equation analytically (although for finite   $K$,  $L$ and $n>1$ we were unable to find an explicit solution, even in the ``simple'' high temperature phase).  The regime $\alpha\rightarrow0$ is equivalent to the infinite temperature regime, where  T cells are insensitive  to the Ag stimulation.  

%Future work:
In future studies we plan to compute phase diagrams for the model in the  fast T-clone equilibration regime with non-integer $n\in\mathbb{R}^+$.  A good starting point here would be to consider the regular case (see \ref{app:ferro-n-real}).  Also, it would be interesting to consider  the case where the fraction of T-regulator clones  is fixed and all T-clones  could be active or inactive, i.e. the case of  $\sigma_i\in\{0,1\}$ and $\xi_i\in\{-1,1\}$. Here one could assume that that the networks and interactions are random, and compute the phase diagrams by solving equation (\ref{eq:P[h]-pop-dyn}) by population dynamics.  In order to model the process of affinity maturation, a further important ingredient of the adaptive immune system  which has not been included  yet,  one can also assume that,  in addition to B clones and T clones, also the interactions $J_\mu$ evolve in time, and use the slow (or fast) variable assumptions to compute stationary distributions of the more complicated process.  Furthermore, the assumption of separated  time-scales  can be relaxed,  
but to make progress in this purely dynamical scenario would require application of  sophisticated analytical tools such as the (dynamical) path integral method~\cite{Mimura2009} or the dynamical replica theory~\cite{Mozeika2009}. However, the most important next step would be to connect the theoretical framework developed in this article with concrete experimental data. %~\cite{Desponds2016} ("power laws" in clonal data )~\cite{Stubbington2016} (single-cell sequencing)~\cite{Best2015} (PCR bias)

\section*{Acknowledgements}
The authors are indebted to Deborah Dunn-Walters, Franca Fraternali, Victoria Martin and Joselli Silva O'Hare for their invaluable assistance and very enlightening discussions.  We also would like to thank Alessia Annibale,  Adriano Barra,  Elena Agliari, Silvia Bartolucci and Daniele Tantari. This work was supported by the United Kingdom Research Councils BBSRC (BB/G017190/1) and MRC (MR/L01257X/1).

\appendix

\section{Analysis of the function $\langle\rme^{F/n\rho}\rangle_\beta$\label{app:<exp(F)>}}
In this section we study the behaviour as a function of $\beta$ of the following average:
%\label{eq:P(s)-integer-n-Ising}
\begin{eqnarray}
\langle\rme^{F/n\rho}\rangle_\beta &=&  \sum_{\{\bsigma^\alpha\}}P(\bsigma^1,\ldots, \bsigma^n)\frac{1}{M} \sum_{\nu=1}^{M}\rme^{\frac{1}{n\rho}\sum_{\alpha=1}^{n}\!F_\nu(\bsigma^\alpha)}\label{def:<exp(F)>}\\
&=&\!\frac{1}{M}\!\sum_{\nu=1}^{M}\!
\frac{ \sum_{\{\bsigma^\alpha\}}\! \rme^{\frac{\beta   }{2n\rho}\!\sum_{\mu=1}^{M}\left(\sum_{\alpha=1}^{n}\!F_\mu(\bsigma^\alpha)\right)^2 }    \!\!\rme^{\frac{1}{n\rho}\sum_{\alpha=1}^{n}\!F_\nu(\bsigma^\alpha)} }{ \sum_{\{\tilde\bsigma^\alpha\}}    \rme^{\frac{\beta   }{2n\rho}\sum_{\mu=1}^{M}\left(\sum_{\alpha=1}^{n}F_\mu(\tilde{\bsigma}^\alpha) \right)^2}         } \nonumber.
\end{eqnarray}
Here $F_\mu(\bsigma^\alpha)=J_\mu(\sum_{i\in\partial\mu}\sigma_i^\alpha +\theta_\mu)$ with $\sigma_i^\alpha\in\{-1,1\}$, and we have $J_\mu\geq0$ and $\theta_\mu\geq0$ for all $\mu$.  In particular we are interested in the derivative 
\begin{eqnarray}
2n\rho\frac{\partial}{\partial\beta}\langle\rme^{F/n\rho}\rangle_\beta &=&
\frac{1}{ M}\sum_{\nu,\mu}\Big\{\Big\langle  \Big(\sum_{\alpha=1}^{n}\!F_\mu(\bsigma^\alpha)\Big)^2  \rme^{\frac{1}{n\rho}\sum_{\alpha=1}^{n}\!F_\nu(\bsigma^\alpha)}\Big\rangle\nonumber\\ 
&&~~~-\Big\langle  \Big(\sum_{\alpha=1}^{n}\!F_\mu(\bsigma^\alpha)\Big)^2\Big\rangle  \Big\langle\rme^{\frac{1}{n\rho}\sum_{\alpha=1}^{n}\!F_\nu(\bsigma^\alpha)}\Big\rangle\Big\}.\label{eq:d<exp(F)>-1}
\end{eqnarray}
Using Taylor's expansion of the exponential, 
\begin{eqnarray}
\rme^{\frac{1}{n\rho}\sum_{\alpha=1}^{n}\!F_\nu(\bsigma^\alpha)}&=&\sum_{\ell\geq0}\frac{1}{(n\rho)^\ell \ell!}\Big(\sum_{\alpha=1}^{n}\!F_\nu(\bsigma^\alpha)\Big)^\ell\label{eq:taylor}
\end{eqnarray}
this derivative can be written as the infinite sum 
\begin{eqnarray}
\hspace*{-5mm}
2n\rho\frac{\partial}{\partial\beta}\langle\rme^{F/n\rho}\rangle_\beta 
&=& \sum_{\ell\geq0}\frac{1}{(n\rho)^\ell \ell!}
\frac{1}{ M}\sum_{\nu,\mu}\Big\{
\Big\langle  \Big(\sum_{\alpha=1}^{n}\!F_\mu(\bsigma^\alpha)\Big)^2 \Big(\sum_{\alpha=1}^{n}\!F_\nu(\bsigma^\alpha)\Big)^\ell\Big\rangle\nonumber\\ 
&&-\Big\langle  \Big(\sum_{\alpha=1}^{n}\!F_\mu(\bsigma^\alpha)\Big)^2\Big\rangle  \Big\langle \Big(\sum_{\alpha=1}^{n}\!F_\nu(\bsigma^\alpha)\Big)^\ell \Big\rangle\Big\}.\label{eq:d<exp(F)>-2}
\end{eqnarray}
Finally, we rewrite the products $\left(\sum_{\alpha=1}^{n}\!F_\nu(\bsigma^\alpha)\right)^\ell =\sum_{\alpha_1,\ldots,\alpha_\ell\leq n}\prod_{j=1}^\ell F_\nu(\bsigma^{\alpha_j}) $ as sums, using the identity
\begin{eqnarray}
\prod_{j=1}^{\ell}  \left(x_j+\theta \right)&=&\sum_{\mathrm{S}_\ell\subseteq[\ell]}  \theta^{\ell-\vert\mathrm{S}_\ell\vert}\prod_{j\in \mathrm{S}_\ell}x_j, \label{eq:prod}
\end{eqnarray}
where $\mathrm{S}_\ell$ is a subset of the set $[\ell]=\{1,\ldots,\ell\}$. This leads us to the equation
\begin{eqnarray}
\hspace*{-10mm}
2n\rho\frac{\partial}{\partial\beta}\langle\rme^{F/n\rho}\rangle_\beta 
&=& \sum_{\ell\geq0}\frac{1}{(n\rho)^\ell \ell!}
\sum_{\mathrm{S}_2\subseteq[2]}\sum_{\mathrm{S}_\ell\subseteq[\ell]}
\frac{1}{ M}\sum_{\nu,\mu}J_\mu^2\,J_\nu^\ell\,\theta_\mu^{2-\vert\mathrm{S}_2\vert}\,\theta_\nu^{\ell-\vert\mathrm{S}_\ell\vert}\nonumber\\
&&\times\sum_{\gamma_1, \gamma_2}\sum_{\alpha_1,\ldots,\alpha_\ell=1}^n\Bigg\{
\Big\langle  \Big(\prod_{k\in\mathrm{S}_2} \sum_{i_k\in\partial\mu}\sigma_{i_k}^{\gamma_k}\Big)  \Big( \prod_{j\in\mathrm{S}_\ell} \sum_{i_j\in\partial\nu}\sigma_{i_j}^{\alpha_j}  \Big)    \Big\rangle\nonumber\\ 
&&~~~~~-\Big\langle  \Big(\prod_{k\in\mathrm{S}_2} \sum_{i_k\in\partial\mu}\sigma_{i_k}^{\gamma_k}\Big)   \Big\rangle  \Big\langle   \Big( \prod_{j\in\mathrm{S}_\ell} \sum_{i_j\in\partial\nu}\sigma_{i_j}^{\alpha_j}  \Big)    \Big\rangle
\Bigg\}.\label{eq:d<exp(F)>-3}
\end{eqnarray}
Now, by the GKS theorem~\cite{Domb1972}, the correlation terms in the above sum are  positive,  from which it follows that $\frac{\partial}{\partial\beta}\langle\rme^{F/n\rho}\rangle_\beta\geq0$ and hence the average $\langle\rme^{F/n\rho}\rangle_\beta$ is a monotonic non-decreasing function of $\beta$. Furthermore, for  $J_\mu=J\sum_{\nu\leq M} S_{\mu\nu} a_\mu$ the average $\langle\rme^{F/n\rho}\rangle_\beta$ is also a monotonic non-decreasing function of $J$. To show this we  consider the derivative
\begin{eqnarray}
J\frac{\partial}{\partial J}\langle\rme^{F/n\rho}\rangle_\beta&=&2\beta\frac{\partial}{\partial \beta}\langle\rme^{F/n\rho}\rangle_\beta\nonumber\\
&&+ \frac{1}{M} \sum_{\nu=1}^{M}\left\langle\rme^{\frac{1}{n\rho}\sum_{\alpha=1}^{n}\!F_\nu(\bsigma^\alpha)}\frac{1}{n\rho}\sum_{\alpha=1}^{n}\!F_\nu(\bsigma^\alpha)\right\rangle\label{eq:d<exp(F)>/dJ}.
\end{eqnarray}
Now the first term on the RHS of above is positive by the previous argument for $\frac{\partial}{\partial \beta}\langle\rme^{F/n\rho}\rangle_\beta$ and the second term is also positive by a similar argument which uses Taylor expansion (\ref{eq:taylor}) and the GKS theorem. 

The distribution $P(\bsigma^1,\ldots, \bsigma^n)$ used in definition (\ref{def:<exp(F)>}) can be written in the canonical form $P(\bsigma^1,\ldots, \bsigma^n)\propto\exp[-\beta E(\bsigma^1,\ldots, \bsigma^n)]$, where $E(\bsigma^1,\ldots, \bsigma^n)=-(2n\rho)^{-1}\sum_{\mu=1}^{M}(\sum_{\alpha=1}^{n}F_\mu(\bsigma^\alpha))^2$ is the corresponding energy function. Then the specific heat (density) $C(\beta)=N^{-1}\frac{\rmd}{\rmd T} \langle E(\bsigma^1,\ldots, \bsigma^n)\rangle$, where $T=\beta^{-1}$, is given by
\begin{eqnarray}
\hspace*{-5mm}
C(\beta)&=&\frac{\beta^2}{N} \left\{  \left\langle E^2(\bsigma^1,\ldots, \bsigma^n) \right\rangle   - \left\langle E(\bsigma^1,\ldots, \bsigma^n) \right\rangle^2\right\}\nonumber\\
&=&\Big(\frac{\beta}{2n\rho}\Big)^{\!2} \frac{1}{N}\sum_{\mu=1}^{M}\Bigg\{ \Big\langle \Big(\sum_{\alpha=1}^{n}\!F_\mu(\bsigma^\alpha)\Big)^{\!4}\Big\rangle 
-   \Big\langle\Big(\sum_{\alpha=1}^{n}\!F_\mu(\bsigma^\alpha)\Big)^{\!2}\Big\rangle^2\Bigg\}.
\label{eq:C}
\end{eqnarray}
Comparing this with equation (\ref{eq:d<exp(F)>-2}) reveals that  $2n\rho\frac{\partial}{\partial\beta}\langle\rme^{F/n\rho}\rangle_\beta=\frac{2N}{\beta^2 M}C(\beta) + \cdots$. As $N\rightarrow\infty$, with $0<N/M<\infty$,  the specific heat $C(\beta)$ will  diverging in some systems, such as ferromagnetic Ising models on  $d$-dimensional lattice, whereas in others, such as ferromagnetic Ising models on random trees, it will jump when $\beta\rightarrow\beta_c^-$ or $\beta\rightarrow\beta_c^+$.

\section{Computation on a factor tree for  $n\in \mathbb{Z}^{+}$\label{app:BP-integer-n}    }
\begin{figure}[t]
%\vspace*{-7mm}
\setlength{\unitlength}{1mm}
\begin{center}{
%\hspace*{10mm}
\begin{picture}(100,73)
\put(0,0){\includegraphics[height=75\unitlength]{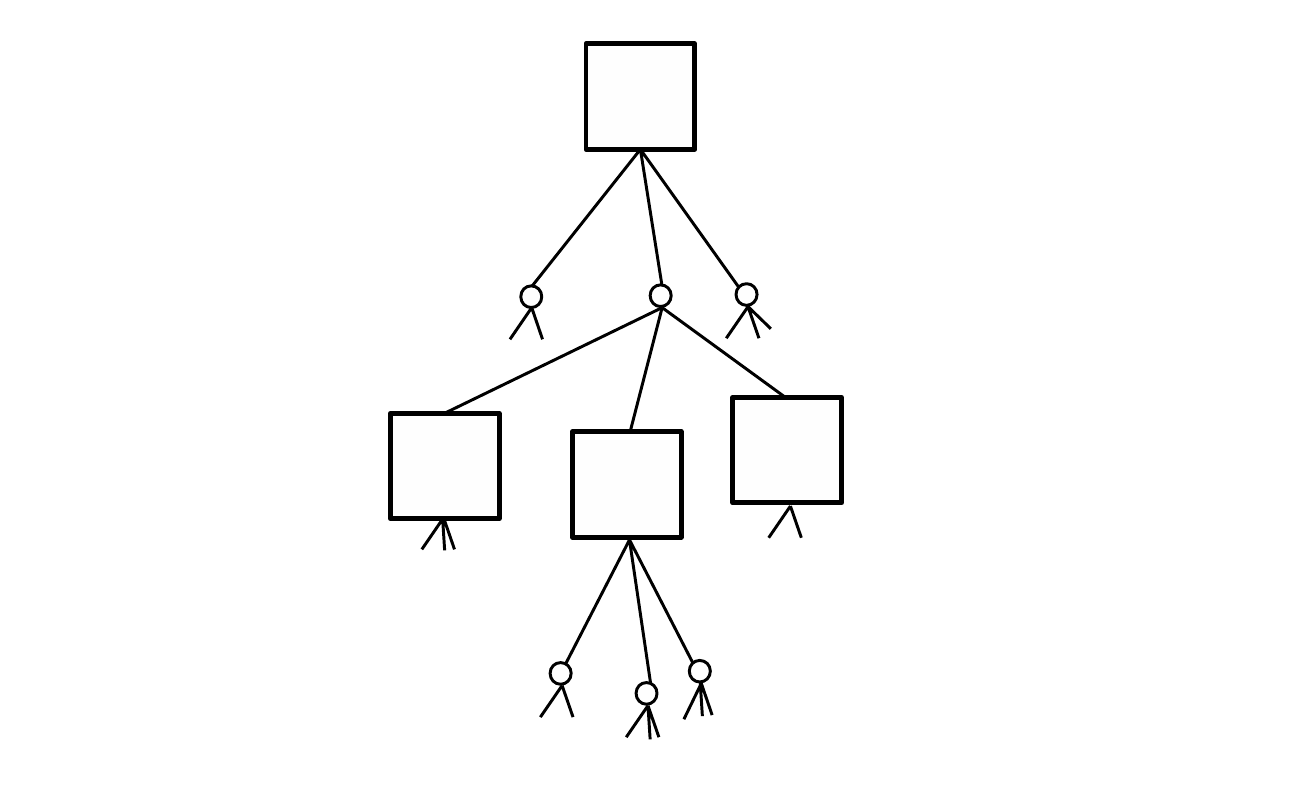}}
\put(60,64){$\mu$}
\put(58,47){$\bsigma_i$}
\put(59,27){$\nu$}
\put(57,10){$\bsigma_j$}
\end{picture}
}\end{center}
\vspace*{-6mm}
\caption{The interaction topology of a replicated spin system on a factor-tree $\mathcal{T}_\mu$, rooted at factor-node $\mu$. All spins are represented by circular ``variable'' nodes, and each term in the  ``Boltzmann factor'' $\exp[\frac{\beta   }{2n\rho}\sum_{\nu\in\mathcal{T}_\mu}(\sum_{\alpha=1}^{n}F_\nu(\bsigma^\alpha))^2]=\prod_{\nu\in\mathcal{T}_\mu}F_\nu(\{\bsigma_j:j\in\partial\nu\})$  corresponds to a square ``factor'' node. A link between variable node $\ell$ and factor node $\nu$  implies that $\bsigma_\ell$ acts as an argument of $F_\nu$.}
\label{figure:factor-tree-integer-n} 
\end{figure}

Let us assume that the system (\ref{eq:P(s)-integer-n-Ising}) is defined on a factor tree $\mathcal{T}_\mu(r)$ of radius $r$ rooted at factor-node $\mu$  (see Figure \ref{figure:factor-tree-integer-n}).  
Then the distribution of (replicated) fields on $\mu$ 
\begin{eqnarray}
P_\mu(F) &=&  \sum_{\{\bsigma^\alpha\}}P(\bsigma^1,\ldots, \bsigma^n)\,\delta\!\Big(F- \sum_{\alpha=1}^{n}F_\mu(\bsigma^\alpha)\Big)
\end{eqnarray}
can be computed recursively as follows. Firstly, we compute the ``partition'' function 
\begin{eqnarray}
\hspace*{-10mm}
Z_\mu(F)&=&\sum_{\{\bsigma_i:~i\in\mathcal{T}_\mu\}} \rme^{\frac{\beta   }{2n\rho}\sum_{\nu\in\mathcal{T}_\mu}\left(\sum_{\alpha=1}^{n}F_\nu(\bsigma^\alpha)\right)^2 }  \delta\Big(F-\sum_{\alpha=1}^{n}F_\mu(\bsigma^\alpha)\Big)\nonumber\\
&=&\rme^{\frac{\beta   }{2n\rho}F^2 }\!\!\sum_{\{\bsigma_i:~i\in\mathcal{T}_\mu\}} 
\!\rme^{\frac{\beta   }{2n\rho}\sum_{\nu\in\mathcal{T}_\mu\setminus\mu}\left(\sum_{\alpha=1}^{n}F_\nu(\bsigma^\alpha)\right)^2 }\delta\Big(F\!-\! \sum_{\alpha=1}^{n}F_\mu(\bsigma^\alpha)\Big)\nonumber\\
&=& \rme^{\frac{\beta   }{2n\rho}F^2 }\!\Bigg\{ \prod_{i\in\mu}\sum_{\{\bsigma_j:j\in\mathcal{T}_i\}} \!\rme^{\frac{\beta   }{2n\rho}\sum_{\nu\in\mathcal{T}_i}\left(\sum_{\alpha=1}^{n}F_\nu(\bsigma^\alpha)\right)^2}\!\Bigg\}
\delta\Big(F\!-\! \sum_{\alpha=1}^{n}F_\mu(\bsigma^\alpha)\Big)\nonumber\\
&=&\rme^{\frac{\beta   }{2n\rho}F^2 } \!\! \sum_{\{\bsigma_i:i\in\partial\mu\}}    \Bigg\{ \prod_{i\in\mu} Z_{\mu i}[\bsigma_i]  \Bigg\}  ~\delta\Big(F\!-\! \sum_{\alpha=1}^{n}F_\mu(\bsigma^\alpha)\Big)
\label{eq:Z-integer-n}, 
\end{eqnarray}
In above we defined the factor tree $\mathcal{T}_i$ (of radius $r\!-\!1$), rooted at variable node $i$. The partition function $Z_\mu(F)$ can be used to construct the distribution of fields
\begin{eqnarray}
P_\mu(F)&=&  \frac{Z_\mu(F)}{\int Z_\mu(\tilde{F})\,\rmd\tilde{F}}\label{def:P(F)-integer-n}\\
&=&    \frac{\rme^{\frac{\beta   }{2n\rho}F^2 } \sum_{\{\bsigma_i:i\in\partial\mu\}}  \left\{ \prod_{i\in\mu} Z_{\mu i}[\bsigma_i]  \right\} \delta\left(F- \sum_{\alpha=1}^{n}F_\mu(\bsigma^\alpha)\right)}
{ \sum_{\{\tilde{\bsigma}_i:i\in\partial\mu\}}  \left\{ \prod_{i\in\mu} Z_{\mu i}[\tilde{\bsigma}_i]  \right\}     \rme^{\frac{\beta   }{2n\rho}\left(  \sum_{\alpha=1}^{n}F_\mu(\tilde{\bsigma}^\alpha)     \right)^2 }                     } 
\nonumber
\end{eqnarray}
In the same way we can define the ``cavity'' distribution, $P_{\mu i}[\bsigma]$, corresponding to the topology that would be found if the edge $(i, \mu)$ were removed (see Figure \ref{figure:factor-tree-integer-n}), 
\begin{eqnarray}
P_{\mu i}[\bsigma]&=&  \frac{Z_{\mu i}[\bsigma]}{     \sum_{\tilde{\bsigma}}  Z_{\mu i}[\tilde{\bsigma}]         }, \label{def:P-cavity-spin-integer-n}
\end{eqnarray}
and thereby  obtain the equation 
\begin{eqnarray}
P_\mu(F)&=&\Big\langle   \delta\Big(F-\sum_{\alpha=1}^nF_\mu(\bsigma^\alpha\Big)\Big\rangle_n \label{eq:P(F)-integer-n-cavity}\\
&=&
\frac{\sum_{\{\bsigma_i\}} 
 \prod_{i\in \partial \mu}  P_{\mu i} [\bsigma_i] \, \rme^{   \frac{\beta}{2n\rho}   F^2  } \,   \delta\left(F-\sum_{\alpha=1}^n F_\mu(\bsigma^\alpha)\right)}{
 \sum_{\{\tilde{\bsigma}_i\}}   \prod_{i\in \partial \mu}   P_{\mu i} [\tilde{\bsigma}_i] \rme^{ 
 \frac{\beta}{2n\rho}  \left(  \sum_{\alpha=1}^n F_\mu(\tilde{\bsigma}^\alpha)    \right)^2           }}\nonumber. 
\end{eqnarray}
 We note that in order to distinguish   ``cavity'' distributions, such as $P_{\mu i}[\bsigma]$, from the corresponding distributions on graphs with all links intact, such as $P_{ i} (\bsigma)$, we will use square brackets $[\ldots]$ in referring to the former,  throughout this paper. 
 
We next  we compute the cavity partition function 
\begin{eqnarray}
\hspace*{-10mm}
Z_{\mu i}[\bsigma]
&=& \sum_{\{\bsigma_j:j\in\mathcal{T}_i\}} \rme^{\frac{\beta   }{2n\rho}\sum_{\nu\in\mathcal{T}_i}\left(\sum_{\alpha=1}^{n}F_\nu(\bsigma^\alpha)\right)^2 }\delta_{\bsigma;\bsigma_i}\nonumber\\
&=& \sum_{\{\bsigma_j:j\in\mathcal{T}_i\}} \rme^{\frac{\beta   }{2n\rho}\sum_{\nu\in\partial j\setminus\mu}\left(\sum_{\alpha=1}^{n}F_\nu(\bsigma^\alpha)\right)^2 } \nonumber\\
&&\times\prod_{\nu\in\partial i\setminus\mu} \prod_{j\in\partial\nu\setminus i}  \rme^{\frac{\beta   }{2n\rho}\sum_{\hat{\nu}\in \mathcal{T}_j}\left(\sum_{\alpha=1}^{n}F_{\hat{\nu}}(\bsigma^\alpha)\right)^2 }\delta_{\bsigma;\bsigma_i}\nonumber\\
&=& \sum_{\{\bsigma_j\}} \rme^{\frac{\beta   }{2n\rho}\sum_{\nu\in\partial i\setminus\mu}\left(\sum_{\alpha=1}^{n}F_\nu(\bsigma^\alpha)\right)^2 } \Bigg\{
\!\prod_{\nu\in\partial i\setminus\mu} \prod_{j\in\partial\nu\setminus i} Z_{\nu j}[\bsigma_j]\!\Bigg\} \delta_{\bsigma;\bsigma_i}
\nonumber
\\[-3mm]
&&
\label{eq:Z-cavity-integer-n}
\end{eqnarray}
Using this expression in definition (\ref{def:P-cavity-spin-integer-n}) gives us equation 
\begin{eqnarray}
\hspace*{-10mm}
 P_{\mu i} [\bsigma_i]&=&  \sum_{\{\bsigma_j\}}\! \Bigg\{\!\prod_{\nu\in \partial i\setminus \mu }\prod_{j\in \partial \nu \setminus i}\!   P_{\nu j} [\bsigma_j]\!\Bigg\}\!  \rme^{        \frac{\beta}{2n\rho} \! \sum_{\nu\in \partial i\setminus \mu} \! \left(\sum_{\alpha=1}^n \!F_\nu(\bsigma^\alpha) \right)^2 }\label{eq:P-cavity-spin-integer-n}\\
 \hspace*{-10mm}
 &&\times \Bigg[
    \sum_{\tilde{\bsigma}_i}\!\sum_{\{\tilde{\bsigma}_j\}}\!      
    \Bigg\{\!\prod_{\nu\in \partial i\setminus \mu }\prod_{j\in \partial \nu \setminus i}\!   P_{\nu j} [\tilde{\bsigma}_j]\!\Bigg\}  \rme^{        \frac{\beta}{2n\rho} \! \sum_{\nu\in \partial i\setminus \mu}\!  \left(\sum_{\alpha=1}^n \!F_\nu(\tilde{\bsigma}^\alpha) \right)^2 } \Bigg]^{-1}\!
.\nonumber
\end{eqnarray}

Furthermore,  we note that the partition function
\begin{eqnarray}
Z_i(\bsigma)&=&\sum_{\{\bsigma_j:j\in\mathcal{T}_i\}} \rme^{\frac{\beta   }{2n\rho}\sum_{\nu\in\mathcal{T}_i}\left(\sum_{\alpha=1}^{n}F_\nu(\bsigma^\alpha)\right)^2 }   \delta_{ \bsigma; \bsigma_i}\nonumber\\
&=& \sum_{\{\bsigma_j\}}  \Bigg\{  \prod_{\mu\in\partial i}  \prod_{j\in\partial\mu\setminus i} Z_{\mu j}[\bsigma_j]  \Bigg\}   \nonumber\\
&&~~\times  \rme^{\frac{\beta   }{2n\rho}\sum_{\mu\in\partial i}  J_\mu^2\left(\sum_{j\in\partial\mu\setminus i} \xi_j^\mu \sum_{\alpha=1}^{n}\sigma_j^\alpha +  \xi_i^\mu \sum_{\alpha=1}^{n}\sigma^\alpha +\theta_\mu )\right)^2 }
\label{eq:Zi-integer-n},
\end{eqnarray}
which is defined on a factor-tree $\mathcal{T}_i(R)$ rooted at the variable node $i$,  can be used to construct the distribution 
\begin{eqnarray}
P_i(\bsigma)&=&\frac{Z_i(\bsigma)}{\sum_{\tilde{\bsigma}}Z_i(\tilde{\bsigma})}
\label{def:Pi-integer-n}
\end{eqnarray}
which  gives  the (local) magnetization equation
\begin{eqnarray}
\langle\sigma_i\rangle&=& \sum_\bsigma P_i(\bsigma)\frac{1}{n}\sum_{\alpha=1}^{n}\sigma^\alpha
\nonumber\\
&=&\sum_{\{\bsigma_j\}} \Bigg\{  \prod_{\mu\in\partial i}  \prod_{j\in\partial\mu\setminus i} P_{\mu j}[\bsigma_j]  \Bigg\}  \sum_\bsigma \frac{1}{n}\sum_{\alpha=1}^{n}\sigma^\alpha
\nonumber\\
&&~~\times \rme^{\frac{\beta   }{2n\rho}\sum_{\mu\in\partial i}J_\mu^2\left(\sum_{j\in\partial\mu\setminus i} \xi_j^\mu \sum_{\alpha=1}^{n}\sigma_j^\alpha +  \xi_i^\mu \sum_{\alpha=1}^{n}\sigma^\alpha  +\theta_\mu)\right)^2 }\nonumber\\ 
&&~~\times\Bigg\{\sum_{\{\tilde{\bsigma}_j\}}  \left\{  \prod_{\mu\in\partial i}  \prod_{j\in\partial\mu\setminus i} P_{\mu j}[\tilde{\bsigma}_j]  \right\}\sum_{\tilde{\bsigma}}  
\label{eq:m-integer-n}
\\
&&~~~~~~~\times  \rme^{\frac{\beta   }{2n\rho}\sum_{\mu\in\partial i}J_\mu^2\left(\sum_{j\in\partial\mu\setminus i} \xi_j^\mu \sum_{\alpha=1}^{n}\tilde{\sigma}_j^\alpha +  \xi_i^\mu \sum_{\alpha=1}^{n}\tilde{\sigma}^\alpha +\theta_\mu)\right)^2 }  \Bigg\}^{-1}
\nonumber
\end{eqnarray}
The structure of the equations derived in this section  is not affected by the choice for  the $\sigma_i^\alpha$ variables, but in the Ising  case $\sigma_i^\alpha\in\{-1,1\}$ with $\theta_\mu=0$ the recursion (\ref{eq:P-cavity-spin-integer-n}) preserves the spin-reversal symmetry, i.e. the  equality  $P_{\nu\ell}[\bsigma]=P_{\nu\ell}[-\bsigma]$ implies that $P_{\mu j} [\bsigma]=P_{\mu j} [-\bsigma]$.  For $n=1$ this implies that $P_{\nu\ell}[\sigma]=\frac{1}{2}$ is a solution of this map for any $\beta$, but for $n>1$ the uniform distribution $P_{\nu\ell}[\sigma]=\frac{1}{2^n}$ is a solution of (\ref{eq:P-cavity-spin-integer-n}) only when $\beta=0$.  The consequence of the symmetry $P_{\mu j} [\bsigma]=P_{\mu j} [-\bsigma]$  is that the local magnetization becomes $\langle\sigma_i\rangle=0$, and that the distribution (\ref{eq:P(F)-integer-n-cavity}) is  symmetric, i.e. $P_\mu(F)=P_\mu(-F)$. 

Finally, we note that the equations (\ref{eq:P(F)-integer-n-cavity}), (\ref{eq:P-cavity-spin-integer-n}) and (\ref{eq:m-integer-n})  can be simplified if  we define the replica ``magnetization'' distribution
\begin{eqnarray}
 P[h] &=& \sum_{\bsigma}P [\bsigma] \delta_{h;\sum_{\alpha=1}^n\sigma^\alpha }\label{def:P[h]}. 
\end{eqnarray}
Using above definition in these equations gives the equations (\ref{eq:P(F)-using-P[h]}), (\ref{eq:P[h]}) and (\ref{eq:m-using-P[h]}).

\section{Computation on a factor tree for  $n\in \mathbb{R}^{+}$  \label{app:BP-real-n} }
\begin{figure}[t]
 %\vspace*{-7mm}
 \setlength{\unitlength}{1mm}
 \begin{center}{
 \hspace*{50mm}
 \begin{picture}(100,75)
 \put(0,0){\includegraphics[height=75\unitlength]{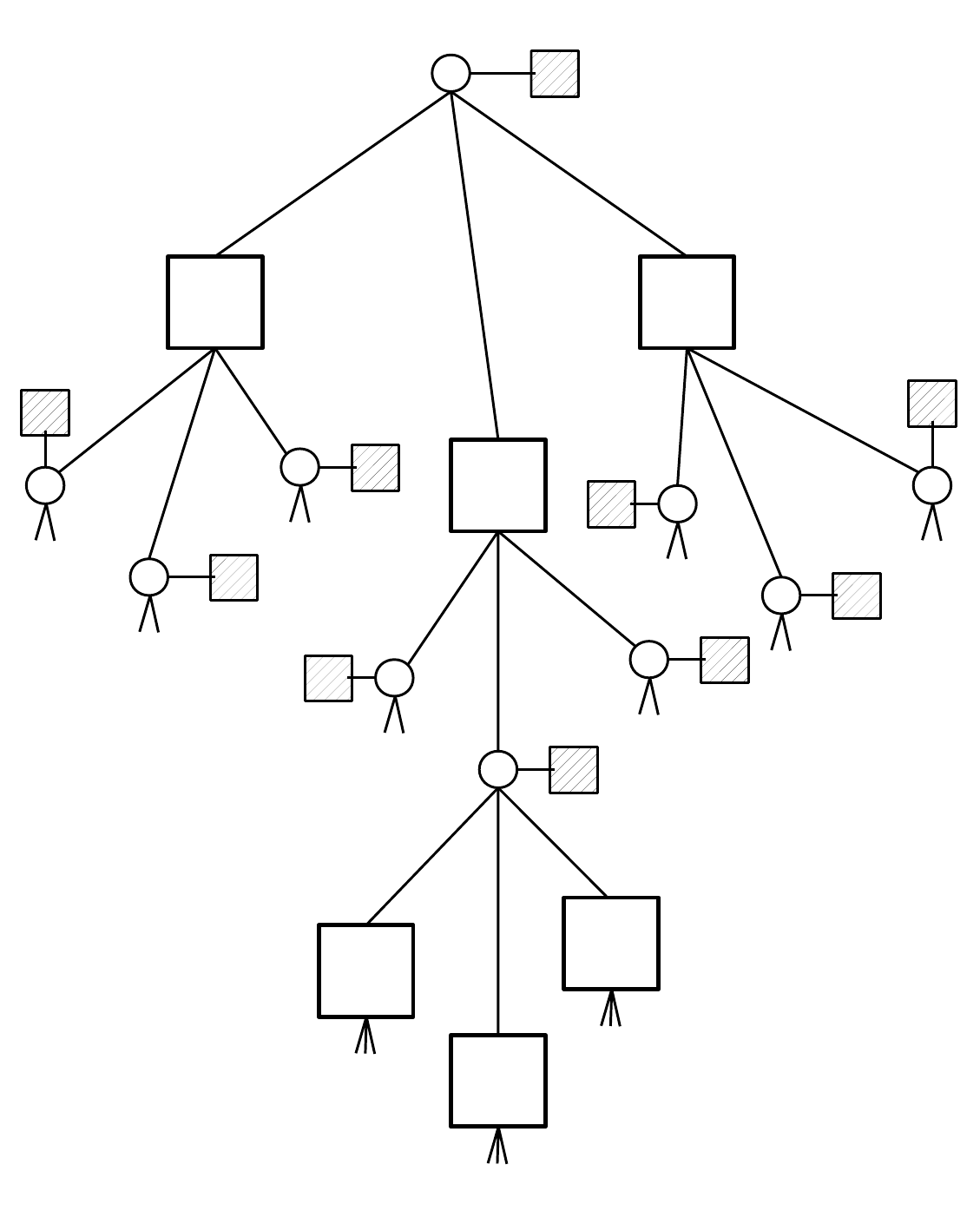}}
 \put(23,70){$b_\mu$} 
 \put(30,44){$i$}
 \put(26,27){$b_\nu$}
 \put(30,8){$j$}
 \end{picture}
 }\end{center}
 \vspace*{-6mm}
 \caption{Interaction topology of a replicated system of  spins on a factor-tree $\mathcal{T}_\mu$, rooted at variable-node $\mu$. Spins are represented by circular ``variable'' nodes, and each term in the  Boltzmann factor  $\exp[ -\frac{1}{2}\rho n \beta \sum_{\nu\in\mathcal{T}_\mu} \!(b-\!J_{\nu}\theta_{\nu}/\rho)^2 + n\sum_{i\in\mathcal{T}_\mu}\log 2\cosh( \beta\sum_{\nu\in\partial i} J_{\nu}\xi_i^{\nu} b_\nu)]=  \{\prod_{\nu\in\mathcal{T}_\mu} f(b_\nu) \}\{\prod_{i\in\mathcal{T}_\mu} F_i(\{b_\nu\!\!: \nu\in\partial i\})\}$  corresponds to a small filled square ``factor'' node (when representing $f(b_\nu)$) or large unfilled one (when representing $F_i$). A link between variable node $\nu$ and factor node $i$  implies that $b_\nu$ acts as an argument of $F_i$.}
 \label{figure:factor-tree-real-n} 
 \end{figure}

Let us assume that the distribution (\ref{eq:Prob(b)-real-n}),  as given by   
\begin{eqnarray}
P(\bb)&=& \frac{1}{Z} \rme^{   -\frac{1}{2}\rho n \beta \sum_{\mu=1}^{M}(b_\mu -\frac{J_\mu \theta_\mu}{\rho})^2 +n\sum_{i=1}^{N}\log2 \cosh( \beta\sum_{\mu\in\partial i} b_\mu J_\mu \xi_i^{\mu}) }  \label{eq:Prob(b)-real-n-Ising}
\end{eqnarray}
when adopted to the  Ising case $\sigma_i\in\{-1,1\}$, is defined on a factor-tree $\mathcal{T}_\mu(r)$ of radius $r$ rooted at variable-node $\mu$  (see Figure \ref{figure:factor-tree-real-n}). We note that  all equations  in this section  are derived for  $\sigma_i\in\{-1,1\}$, but they can be used trivially also for  $\sigma_i\in\{0,1\}$, by  making the transformations  $2\cosh(x)\rightarrow1+\rme^{x}$ and $2\sinh(x)\rightarrow\rme^{x}$.
To compute the marginal distribution
\begin{eqnarray}
P_\mu(b) &=& \int\!\rmd\bb~ P(\bb)\,\delta\!\left(b- b_\mu\right)
\end{eqnarray}
we consider  the partition function 
\begin{eqnarray}
Z_\mu(b)&=&\Bigg\{\prod_{\nu\in\mathcal{T}_\mu}\int\!\rmd b_\nu \Bigg\} \delta\left(b-b_\mu\right) \rme^{ -\frac{1}{2}\rho n \beta \sum_{\nu\in\mathcal{T}_\mu}  \left(b_\nu-\frac{J_{\nu}}{\rho}\theta_{\nu}\right)^2} \nonumber\\
&&\times \Bigg\{\prod_{i\in\mathcal{T}_\mu} 2\cosh\Big( \beta\sum_{\nu\in\partial i}J_{\nu} \xi_i^{\nu}b_\nu \Big)\Bigg\}^n\nonumber\\
&=&\rme^{ -\frac{1}{2}\rho n \beta \left(b-\frac{J_{\mu}}{\rho}\theta_{\mu}\right)^2}\Bigg\{\prod_{\nu\in\mathcal{T}_\mu}\int\!\rmd b_\nu \Bigg\}\delta\left(b-b_\mu\right)\nonumber\\
&&
\times \Bigg\{\prod_{i\in\partial\mu} 2\cosh \Big( \beta\sum_{\nu\in\partial i} J_{\nu}\xi_i^{\nu} b_\nu \Big)\Bigg\}^{\!n}
\!\rme^{ -\frac{1}{2}\rho n \beta \sum_{\nu\in\mathcal{T}_\mu\setminus\mu}\left(b_\nu-\frac{J_{\nu}}{\rho}\theta_{\nu}\right)^2 } \nonumber\\
&&\times\Bigg\{\prod_{i\in\mathcal{T}_\mu\setminus\partial\mu} 2\cosh \left( \beta\sum_{\nu\in\partial i} J_{\nu}\xi_i^{\nu} b_\nu \right)\Bigg\}^{\!n}\nonumber\\
&=&
\rme^{ -\frac{1}{2}\rho n \beta \left(b-\frac{J_{\mu}}{\rho}\theta_{\mu}\right)^2}\Bigg\{\prod_{\nu\in\mathcal{T}_\mu}\int\!\rmd b_\nu \Bigg\}\delta\left(b-b_\mu\right)\nonumber\\
&&
\times\Bigg\{\prod_{i\in\partial\mu} 2\cosh \Big( \beta\sum_{\nu\in\partial i} J_{\nu}\xi_i^{\nu} b_\nu \Big)\Bigg\}^{\!n}\nonumber\\
&&\times\prod_{i\in\partial\mu} \prod_{\nu\in\partial i\setminus\mu} \Bigg[\rme^{ -\frac{1}{2}\rho n \beta \sum_{\hat{\nu}\in\mathcal{T}_\nu}\left(b_{\hat\nu}-\frac{J_{\hat\nu}}{\rho}\theta_{\hat\nu}\right)^2 }\nonumber\\
&&\times \Bigg\{\prod_{j\in\mathcal{T}_\nu } 2\cosh \Big( \beta\sum_{\hat{\nu}\in\partial j}  J_{\hat{\nu}}   \xi_j^{\hat{\nu}}  b_{\hat{\nu}}\Big)\Bigg\}^n\Bigg]\nonumber\\
&=&\rme^{ -\frac{1}{2}\rho n \beta \left(b-\frac{J_{\mu}}{\rho}\theta_{\mu}\right)^2}\Bigg\{ \prod_{i\in\partial\mu} \prod_{\nu\in\partial i\setminus\mu} \int\!\rmd b_\nu~ Z_{\nu i}[b_\nu]\Bigg\} \nonumber\\
&&
\times\int\!\rmd b_\mu \Bigg\{\prod_{i\in\partial\mu}2\cosh \Big( \beta\sum_{\nu\in\partial i} J_{\nu} \xi_i^{\nu} b_\nu \Big)\Bigg\}^{\!n} \delta\left(b-b_\mu\right)
\label{eq:Z-real-n}
\end{eqnarray}
 which subsequently gives us the marginal $P_\mu(b)$ via the equation
\begin{eqnarray}
P_\mu(b)&=&  \frac{Z_\mu(b)}{\int_{-\infty}^{\infty}\rmd\tilde{b}\,Z_\mu(\tilde{b})}\label{def:P(b)-real-n}\\
&=& \Bigg\{ \prod_{i\in\partial\mu} \prod_{\nu\in\partial i\setminus\mu} \int\!\rmd b_\nu~ Z_{\nu i}[b_\nu]\Bigg\}    \rme^{ -\frac{1}{2}\rho n \beta (b-\frac{J_\mu}{\rho} \theta_\mu  )^2}  \nonumber\\
&&
\times   \int\! \rmd b_\mu \Bigg\{\prod_{i\in\partial\mu}2\cosh \Big( \beta\sum_{\nu\in\partial i}J_\nu \xi_i^{\nu}b_\nu \Big)\Bigg\}^{\!n} \delta\left(b-b_\mu\right)\nonumber\\
&&
\times\Bigg[\Bigg\{ \prod_{i\in\partial\mu} \prod_{\nu\in\partial i\setminus\mu} \int\!\rmd \tilde{b}_\nu~ Z_{\nu i}[\tilde{b}_\nu]\Bigg\}  \int\!\rmd \tilde{b}_\mu ~\rme^{ -\frac{1}{2}\rho n \beta (\tilde{b}_\mu  - \frac{J_\mu}{\rho} \theta_\mu   ) ^2}\nonumber\\
&&\times \Bigg\{\prod_{i\in\partial\mu}2\cosh \Big( \beta\sum_{\nu\in\partial i}J_\nu \xi_i^{\nu}\tilde{b}_\nu \Big)\Bigg\}^{\!n} ~\Bigg]^{-1}\nonumber 
\end{eqnarray}
This immediately gives  us the equation 
\begin{eqnarray}
P_{\mu}(b)&=& 
\Bigg\{\!\prod_{i\in\partial \mu}\prod_{\nu\in\partial i\setminus\mu}  \int_{-\infty}^{\infty} \!P_{i\nu} [b_\nu] \rmd b_\nu\!\Bigg\} \rme^{ -\frac{1}{2}\rho n \beta \left(b-\frac{J_\mu}{\rho}\theta_\mu\right)^2}
\label{eq:P(b)-real-n}\\
&&~\times \Bigg\{\!\prod_{i\in\partial \mu} \!2\cosh \Big( \beta\sum_{\nu\in\partial i\setminus\mu}   J_{\nu}\xi_i^{\nu} b_\nu +\beta J_{\mu}\xi_i^{\mu}b \Big)\!\Bigg\}^n\nonumber\\
 &&~\times\Bigg[ 
 \Bigg\{\!\prod_{i\in\partial \mu}\prod_{\nu\in\partial i\setminus\mu}  \int_{-\infty}^{\infty} P_{i\nu} [\tilde{b}_\nu] \rmd \tilde{b}_\nu\!\Bigg\} \int_{-\infty}^{\infty}\rmd\tilde{b} \, \rme^{ -\frac{1}{2}\rho n \beta \left(\tilde b-\frac{J_\mu}{\rho}\theta_\mu\right)^2}\nonumber\\
&&~~\times \Bigg\{\!\prod_{i\in\partial \mu} 2\cosh \Big( \beta\sum_{\nu\in\partial i\setminus\mu}   J_{\nu}\xi_i^{\nu} \tilde{b}_\nu  +\beta J_{\mu} \xi_i^{\mu}\tilde{b}  \Big)\!\Bigg\}^n\Bigg]^{-1}\nonumber, 
\end{eqnarray}
if we insert the definition
\begin{eqnarray}
P_{\nu i}[b]&=&  \frac{Z_{\nu i}[b]}{ \int_{-\infty}^\infty\!  \rmd \tilde{b}  ~ Z_{\nu i}[\tilde{b}]      }\label{def:P-cavity-real-n}.
\end{eqnarray}

In order to derive an equation for the distribution $P_{\nu i}[b]$ we compute the partition function,  associated  with the factor-tree $\mathcal{T}_\nu$ of radius $r\!-\!1$ (see Figure \ref{figure:factor-tree-real-n}), as follows
\begin{eqnarray}
Z_{\nu i}[b]&=&\Bigg\{\prod_{{  \hat{\nu}    }\in\mathcal{T}_\nu}\int_{-\infty}^{\infty}\rmd b_{\hat{\nu}} \Bigg\} \delta\left(b-b_\nu\right)  
\nonumber\\
&&\times \rme^{ -\frac{1}{2}\rho n \beta \sum_{\hat{\nu}\in\mathcal{T}_\nu}   (b_{\hat\nu} - \frac{J_{\hat\nu}}{\rho} \theta_{\hat\nu})^2 + n\sum_{j\in\mathcal{T}_\nu}\log 2\cosh \left( \beta\sum_{\hat{\nu}\in\partial j} J_{\hat{\nu}}\xi_j^{\hat{\nu}} b_{\hat{\nu}}    \right)}\nonumber\\
&=& \Bigg\{\prod_{\hat{\nu}\in\mathcal{T}_\nu}\int_{-\infty}^{\infty}\rmd b_{\hat{\nu}} \Bigg\} \rme^{ -\frac{1}{2}\rho n \beta (b-\frac{J_{\nu}}{\rho} \theta_{\nu} )^2}\delta\left(b-b_\nu\right)\nonumber\\
&&\times\Bigg\{ \prod_{j\in\partial\nu\setminus i}2\cosh \Big( \beta\sum_{   \hat{\nu}    \in\partial j} J_{ \hat{\nu} }\xi_j^{ \hat{\nu} }b_{ \hat{\nu} } \Big)\Bigg\}^{\!n}
  \nonumber\\
&&\times   \prod_{j\in\partial\nu\setminus i}   \prod_{\hat{\nu}     \in\partial j\setminus \nu} \Bigg[\rme^{ -\frac{1}{2}\rho n \beta \sum_{v\in\mathcal{T}_{ \hat{\nu}}}   (b_v - \frac{J_{v}}{\rho} \theta_{v})^2 } \nonumber\\
&&\times\Bigg\{\prod_{\ell\in\mathcal{T}_{ \hat{\nu}}}\log 2\cosh \Big( \beta\sum_{v\in\partial \ell} J_v \xi_\ell^{v}  b_v      \Big)\Bigg\}^{\!n}~\Bigg]\nonumber\\
&=& \Bigg\{    \prod_{j\in\partial\nu\setminus i}   \prod_{\hat{\nu}     \in\partial j\setminus \nu}           \int\!\rmd b_{\hat{\nu}}~ Z_{\hat{\nu} j}[b_{\hat{\nu}}]  \Bigg\}   \rme^{ -\frac{1}{2}\rho n \beta (b-\frac{J_{\nu}}{\rho} \theta_{\nu} )^2}        \nonumber\\
&&\times
\Bigg\{ \prod_{j\in\partial\nu\setminus i}2\cosh \Big( \beta\sum_{   \hat{\nu}    \in\partial j\setminus\nu} J_{ \hat{\nu} }\xi_j^{ \hat{\nu} } b_{ \hat{\nu} }  + \beta J_{ \nu}\xi_j^{ \nu} b\Big)\Bigg\}^{\!n}
\end{eqnarray}
Using this expression in definition  (\ref{def:P-cavity-real-n}) gives us  the recursive equation
\begin{eqnarray}
P_{\nu i}[b]&=&   \Bigg\{    \prod_{j\in\partial\nu\setminus i}   \prod_{\hat{\nu}     \in\partial j\setminus \nu}           \int_{-\infty}^{\infty} P_{\hat{\nu} j}[b_{\hat{\nu}}]\rmd b_{\hat{\nu}} \Bigg\}   \rme^{ -\frac{1}{2}\rho n \beta \left(b-\frac{J_\nu}{\rho}\theta_\nu\right)^2}        \label{eq:P-cavity-real-n}    \\
&&\times
\Bigg\{ \prod_{j\in\partial\nu\setminus i}2\cosh \left( \beta \sum_{   \hat{\nu}    \in\partial j\setminus\nu} J_{ \hat{\nu} }\xi_j^{ \hat{\nu} } b_{ \hat{\nu} } + \beta J_{ \nu}\xi_j^{ \nu}b\right)\Bigg\}^n \nonumber\\
&&\times\Bigg[ \Bigg\{    \prod_{j\in\partial\nu\setminus i}   \prod_{\hat{\nu}     \in\partial j\setminus \nu}           \int_{-\infty}^{\infty}P_{\hat{\nu} j}[\tilde{b}_{\hat{\nu}}] \rmd \tilde{b}_{\hat{\nu}} \Bigg\}    \int_{-\infty}^{\infty}\rmd\tilde{b}\,\rme^{ -\frac{1}{2}\rho n \beta \left(\tilde b-\frac{J_\nu}{\rho}\theta_\nu\right)^2}\nonumber    \\
&&\times
\Bigg\{ \prod_{j\in\partial\nu\setminus i}2\cosh \left( \beta\sum_{   \hat{\nu}    \in\partial j\setminus\nu}  J_{ \hat{\nu} } \xi_j^{ \hat{\nu} } b_{ \hat{\nu} }  + \beta J_{ \nu}\xi_j^{ \nu}\tilde{b}\right)\Bigg\}^n \Bigg]^{-1}.\nonumber
\end{eqnarray}
For $n\in \mathbb{Z}^{+}$, we can use the above equation,  together with the identity $2^n\cosh^n(x)=\sum_{\sigma^1,\ldots, \sigma^n}\rme^{x\sum_{\alpha=1}^n \sigma^\alpha}$, and the definition
\begin{eqnarray}
\hspace*{-5mm}
 P_{ i\mu}[\bsigma_i]&=&\frac{  \left\{\prod_{\nu\in\partial i\setminus\mu}  \int_{-\infty}^{\infty} \!P_{i\nu} [b_\nu] \rmd b_\nu\right\} \rme^{ \beta \sum_{   \nu    \in\partial i\setminus\mu}J_{ \nu }\xi_i^{ \nu } b_{ \nu }\sum_{\alpha=1}^n \sigma_i^\alpha}    }{ \left\{\!\prod_{\nu\in\partial i\setminus\mu}\!  \int_{-\infty}^{\infty} \!P_{i\nu} [\tilde{b}_\nu] \rmd \tilde{b}_\nu\!\right\} \!\left[2\cosh\!\left(\!\beta \sum_{   \nu    \in\partial i\setminus\mu}\!J_{ \nu }\xi_i^{ \nu } \tilde{b}_{ \nu }       \right)\!\right]^n}, ~~~~
 \end{eqnarray}
  to recover equation (\ref{eq:P-cavity-spin-integer-n}) and all other equations of $n\in \mathbb{Z}^{+}$ which involve the cavity distribution $P_{ i\mu}[\bsigma_i]$. 

Furthermore, the above approach  can be used to compute any marginal of (\ref{eq:Prob(b)-real-n-Ising}). In particular, the joint distribution of the variables in the set $\{b_\mu : \mu\in\partial i\}$ is derived  by considering the distribution (\ref{eq:Prob(b)-real-n-Ising}) defined on a factor-tree $\mathcal{T}_i$ rooted at factor-node $i$:
\begin{eqnarray}
\hspace*{-5mm}
P_i(\{b_\mu\!\!: \mu\in\partial i\})
&=&\frac{ \left\{ \prod_{  \mu\in\partial i} P_{\mu i}[b_\mu ] \right\}   \cosh^n\! \big( \beta\sum_{   \mu   \in\partial i} b_{ \mu} \xi_i^{ \mu }\big)}
{  \left\{ \prod_{  \mu\in\partial i} \int\!\rmd  \tilde{b}_\mu~ P_{\mu i}[\tilde{b}_\mu ] \right\}    \cosh^n\! \big( \beta\sum_{   \mu   \in\partial i} \tilde{b}_{ \mu} \xi_i^{ \mu }\big)        }
 \label{eq:Pi-real-n}.
\end{eqnarray}
This is then used to compute the local magnetization $\langle\sigma_i\rangle=\sum_\sigma P_i(\sigma)\sigma$ from  the distribution (\ref{eq:Pi(s)-real-n}) which  gives us the equation 
\begin{eqnarray}
\langle\sigma_i\rangle&=& \Bigg\{ \prod_{  \mu\in\partial i} \int_{-\infty}^{\infty}P_{\mu i}[b_\mu ]\rmd  b_\mu \Bigg\} %\rme^{ -\frac{1}{2}\rho n \beta \sum_{\mu\in\partial i}\left(b_\mu  -\frac{J_\mu}{\rho}\theta_\mu \right)^2}
 \cosh^n\big( \beta\sum_{   \mu   \in\partial i} J_{ \mu } \xi_i^{ \mu }b_{ \mu} \big)\nonumber\\
 &&\times \tanh\big( \beta\sum_{   \mu   \in\partial i}J_{ \mu } \xi_i^{ \mu }b_{ \mu} \big)
 \nonumber\\[-1mm]
 &&\times\Bigg[\Bigg\{ \prod_{  \mu\in\partial i} \int_{-\infty}^{\infty}P_{\mu i}[\tilde{b}_\mu ]\rmd  \tilde{b}_\mu \Bigg\}    % \rme^{ -\frac{1}{2}\rho n \beta \sum_{\mu\in\partial i}\left(\tilde{b}_\mu  -\frac{J_\mu}{\rho}\theta_\mu \right)^2} 
   \cosh^n\big( \beta\sum_{   \mu   \in\partial i} J_{ \mu } \xi_i^{ \mu }   \tilde{b}_{ \mu}        \big)\Bigg]^{-1}.\label{eq:m-real-n}
\end{eqnarray}

Finally, to compute the distribution of fields (\ref{def:P(F)-real-n}) we need to know the joint distribution $P_\mu (b, \{b_\nu\})$ of the variable $b=b_\mu$ and of all its immediate neighbours $b_\nu$ (see Figure \ref{figure:factor-tree-real-n}). This distribution can be ``read off'' from equation (\ref{eq:P(b)-real-n}), using the identity  $P_\mu (b)=  \big\{ \int_{-\infty}^{\infty} \rmd  b_\nu\big\}   P_\mu (b, \{b_\nu\})$, which gives 
\begin{eqnarray}
P_{\mu}(F)&=& 
\Bigg\{\!\prod_{i\in\partial \mu}\prod_{\nu\in\partial i\setminus\mu}  \int_{-\infty}^{\infty}\!\!\rmd b_\nu~ P_{i\nu} [b_\nu] \!\Bigg\} \int_{-\infty}^{\infty}\!\rmd b_\mu\rme^{ -\frac{1}{2}\rho n \beta \left(b_\mu-\frac{J_\mu}{\rho}\theta_\mu\right)^2}\label{eq:P(F)-real-n}\\
&&~\times \prod_{i\in\partial \mu} \cosh^n \big( \beta\sum_{\nu\in\partial i}   J_{\nu}\xi_i^{\nu} b_\nu  \big)\nonumber\\
&&\times  
 \sum_{\{\sigma_i\}}    \frac{   \rme^{ \beta \sum_{i\in\partial \mu}   \sigma_i\sum_{\nu\in\partial i} b_\nu J_\nu \xi_i^{\nu} }    }{ \prod_{i\in\partial \mu} 
2\cosh\left(\beta\sum_{\nu\in\partial i} b_\nu J_\nu \xi_i^{\nu} \right)
  } \delta\left(F-F_\mu(\bsigma)\right)
\nonumber\\
&&\times\Bigg[
 \Bigg\{\!\prod_{i\in\partial \mu}\prod_{\nu\in\partial i\setminus\mu}  \int_{-\infty}^{\infty}\!\!\rmd \tilde{b}_\nu~ P_{i\nu} [\tilde{b}_\nu] \!\Bigg\} \int_{-\infty}^{\infty}\rmd\tilde{b}_\mu \, \rme^{ -\frac{1}{2}\rho n \beta \left(\tilde b_\mu-\frac{J_\mu}{\rho}\theta_\mu\right)^2}\nonumber\\
&&~~\times \prod_{i\in\partial \mu} \cosh^n \big( \beta\sum_{\nu\in\partial i}   J_{\nu}\xi_i^{\nu} \tilde{b}_\nu    \big) \Bigg]^{-1} .\nonumber
\end{eqnarray}
Equation (\ref{eq:P-cavity-real-n}) preserves the symmetry $P_{\nu i}[b]=P_{\nu i}[-b]$ when $\theta_\mu=0$. This implies that for such $P_{\nu i}[b]$ the marginal distribution (\ref{eq:P(b)-real-n}) and the distribution of fields (\ref{eq:P(F)-real-n})  are both symmetric  functions, and that the local magnetization (\ref{eq:m-real-n}) is zero.

\section{Analysis  of $n=1$ case \label{app:n=1}}
\begin{figure}[t]
%\vspace*{5mm} \hspace*{35mm} 
\setlength{\unitlength}{0.58mm}
\hspace*{10mm}
\begin{picture}(150,140)
\put(0,0){\includegraphics[height=145\unitlength,width=145\unitlength]{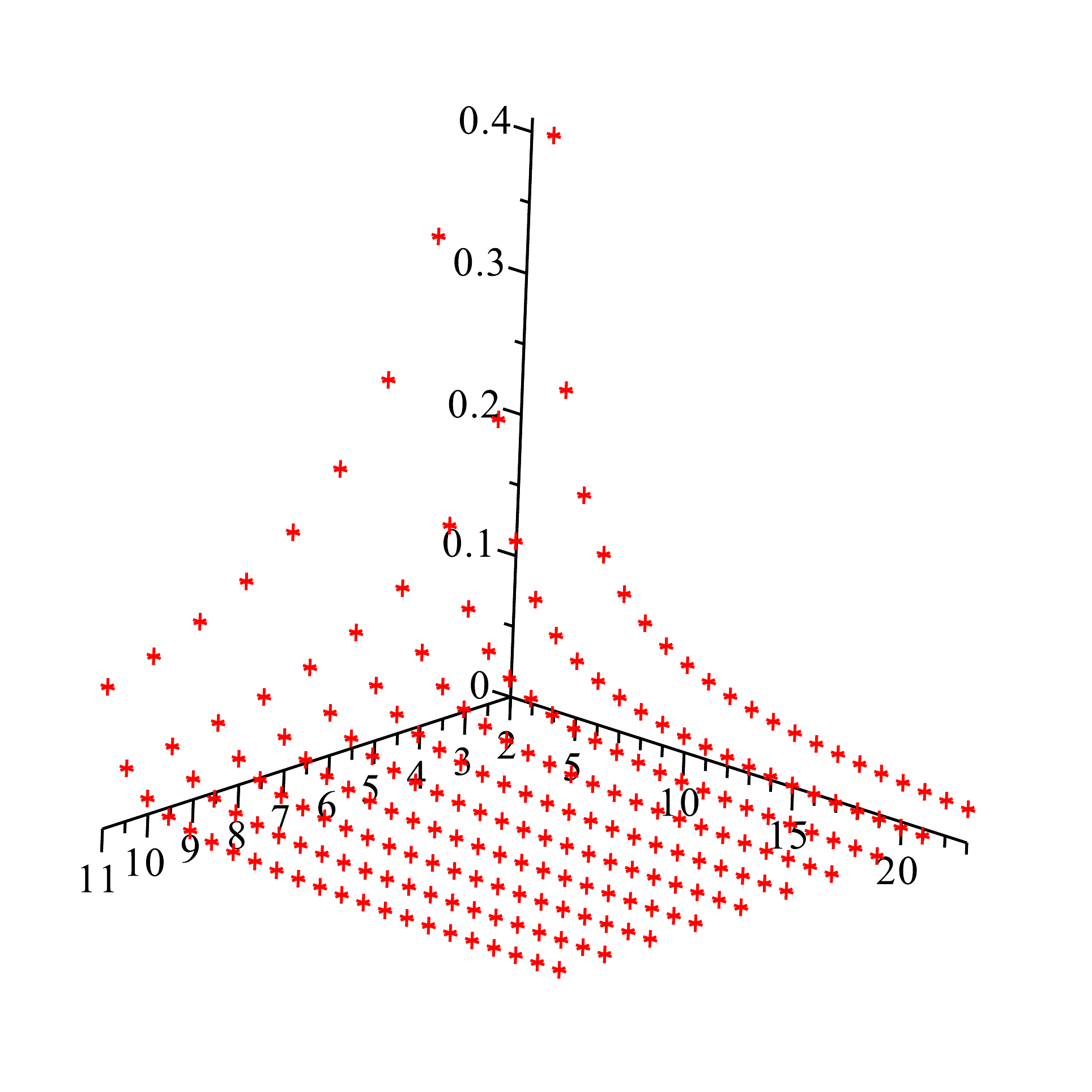}} %,width=145\unitlength
\put(50,120){\small{$\beta_{c}$}} %\small
\put(3,40){\small{$L$}} \put(100,55){\small{$K$}} 
\put(50,10){\small{$m=0$}} \put(80,75){\small{$m\neq0$}} 
\end{picture}
\vspace*{-3mm} 
\caption{Phase diagram of ferromagnetic Ising model  on random $(K, L)$-regular clique graphs.  We plot the critical inverse temperature $\beta_{c}$ as a function of the vertex degree $L$ and the factor degree $K$. The system is in the ferromagnetic $ m\neq0$ phase for $\beta>\beta_c$, and in the paramagnetic $m=0$ phase for $\beta<\beta_c$.} \label{figure:PDn1}
\end{figure}

For $n=1$,  equation  (\ref{eq:P[h]-ferro}) can be simplified further by noticing that here $h\in\{-1,1\}$ and the distribution $ P[h]$, with $h=\sigma$, can be written in the form
\begin{eqnarray}
  P[\sigma]&=&\frac{\rme^{\phi\sigma}}{2\cosh(\phi)}\label{def:phi_n1} .
\end{eqnarray}
where $\phi$ is a ``cavity field'' parameter.  Using 
this  in  (\ref{eq:P[h]-ferro}) gives us the equation 
\begin{eqnarray}
  P[\sigma]&=&\frac{\Big[\sum_{\{\sigma_j\}} \rme^{   \frac{1}{2} \frac{\beta J^2}{\rho} \left(\sum_{j=1}^{K-1} \sigma_j +\sigma\right)^2 + \phi\sum_{j=1}^{K-1} \sigma_j}\Big]^{L-1}} {
 \sum_{\tilde{\sigma}}   \Big[\sum_{\{\tilde{\sigma}_j\}}\rme^{ \frac{1}{2} \frac{\beta J^2}{\rho} \left(\sum_{j=1}^{K-1} \tilde{\sigma}_j +\tilde{\sigma} \right)^2 + \phi\sum_{j=1}^{K-1} \tilde{\sigma}_j  }\Big]^{L-1}                                   }\label{eq:P[h]-ferro-n-1} 
\end{eqnarray}
from which, via the identity  $ \phi=\frac{1}{2} \log( P[+1]/P[-1])$, 
we can derive 
 \begin{eqnarray}
  \phi=\frac{L-1}{2} \log  \left(  
  \frac{  \sum_{\ell=0}^{K-1}  {{K-1}\choose{\ell}} \rme^{  \frac{1}{2}\frac{\beta J^2}{\rho}(2\ell-K+2)^2             +    \phi(2\ell-K+1) }      }                    {                    \sum_{\tilde\ell=0}^{K-1}
   %\binom{K-1}{\tilde\ell}
  {{K-1}\choose{\tilde\ell}}
  \rme^{ \frac{1}{2}\frac{\beta J^2}{\rho}  (2\tilde\ell-K)^2 + \phi(2\tilde\ell-K+1)  }  }                                                          \right)\label{eq:phi}.
 \end{eqnarray}
The solution $\phi$ of the above equation can be used to compute the marginal distribution   
\begin{eqnarray}
  P(\sigma) &=&  \frac{\Big[\sum_{\{\sigma_j\}} \rme^{ \frac{1}{2}\frac{\beta J^2}{\rho} \left(\sum_{j=1}^{K-1} \sigma_j +\sigma\right)^2 + \phi\sum_{j=1}^{K-1} \sigma_j}\Big]^{L}} {
 \sum_{\tilde{\sigma}}   \Big[\sum_{\{\tilde{\sigma}_j\}}\rme^{ \frac{1}{2}\frac{\beta J^2}{\rho} \left(\sum_{j=1}^{K-1} \tilde{\sigma}_j +\tilde{\sigma} \right)^2 + \phi\sum_{j=1}^{K-1} \tilde{\sigma}_j  }\Big]^{L}                                   }\label{eq:P(s)-ferro-n-1} .
\end{eqnarray}
Comparing this with the cavity distribution (\ref{eq:P[h]-ferro-n-1}) gives us the magnetization formula 
\begin{eqnarray}
 m=\sum_{\sigma}P(\sigma)\sigma=\tanh\Big(\frac{L\phi}{L\!-\!1}\Big).\label{eq:m-ferro-n-1}
\end{eqnarray}
Also, using definition (\ref{def:phi}) in the distribution of fields (\ref{eq:P(F)-ferro}) gives us the equation
\begin{eqnarray}
  P(F)  &=&    \frac{ \sum_{\{\sigma_j\}}
    \rme^{ \frac{1}{2}  \frac{\beta }{\rho}  F^2  + \phi \sum_{j=1}^{K} \sigma_j }  \delta\big(F\!-\!J\sum_{j=1}^{K} \sigma_j\big)
 % \frac{F}{J} 
 }{
    \sum_{\{\tilde{\sigma}_j\}}  \rme^{   \frac{1}{2} \frac{\beta J^2}{\rho}(\sum_{j=1}^{K} \tilde{\sigma}_j)^2  + \phi \sum_{j=1}^{K} \tilde{\sigma}_j }}
   \label{eq:P(F)-ferro-n-1}.
\end{eqnarray}
Upon defining the RHS of (\ref{eq:phi}) as $f(\phi)$, 
this equation takes the form $\phi= f(\phi)$. The value $\phi=0$, which corresponds to the paramagnetic (PM)  $m=0$ phase, is always a solution.  However, it becomes unstable at the point where $f^\prime(0)=1$, in which 
\begin{eqnarray}
\hspace*{-10mm}
 f^\prime(0)&=&(L\!-\!1)(K\!-\!1)\frac{\sum_{\{\sigma_j\}} \rme^{   \frac{1}{2} \frac{\beta J^2}{\rho} \left(\sum_{j=1}^{K-1} \sigma_j \right)^2 } \!\sinh\Big(\frac{\beta J^2}{\rho}\sum_{j=1}^{K-1} \sigma_j \Big)\sigma_1} {
 \sum_{\{\tilde{\sigma}_j\}}\rme^{ \frac{1}{2} \frac{\beta J^2}{\rho} \left(\sum_{j=1}^{K-1} \tilde{\sigma}_j \right)^2 }  \! \cosh\Big(\frac{\beta J^2}{\rho}\sum_{j=1}^{K-1} \tilde{\sigma}_j \Big)                              },\label{eq:df0} 
\end{eqnarray}
For $f^\prime(0)>1$  equation (\ref{eq:phi}) has two stable solutions $\phi\neq0$, which correspond to the ferromagnetic (FM) $m\neq0$ phase. Solving the equation $f^\prime(0)=1$ gives us the critical inverse temperature $\beta_c$ ($J=\rho=1$) where the PM to FM transition occurs (see Figure \ref{figure:PDn1}). 
We note that  for $L>2$ and $K=2$  the equation (\ref{eq:df0})  recovers the result $\beta_c=\tanh^{-1}\left(1/(L\!-\!1)\right)$ of ferromagnetic Ising models on Bethe lattices~\cite{Mezard2001}. 

\section{Homogeneous systems on random regular factor-graphs, with$n\in \mathbb{R}^{+}$\label{app:ferro-n-real}}
The equations derived in  \ref{app:BP-real-n} can be easily adopted to study the system (\ref{eq:Prob(b)-real-n-Ising}), which is homogeneous and defined on a random regular factor-graph (see section \ref{section:ferro-n-integer} for details). Here the marginal distribution $P(b)=P_\mu(b)$   is given by
\begin{eqnarray}
\hspace*{-5mm}
P(b)
%&=& 
%\left\{\!\prod_{j=1}^{K}\prod_{\nu=1}^{L-1}  \int_{-\infty}^{\infty} \!P [b_{j\nu}] \rmd b_{j\nu}\!\right\}\label{eq:P(b)-ferro-real-n}\\
%&&~\times  \rme^{ -\frac{1}{2}\rho n \beta b^2}  \left\{\!\prod_{j=1}^{K} \cosh \left( \beta J\sum_{\nu=1}^{L-1} b_{j\nu}  +\beta J b \right)\!\right\}^n\nonumber\\
% &&~\times\Bigg[ 
%\left\{\!\prod_{j=1}^{K}\prod_{\nu=1}^{L-1}  \int_{-\infty}^{\infty} \!P [\tilde{b}_{j\nu}] \rmd \tilde{b}_{j\nu}\!\right\} \int_{-\infty}^{\infty}\rmd\tilde{b} \,\rme^{ -\frac{1}{2}\rho n \beta \tilde{b}^2}\nonumber\\
%&&~~\times \left\{\!\prod_{j=1}^{K} \cosh \left( \beta J\sum_{\nu=1}^{L-1} \tilde{b}_{j\nu}  +\beta J\tilde{b} \right)\!\right\}^n\Bigg]^{-1}\nonumber\\
%
&=& 
 \rme^{ -\frac{1}{2}\rho n \beta b^2} \Bigg[\prod_{\nu=1}^{L-1}  \int\!\rmd b_{\nu} 
  ~P [b_{\nu}]  \cosh^n\!\Big( \beta J \Big(\sum_{\nu=1}^{L-1} b_{\nu}  + b \Big)\Big) \Bigg]^{\!K}\label{eq:P(b)-ferro-real-n}\\
%&&~\times   \left\{\!\prod_{j=1}^{K} \cosh \left( \beta J\sum_{\nu=1}^{L-1} b_{j\nu}  +\beta J b \right)\!\right\}^n\nonumber\\
 &&\times
 \Bigg\{\!
 \int\!\rmd\tilde{b}\,\rme^{ -\frac{1}{2}\rho n \beta \tilde{b}^2}\!
 \Bigg[\prod_{\nu=1}^{L-1}\!  \int\!\rmd \tilde{b}_{\nu}~P [\tilde{b}_{\nu}]  \cosh^n\!\Big(\beta J\Big( \sum_{\nu=1}^{L-1} \!\tilde{b}_{\nu} \! +\!\tilde{b} \Big)\Big)\! \Bigg]^{\!K}\Bigg\}^{\!-1}\nonumber
\end{eqnarray}
and the cavity distribution $P[\,b\,]=P_{\nu i}[\,b\,]$ can be computed recursively via
%
%\begin{eqnarray}
%P[\,b\,]&=&   \left\{    \prod_{j=1}^{K-1}   \prod_{\nu=1}^{L-1}           \int_{-\infty}^{\infty} P[b_{j\nu}]\rmd b_{j\nu} \right\}   \label{eq:P-ferro-cavity-real-n}    \\
%&&\times\rme^{ -\frac{1}{2}\rho n \beta b^2}
%\left\{ \prod_{j=1}^{K-1}\cosh \left( \beta J\sum_{   \nu=1}^{L-1} b_{ j\nu }  + \beta J b\right)\right\}^n \nonumber\\
%%
%&&\times \Bigg[\left\{    \prod_{j=1}^{K-1}   \prod_{\nu=1}^{L-1}           \int_{-\infty}^{\infty} P[\tilde{b}_{j\nu}]\rmd \tilde{b}_{j\nu} \right\}   \nonumber    \\
%&&\times \int_{-\infty}^{\infty}\rmd\tilde{b}\,\rme^{ -\frac{1}{2}\rho n \beta \tilde{b}^2}
%\left\{ \prod_{j=1}^{K-1}\cosh \left( \beta J\sum_{   \nu=1}^{L-1} \tilde{b}_{ j\nu }  + \beta J \tilde{b}\right)\right\}^n\Bigg]^{-1} \nonumber
%%
%\end{eqnarray}
%
%
\begin{eqnarray}
\hspace*{-5mm}
P[b]&=& \frac{Z[b]}{\int\!\rmd\tilde{b}~Z[\tilde{b}]}
\label{eq:P-ferro-cavity-real-n}
\\
\hspace*{-5mm}
Z[b]&=&  \rme^{ -\frac{1}{2}\rho n \beta b^2} \Bigg[\Bigg\{      \prod_{\nu=1}^{L-1}           \int\!\rmd b_{\nu}~ P[b_{\nu}] \Bigg\} \cosh^n\!  \Big(\beta J\Big(\sum_{   \nu=1}^{L-1} b_{ \nu }  \!+\! b\Big)\Big)\Bigg]^{K-1}   \end{eqnarray}
We note that $P(b)$ and $P[b]$ are related by the transformation $K\!-\!1\rightarrow K$. 
Finally, once we know the distribution $P[b]$, then we can also compute  the magnetization  
\begin{eqnarray}
\hspace*{-5mm}
\langle\sigma_i\rangle&=& \frac{\left\{ \prod_{  \mu=1}^{L} \int\!\rmd  b_\mu~ P[b_\mu ] \right\} 
   \cosh^n\! \big( \beta J\sum_{   \mu =1}^{L} b_{ \mu}) }{ 
   \left\{ \prod_{  \mu=1}^{L} \int\!\rmd  \tilde{b}_\mu~ P[\tilde{b}_\mu \right\}  \cosh^n\! \big(\beta J\sum_{   \mu =1}^{L} \tilde{b}_{ \mu} \big)  } \tanh \Big(\beta J\sum_{   \mu  =1}^{L} b_{ \mu} \Big) \label{eq:m-ferro-real-n}
\end{eqnarray}
and the distribution of fields 
\begin{eqnarray}
P(F)&=& 
\Bigg\{\!\prod_{j=1}^{K}\prod_{\nu=1}^{L-1}  \int\!\rmd b_{j\nu}~P [b_{j\nu}] \Bigg\}   \int\!\rmd b~\rme^{ -\frac{1}{2}\rho n \beta b^2} 
\nonumber\\
&&~\times  \Bigg\{ \prod_{j=1}^{K} \cosh \Big( \beta J\Big(\sum_{\nu=1}^{L-1} b_{j\nu}  + b\Big) \Big)\Bigg\}^{\!n}
\nonumber\\
&&\times \sum_{\{\sigma_j\}}\frac{       \rme^{\beta J \sum_{j=1}^{K} \sigma_j\left(\sum_{\nu=1}^{L-1} b_{j\nu}  + b \right)}      }{\prod_{j=1}^{K} \cosh \Big( \beta J\Big(\sum_{\nu=1}^{L-1} b_{j\nu}  +b\Big) \Big)}   \,  \delta\Big(F - J  \sum_{j=1}^{K} \sigma_j\Big)  \nonumber\\
 &&\times\Bigg[ 
\Bigg\{
\prod_{j=1}^{K}\prod_{\nu=1}^{L-1}  \int \!\rmd \tilde{b}_{j\nu}~P [\tilde{b}_{j\nu}] 
\Bigg\} 
\int\!\rmd\tilde{b}~\rme^{ -\frac{1}{2}\rho n \beta \tilde{b}^2}
\nonumber\\
&&
\times \Bigg\{\prod_{j=1}^{K} \cosh \Big( \beta J\Big(\sum_{\nu=1}^{L-1} \tilde{b}_{j\nu}  +\tilde{b}\Big) \Big)\Bigg\}^{\!n}\Bigg]^{-1}
 \label{eq:P(F)-ferro-real-n}
\end{eqnarray}

\section*{References}
%\bibliographystyle{unsrt}
%\bibliography{refs}

\end{document}